\documentclass[12pt]{article}
\usepackage{jheppub}
\usepackage{amsmath, amsthm, amssymb}
\usepackage{tikz}
\usepackage{tikz-cd}
\usepackage{simpler-wick} 
\newcommand{\bR}{\mathbb{R}}
\newcommand{\bC}{\mathbb{C}}
\newcommand{\bCP}{\mathbb{CP}}
\newcommand{\Ff}{\mathrm{Ff}}
\newcommand{\Sb}{\mathrm{Sb}}
\newcommand{\cc}{\mathrm{cc}}
\newcommand{\cD}{{\mathcal D}}
\newcommand{\cW}{{\mathcal W}}
\newcommand{\cC}{{\mathcal C}}
\newcommand{\cV}{{\mathcal V}}
\newcommand{\cL}{{\mathcal L}}
\newcommand{\cP}{{\mathcal P}}
\newcommand{\cH}{{\mathcal H}}
\newcommand{\fhs}{\mathfrak{hs}}
\newcommand{\fsl}{\mathfrak{sl}}
\newcommand{\fsu}{\mathfrak{su}}
\newcommand{\fgl}{\mathfrak{gl}}

\newcommand{\fg}{\mathfrak{g}}
\newcommand{\pa}{\partial}

\DeclareMathOperator{\Tr}{Tr}

\def\curved{\tikz[baseline=.1ex]{
		\draw[ ->] (0,0.32) arc (30:330:0.5);}
}

\usetikzlibrary{arrows.meta}

\usetikzlibrary{decorations.markings}

\author[1]{Davide Gaiotto,}
\author[2]{Keyou Zeng}

\affiliation[1]{Perimeter Institute for Theoretical Physics, 31 Caroline Street North, Waterloo, ON N2L
2Y5, Canada}
\affiliation[2]{
Harvard University,
Center for Mathematical Sciences and Applications,
Cambridge, MA 02138}
\emailAdd{dgaiotto@perimeterinstitute.ca}
\emailAdd{kzeng@cmsa.fas.harvard.edu}

\title{Interface Minimal Model Holography and Topological String Theory.}
\abstract{We study the dynamics of 2d fermions coupled to 3d Chern-Simons gauge fields. For $SU(N)$ gauge group and fermions in the fundamental representation, the resulting interfaces are closely related to $\cW_N$ minimal models. We give an holographic description of the interfaces within the A-model Topological String Theory. The model has exotic integrability properties, which allow us to propose an exact holographic match of all sphere correlation functions of meson operators. This construction embeds Minimal Model Holography in String Theory.}
\begin{document}
\maketitle
\section{Introduction}
Minimal Model Holography \cite{Gaberdiel:2010pz,Gaberdiel:2010ar,Gaberdiel:2011wb,Gaberdiel:2011zw,Chang:2011mz,Gaberdiel:2012ku} is a conjectural holographic duality for the $\cW_N$ minimal models, i.e. the family of diagonal Rational CFTs based on the coset chiral algebra
\begin{equation}
	\frac{SU(N)_\kappa \times SU(N)_1}{SU(N)_{\kappa+1}} \, .
\end{equation}
where $SU(N)_k$ denotes the $SU(N)$ WZW current algebra at level $k \in \mathbb{N}$.\footnote{Aka the simple quotient of an $\fsu(N)$ Kac-Moody chiral algebra, which we denote as $\fsu(N)_k$ and is defined for all $k$.}
These CFTs can be analyzed in a 't Hooft expansion, taking $N$ to be large with fixed 't Hooft coupling $t = \frac{N}{\kappa+N}$. The proposed holographic dual is a ``higher spin gauge theory'' in AdS${}_3$, including higher spin gauge fields dual to the chiral- and anti-chiral algebras of the coset model and a scalar field dual to a non-chiral ``meson'' operator. 

The infinite-dimensional chiral algebras known as $\cW_\infty$ algebras play a crucial role in the proposal. These also occur naturally in the context of Twisted Holography for M-theory \cite{Costello:2016nkh}. This suggests the possibility that Minimal Model Holography may be related to a more conventional form of holography involving String Theory. 
Our goal is to show that this is indeed the case and that a dual Topological String Theory description offers both a surprisingly intuitive description of many features of the model and an exact holographic match of chiral OPE and correlation functions.

We find it useful to study a 3d/2d system which is related to the 2d minimal models by a ``topological manipulation'' \cite{Gaiotto:2020iye}, which does not affect the dynamics of the system but alters its topological properties. The 3d/2d system is a weakly-coupled fixed point of an $SU(N)$ gauge theory which admits a transparent 't Hooft expansion. This brings  Minimal Model Holography into the fold of conventional String Theory-based holography. 

The topological manipulation involves two independent steps. One step adds a $U(1)$ current algebra factor to the minimal model
to embed the $SU(N)_1$ denominator factor into a theory $\Ff_\bC^N$ of $N$ complex fermions, giving the coset
\begin{equation}
	M_{N;k} = \frac{SU(N)_\kappa \times \Ff_\bC^N}{SU(N)_{\kappa+1}} \, .
\end{equation}
A more significant step promotes this system to one of two possible non-chiral interfaces for 3d TFTs: $M^-_{N;\kappa}$ for $SU(N)_\kappa$ Chern Simons gauge theory and $M^+_{N;\kappa}$ for $SU(N)_{\kappa+1}$. This ``topological manipulation'' \cite{Gaiotto:2020iye} addresses a problem with the original Minimal Model Holography setup: the existence of a large collection of operators of very small scaling dimension
\begin{equation} \label{eq:light}
	\frac{C_2(R)}{(\kappa+N)(\kappa+N+1)} 
\end{equation}
arising from the coset of $SU(N)_\kappa$ primary fields. 

\begin{figure}[ht]
\centering
\begin{tikzpicture}[
  >=Latex,
  decoration={markings, mark=at position 0.5 with {\arrow{Latex}}},
  dot/.style={circle, fill, inner sep=1.6pt}
]
\node (L) [dot, label=left:{$\dfrac{\overline{SU(N)}_{\kappa}\times \overline{SU(N)}_{1}}{\overline{SU(N)}_{\kappa+1}}$}] at (0,0) {};
\node (R) [dot, label=right:{$\dfrac{SU(N)_{\kappa}\times SU(N)_{1}}{SU(N)_{\kappa+1}}$}] at (6,0) {};

\draw[postaction={decorate}, bend left=45]  (L) to node[midway, above] {$SU(N)_{\kappa}$} (R);
\draw[postaction={decorate}, bend left=-45] (L) to node[midway, above] {$SU(N)_{1}$}      (R);
\draw[postaction={decorate}] (R) -- node[midway, above=1pt] {$SU(N)_{\kappa+1}$} (L);
\end{tikzpicture}

\begin{tikzpicture}[
  >=Latex,
  decoration={markings, mark=at position 0.5 with {\arrow{Latex}}},
  dot/.style={circle, fill, inner sep=1.6pt}
]
\node (L) [dot, label=left:{$\dfrac{\overline{SU(N)}_{\kappa}\times \overline{\Ff}_{\mathbb{C}}^{N}}{\overline{SU(N)}_{\kappa+1}}$}] at (0,0) {};
\node (R) [dot, label=right:{$\dfrac{SU(N)_{\kappa}\times \Ff_{\mathbb{C}}^{N}}{SU(N)_{\kappa+1}}$}] at (6,0) {};

\draw[postaction={decorate}, bend left=25] (L) to node[midway, above] {$SU(N)_{\kappa}$} (R);
\draw[postaction={decorate}, bend left=25] (R) to node[midway, above] {$SU(N)_{\kappa+1}$} (L);
\end{tikzpicture}
\caption{Two variants of the minimal model RCFT. Top: any RCFT can be ``resolved'' into a pair of chiral and anti-chiral boundary conditions for a 3d TFT. Up to a topological manipulation, we can use a $SU(N)_\kappa \times SU(N)_{1} \times SU(N)_{-\kappa-1}$ 3d Chern-Simons theory and represent each boundary condition as a trivalent junction. Bottom: we replace the $SU(N)_1$ chiral algebra in the coset with $N$ chiral complex fermions $\Ff_\bC$, simplifying the associated 3d TFT to $SU(N)_\kappa \times SU(N)_{-\kappa-1}$ and the junctions to interfaces. We can denote the resulting 2d theory as $M_{N;k}$. Individual interfaces can be engineered by canonically coupling the 3d Chern-Simons gauge fields to the (anti)chiral 2d fermions.}
\label{fig:manione}
\end{figure}
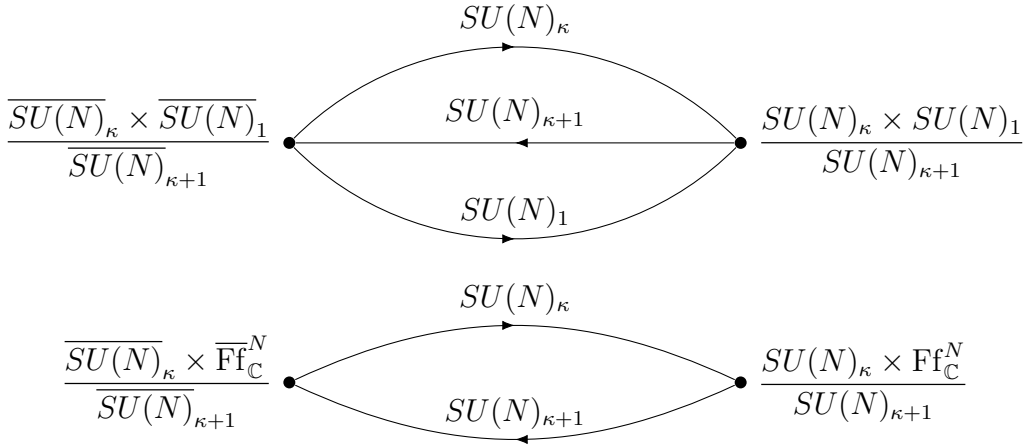

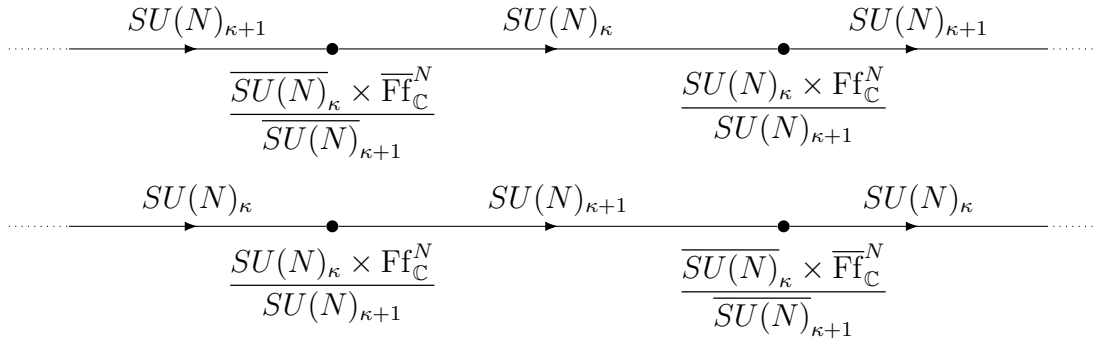
\begin{figure}[ht]
\centering
\begin{tikzpicture}[
  >=Latex,
  decoration={markings, mark=at position 0.5 with {\arrow{Latex}}},
  dot/.style={circle, fill, inner sep=1.6pt}
]
  \def\L{3.5}   
  \def\G{6.0}   

  \node (L) [dot, label=below:{$\dfrac{\overline{SU(N)}_{\kappa}\times \overline{\Ff}_{\mathbb{C}}^{N}}{\overline{SU(N)}_{\kappa+1}}$}] at (0,0) {};
  \node (R) [dot, label=below:{$\dfrac{SU(N)_{\kappa}\times \Ff_{\mathbb{C}}^{N}}{SU(N)_{\kappa+1}}$}]    at (\G,0) {};

  \draw[dotted] (-\L-0.8,0) -- (-\L,0);
  \draw[postaction={decorate}] (-\L,0) -- (L)
      node[midway,above] {$SU(N)_{\kappa+1}$};

  \draw[postaction={decorate}] (R) -- (\G+\L,0)
      node[midway,above] {$SU(N)_{\kappa+1}$};
  \draw[dotted] (\G+\L,0) -- (\G+\L+0.8,0);

  \draw[postaction={decorate}] (L) -- node[midway,above] {$SU(N)_{\kappa}$} (R);
\end{tikzpicture}
\begin{tikzpicture}[
  >=Latex,
  decoration={markings, mark=at position 0.5 with {\arrow{Latex}}},
  dot/.style={circle, fill, inner sep=1.6pt}
]
  \def\L{3.5}
  \def\G{6.0}

  \node (L) [dot, label=below:{$\dfrac{SU(N)_{\kappa}\times \Ff_{\mathbb{C}}^{N}}{SU(N)_{\kappa+1}}$}]    at (0,0) {};
  \node (R) [dot, label=below:{$\dfrac{\overline{SU(N)}_{\kappa}\times \overline{\Ff}_{\mathbb{C}}^{N}}{\overline{SU(N)}_{\kappa+1}}$}] at (\G,0) {};

  \draw[dotted] (-\L-0.8,0) -- (-\L,0);
  \draw[postaction={decorate}] (-\L,0) -- (L)
      node[midway,above] {$SU(N)_{\kappa}$};

  \draw[postaction={decorate}] (R) -- (\G+\L,0)
      node[midway,above] {$SU(N)_{\kappa}$};
  \draw[dotted] (\G+\L,0) -- (\G+\L+0.8,0);

  \draw[postaction={decorate}] (L) -- node[midway,above] {$SU(N)_{\kappa+1}$} (R);
\end{tikzpicture}
\caption{Two interfaces related to $M_{N;k}$ by topological manipulations. Top: the interface $M^+_{N;\kappa}$ connects the chiral and anti-chiral interfaces through the $SU(N)_\kappa$ TFT to produce an interface in $SU(N)_{\kappa+1}$ CS theory. Bottom: the interface $M^-_{N;\kappa}$ connects the chiral and anti-chiral interfaces through the $SU(N)_{\kappa+1}$ TFT to produce an interface in $SU(N)_{\kappa}$ CS theory. If we shrink the intermediate slabs to zero size, both $M^+_{N;\kappa-1}$ and $M^-_{N;\kappa}$ become non-chiral RCFT interfaces in $SU(N)_\kappa$ Chern-Simons theory. They are scale-invariant fixed points for a system of non-chiral 2d fermions coupled to $SU(N)_\kappa$ Chern-Simons gauge fields. This and later figures omit the 2d space-time directions.}
\label{fig:manitwo}
\end{figure}

The reference \cite{Chang:2011mz} proposed that one should isolate two sub-sectors of the coset CFT, each of which enjoys a separate holographic duality to the same higher spin theory, with two distinct boundary conditions for the bulk scalar field. Our topological manipulation precisely identifies these sub-sectors with the full set of local operators available at $M^\pm_{N;\kappa}$. Our manipulation is reversible: one can compactify the topological direction transverse to the interface in order to recover the original 2d system. See Figures \ref{fig:manione} and \ref{fig:manitwo} for a pictorial depiction of our constructions. 

We then identify $M^-_{N;\kappa}$ and $M^+_{N;\kappa-1}$ as two scale-invariant fixed points of an interface defined by coupling the 3d $SU(N)_\kappa$ gauge fields to a set of $N$ 2d complex chiral fermions and a set of $N$ anti-chiral fermions. This gauge theory has a marginal four-fermion coupling whose $\beta$-function is corrected by the Chern-Simons interactions. The RG flow between the fixed points is (a topological manipulation of) the conventional RG flow between consecutive minimal models. At large $\kappa$, the flow is perturbative and the four-fermion coupling is of order $\kappa^{-1}$. 

Minimal Model Holography involves a large $N$ and $\kappa$ limit keeping $t=\frac{N}{\kappa+N}$ fixed. We recognize it as the standard 't Hooft limit for the 3d/2d gauge theory. In particular, the minimal model correlation functions which survive the topological manipulation are computed by gauge theory Feynman diagrams which can be organized into a standard 't Hooft expansion: Minimal Model Holography must fit into the standard framework of String Theory-based holography.  

A general feature of the 't Hooft expansion is that the addition of (anti)fundamental degrees of freedom to an $SU(N)$ gauge theory does not change the dual closed string theory.
Instead, it adds probe D-branes and open string sectors dual to meson operators \cite{Karch:2002sh}. See \cite{Aharony:2019suq} for an inspiring application of that principle in a similar context. The three-dimensional $SU(N)_\kappa$ Chern-Simons gauge theory has a well-known holographic dual involving the A-model Topological Strings \cite{Gopakumar:1998ki,Ooguri:1999bv,Witten:1992fb} on a six-dimensional symplectic manifold. 
Two-dimensional chiral and anti-chiral fermions can be added to the A-model setup with the help of five-dimensional ``coisotropic'' and ``anti-coisotropic'' D-branes \cite{Aganagic:2017tvx,Aharony:2019suq}. We thus have all the ingredients necessary to propose an holographic description of the interfaces: the combination of a coisotropic and an anti-coisotropic D-branes in the A-model background, reaching the boundary at the location of the interface. 

Although coisotropic and anti-coisotropic D-branes are relatively well-understood, open strings stretched between them are less well-studied. In particular, the world-volume theory of a coincident pair of such D-branes has not yet been written down. This complicates the analysis of the meson operator dual to these open string modes. The situation can be simplified by a further topological manipulation which separates the chiral and anti-chiral degrees of freedom in space, as in Figure \ref{fig:manitwo}, without changing the correlation functions. The manipulation splits non-chiral mesons into chiral and anti-chiral halves connected by a topological Wilson line. The D-branes also separate in the dual geometry, so that the non-chiral meson is now dual to a macroscopic open string stretched between the D-branes. 

We analyze in detail the holographic duality between the individual chiral and anti-chiral interfaces and the (anti)coisotropic D-branes. The world-volume theory of individual (or multiple) coisotropic D-branes is a non-commutative 5d Chern-Simons theory which was studied at length in the context of twisted M-theory \cite{Costello:2016nkh,Costello:2017fbo} and is endowed with a rich collection of algebraic and integrable structures \cite{Gaiotto:2019wcc,Gaiotto:2020dsq,Gaiotto:2023ynn}. These allow us to match sphere correlation functions of chiral mesons exactly and at finite $N$ with dual holographic calculations. 

The cost of separating the individual interfaces/D-branes is that we cannot discuss, say, the effect of a mass parameter or the RG flow between the two critical interfaces: the associated non-chiral operators are not local until the chiral and anti-chiral interfaces are re-united. We leave these problems and the local formulation of the coisotropic-anti-coisotropic world-volume theory to future work. 

Although we formulate our result in the language of 2d chiral interfaces and the A-model, we could use an analytic continuation to drop the quantization condition on $\kappa$. The resulting $\cW_N$ chiral algebra and its correlation functions/conformal blocks form a protected subsector of the $(2,0)$ six-dimensional SCFT correlation functions \cite{Alday:2009aq,Wyllard:2009hg,Nekrasov:2010ka,Beem:2014kka,Gaiotto:2017euk}. The holographic dual geometry can be recast as a background for twisted M-theory, possibly understood as a protected subsector of M-theory on AdS${}_7 \times$S${}^4$. Our work thus sharpens the twisted M-theory holographic dictionary proposed by \cite{Costello:2016nkh}: we give a fully back-reacted geometry dual to $\cW_N$ sphere correlation functions. It would be interesting to do the same for the protected correlation functions of the M2 brane 3d SCFT \cite{Costello:2017fbo,Gaiotto:2020vqj}.

To summarize, we study a holographic correspondence between the free-fermion chiral interface in $3d$ $SU(N)$ Chern--Simons theory and a $5d$ non-commutative Chern--Simons theory defined on the backreacted geometry. We show that this duality embeds into standard Minimal Model Holography on both sides. The nontrivial holographic checks we can access and verify include:
\begin{itemize}
    \item Global symmetry algebra. On the gauge theory side, the wedge algebra of $\cW_\infty$ is known to be the higher-spin algebra $\mathfrak{hs}[t]$. On the holographic side, we identify the same symmetry with the classical global symmetries of $5d$ non-commutative Chern-Simons theory.
    \item Chiral OPEs. We compare the planar limit of the chiral algebra OPE with an holographic calculation, analyzing the boundary algebra of the $3d$ theory obtained via dimensional reduction of the $5d$ non-commutative Chern--Simons theory. A all order match in this planar/semi-classical limit is achieve after a non-linear transformation of boundary modes.
    \item Non-planar deformation of the symmetry algebra. On both sides, we identify and match the full non-planar/quantum algebra governing the sphere correlation functions of  specific coset primaries. Fusing primaries into the vacuum channel, we find that sphere correlators of chiral mesons satisfy an infinite tower of commuting differential constraints which appear to determine them uniquely. As a consequence, we establish a complete holographic match of all sphere correlation functions of chiral mesons. 
    This result extends the global symmetry algebra and OPE holographic match to all orders in the 't Hooft expansion.
\end{itemize}

\subsection{Structure of the Paper}
Section \ref{sec:general} develops a general theory of interfaces in 3d Chern-Simons gauge theory, their topological manipulations and RG flows. It should be of interest beyond the scope of Minimal Models and Holography. Section \ref{sec:chiral} describes the meson operators available on chiral interfaces and computes all of their correlation functions on $\bC P^1$. Section \ref{sec:coiso} describes the worldvolume theory of a coisotropic brane and Section \ref{sec:hskk} formulates a Kaluza-Klein reduction appropriate for an holographic dictionary. We also compare our construction with the conventional Minimal Model holography proposal. Section \ref{sec:Wilson} adds to the setup Wilson lines dual to probe D-branes.

\section{The dynamics of non-chiral interfaces}
\label{sec:general}
Chiral boundary conditions and interfaces in Chern-Simons theory and more general 3d TFT are very well-studied. Indeed, they were instrumental to the understanding of 3d TFT \cite{Witten:1988hf,Moore:1988uz,Moore:1988qv,Elitzur:1989nr,TURAEV1992865,turaev1994quantum,Reshetikhin:1991tc}: 3d TFTs are in one-to-one correspondence with Modular Tensor Categories encoding their anyons/topological line defects, and the category of modules $\mathrm{Mod}_V$ for a Rational Vertex Operator Algebra $V$ is an MTC \cite{Huang:2004dg}.
Physically, each RVOA can be realized as the algebra of boundary local operators for a specific chiral boundary condition for the 3d TFT associated to $\mathrm{Mod}_V$. The VOA modules themselves are identified with the spaces of local operators which can be found as boundary endpoints of the corresponding bulk anyons.

Conversely, the non-chiral 2d Rational Conformal Field Theories which can be built by pairing up chiral and anti-chiral RVOAs are in one-to-one correspondence with topological interfaces in the associated 3d TFT via a ``slab construction'', which sandwiches the topological interface between the canonical chiral and anti-chiral boundary conditions \cite{Fuchs:2002cm,Kapustin:2010hk}. The slab construction makes non-chiral operators in the RCFT non-local: they lift to bulk anyons stretched between the two boundaries. In particular, locality forbids relevant deformations of the 2d RCFT, which only become available if the slab is squeezed back to a 2d system \cite{Gaiotto:2020iye}.

These canonical chiral boundary conditions are by no means the only chiral boundaries or interfaces available in a 3d TFT. Intuitively, any 3d TFT which admits a topological interface to $\mathrm{Mod}_V$ will admit a chiral boundary condition supporting $V$, defined by the composition of the interface and the canonical chiral boundary condition \cite{Kapustin:2010if}.
A folding trick gives access to chiral interfaces as well. A typical example we will review momentarily is that of a $G_{k_G}/H_{k_H}$ coset chiral algebra: we can find the chiral algebra at an interface between $G_{k_G}$ and $H_{k_H}$ Chern-Simons gauge theories. For example, the Ising model chiral algebra can be found at the boundary of an ``Ising TFT'', but also 
at an interface between $SU(2)_2$ and $SU(2)_1 \times SU(2)_1$ Chern-Simons theories. 

The availability of chiral and anti-chiral interfaces allows for intricate constructions, where multiple interfaces are stacked together along a topological direction \cite{Fuchs:2012dt,Gaiotto:2020iye}. This gives alternative constructions for RCFTs, but also for non-chiral ``Rational Conformal Interfaces'' or ``Rational Conformal Boundaries'' of 3d TFTs. 

Non-chiral deformations of such system only become available if chiral and non-chiral interfaces are squeezed back together to make the associated operators local \cite{Frohlich:2006ch}. If the slab has multiple layers, we may choose to only squeeze together some consecutive subset of layers, making only certain non-chiral operators local. Locality implies that the RG flows triggered by these operators must keep within this sub-system. This observation can place interesting constraints on the full phase space of the system. We will now see some relevant examples. 

\subsection{Gauging and Cosets}
Consider some 2d CFT $T_k$, not necessarily chiral, endowed with a chiral $G$ Kac-Moody symmetry at level $k$. By definition, the level $k$ is integral. The theory $T_k$ can always be lifted to a slab based on the $G_k$ Chern-Simons theory: one boundary supports the chiral WZW model $G_k$ and the other boundary supports the ``coset'' $T_k/G_k$ consisting of all local operators in $T_k$ which are local with the Kac-Moody currents. See Figure \ref{fig:slabs1}.
\usetikzlibrary{matrix}

\begin{figure}[ht]
\centering
\begin{tikzpicture}
  \matrix[column sep=14mm,row sep=8mm] {
    \node (a11) {\tikz{\useasboundingbox (-0.3,-1.0) rectangle (3.3,1.0);
      \fill (0,0) circle[radius=1.6pt] node[below=2pt] {$T_k/G_k$};
      \fill (3,0) circle[radius=1.6pt] node[below=2pt] {$G_k$};
      \draw[-{Latex}] (0,0) -- (1.5,0);
      \draw (1.5,0) -- node[above,pos=0] {$G_k$} (3,0); }}; &
    \node (a12) {\tikz{\useasboundingbox (-0.3,-1.0) rectangle (3.3,1.0);
      \fill (0,0) circle[radius=1.6pt] node[below=2pt] {$\overline G_k$};
      \fill (3,0) circle[radius=1.6pt] node[below=2pt] {$\overline T_k/\overline G_k$};
      \draw[-{Latex}] (0,0) -- (1.5,0);
      \draw (1.5,0) -- node[above,pos=0] {$G_k$} (3,0); }}; \\
    \node (a21) {\tikz{\useasboundingbox (-0.3,-1.0) rectangle (3.3,1.0);
      \fill (0,0) circle[radius=1.6pt] node[below=2pt] {$\overline G_k$};
      \fill (3,0) circle[radius=1.6pt] node[below=2pt] {$G_k$};
      \draw[-{Latex}] (0,0) -- (1.5,0);
      \draw (1.5,0) -- node[above,pos=0] {$G_k$} (3,0); }}; &
    \node (a22) {\tikz{\useasboundingbox (-0.3,-1.0) rectangle (3.3,1.0);
      \fill (0,0) circle[radius=1.6pt] node[below=2pt] {$T_k/G_k$};
      \fill (3,0) circle[radius=1.6pt] node[below=2pt] {$\overline T_k/\overline G_k$};
      \draw[-{Latex}] (0,0) -- (1.5,0);
      \draw (1.5,0) -- node[above,pos=0] {$G_k$} (3,0); }}; \\
  };
\end{tikzpicture}
\caption{A collection of ``slab'' configurations producing a 2d system from a 3d $G_k$ Chern-Simons theory. The 2d directions are not drawn. Top left: a 2d theory $T_k$ with chiral Kac-Moody $G$ symmetry at level $k>0$ can be decomposed into a slab with a chiral Dirichlet boundary supporting a $G_k$ WZW chiral algebra and enriched Neumann supporting a coset $T_k/G_k$. Top right: the analogous resolution of a 2d theory $\overline T_k$ with anti-chiral Kac-Moody $G$ symmetry. Bottom left: two Dirichlet b.c. engineer a non-chiral WZW model. Bottom right: two enriched Neumann engineer a non-chiral coset. We expect the coset to be the IR limit of both a 2d $G$ gauge theory with matter $T_k \times \overline T_k$ and of an asymptotically free $J \cdot \bar J$ deformation of $T_k \times \overline T_k$.}
\label{fig:slabs1}
\end{figure}
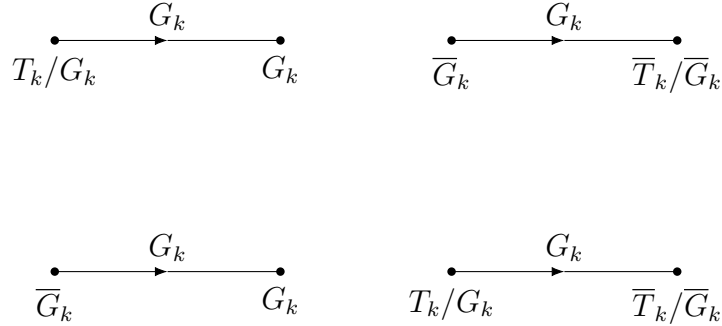
In gauge theory language, the former is a Dirichlet boundary condition and the latter a Neumann boundary condition enriched by $T_k$. We will denote this statement as 
\begin{equation}
    T_k \simeq G_k \times_{G_k} \frac{T_k}{G_k} \, ,
\end{equation}
where the product subscript denotes coupling via the 3d $G_k$ Chern-Simons TFT and the factors indicate the two boundary theories.  

\begin{figure}[ht]
\centering
\begin{tikzpicture}[
  >=Latex,
  decoration={markings, mark=at position 0.5 with {\arrow{Latex}}},
  every node/.style={font=\normalsize}
]
  \def\L{6.0}   
  \def\H{1.5}   

  \draw[dotted] (-\L-0.8,0) -- (-\L,0);
  \draw[postaction={decorate}] (-\L,0) -- (0,0)
      node[midway,above] {$G_{\kappa}$};

  \draw[postaction={decorate}] (0,0) -- (\L,0)
      node[midway,above] {$G_{\kappa + k}$};
  \draw[dotted] (\L,0) -- (\L+0.8,0);

  \draw[postaction={decorate}] (0,-\H) -- (0,0)
      node[midway,left=2pt] {$G_k$};

  \fill (0,0) circle[radius=1.6pt]
        node[above=4pt] {$\displaystyle \frac{G_{\kappa}\times G_k}{G_{\kappa + k}}$};

  \fill (0,-\H) circle[radius=1.6pt]
        node[below=3pt] {$T_k/G_k$};
\end{tikzpicture}
\caption{A proposed resolution of the interface between $G_\kappa$ and $G_{\kappa+k}$ Chern-Simons theories defined by coupling the 3d gauge fields to $T_k$. We assume $\kappa>0$ here. Shrinking the segment gives a $\frac{G_{\kappa}\times T_k}{G_{\kappa + k}}$ coset at the interface.}
\label{fig:slabs2}
\end{figure}
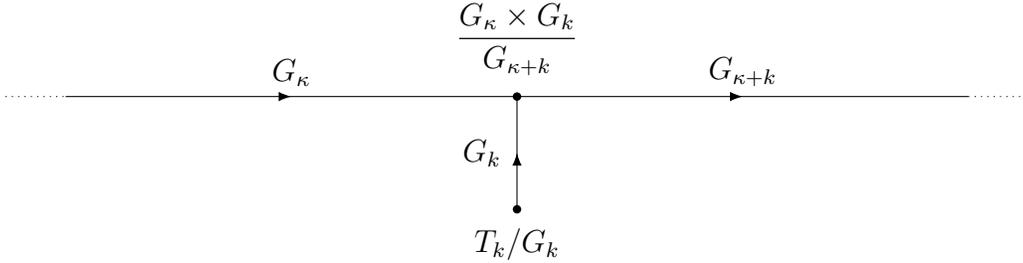

It is also very natural to place $T_k$ at an interface, and couple it to 3d $G$ Chern-Simons gauge fields. The CS level will have to jump by $k$ across the interface. Denote the levels on the two sides as $\kappa$ and $\kappa + k$. The resulting dynamics depends in an interesting way on the value of $\kappa$. For positive $\kappa$, we propose that the interface will produce a richer coset: 
\begin{equation}
    \frac{G_{\kappa} \times T_k}{G_{\kappa + k}} \, .
\end{equation}
This is a potentially non-chiral system whose local operators are combinations of operators in $T_k$ and currents in $G_{\kappa}$ which are local with the total $G_{\kappa + k}$ currents. In the weak coupling limit $\kappa \to \infty$, the central charges of this system approach these of $T_k$ and the coset simplifies to the collection of $G$-invariant operators in $T_k$. This makes the proposal plausible. See Figure \ref{fig:slabs2}.

For non-chiral $T_k$, this statement is a bit more subtle than one may guess. In principle, coupling to the 3d gauge fields should be accompanied by a scheme-dependent generic relevant deformation of $T_k$. Splitting $T_k$ into $G_k \times_{G_k} \frac{T_k}{G_k}$ gives a scheme where 
such deformations can be naturally turned off, as they live at a different location than the coupling to the 3d gauge fields. See Figure \ref{fig:slabs2}. Then 
\begin{equation}
    \frac{G_{\kappa} \times T_k}{G_{\kappa + k}} \simeq \frac{G_{\kappa} \times G_k}{G_{\kappa + k}} \times_{G_k} \frac{T_k}{G_k}
\end{equation}
reduces the claim to a claim about a {\it chiral junction} between $G_{\kappa + k}$ and $G_{\kappa} \times G_k$ Chern-Simons theories, defined by boundary conditions
\begin{align}
    &\kappa A_z^{(\kappa)} + k A_z^{(k)}= (\kappa+k) A_z^{(\kappa+k)} \cr
    & A_{\bar z}^{(\kappa)}= A_{\bar z}^{(k)} = A_{\bar z}^{(\kappa+k)} \, .
\end{align}
This junction should support the coset chiral algebra $\frac{G_{\kappa} \times G_k}{G_{\kappa + k}}$ if $k>0$ and $\kappa>0$. 

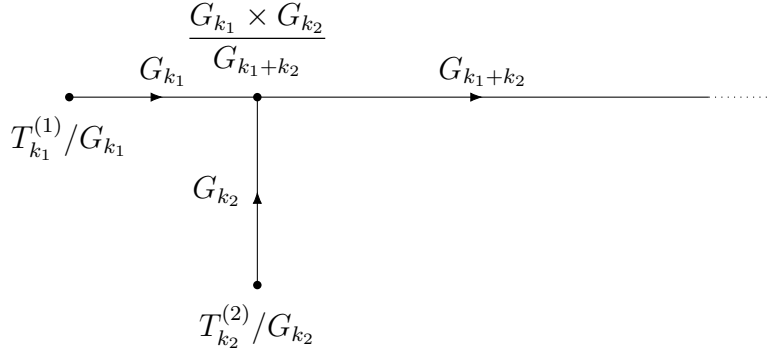
\begin{figure}[ht]
\centering
\begin{tikzpicture}[
  >=Latex,
  decoration={markings, mark=at position 0.5 with {\arrow{Latex}}},
  every node/.style={font=\normalsize}
]
  \def\L{6.0}   
  \def\H{2.5}   

  \fill (-\H,0) circle[radius=1.6pt]
        node[below=3pt] {$T^{(1)}_{k_1}/G_{k_1}$};
  \draw[postaction={decorate}] (-\H,0) -- (0,0)
      node[midway,above] {$G_{k_1}$};

  \draw[postaction={decorate}] (0,0) -- (\L,0)
      node[midway,above] {$G_{k_1 + k_2}$};
  \draw[dotted] (\L,0) -- (\L+0.8,0);

  \draw[postaction={decorate}] (0,-\H) -- (0,0)
      node[midway,left=2pt] {$G_{k_2}$};
  \fill (0,-\H) circle[radius=1.6pt]
        node[below=3pt] {$T^{(2)}_{k_2}/G_{k_2}$};

  \fill (0,0) circle[radius=1.6pt]
        node[above=4pt] {$\displaystyle \frac{G_{k_1}\times G_{k_2}}{G_{k_1 + k_2}}$};
\end{tikzpicture}
\caption{A resolution of an enriched Neumann b.c. for $G_{k_1+k_2}$ CS theory coupled to $T^{(1)}_{k_1}\times T^{(2)}_{k_2}$. Shrinking the segments reproduces the coset $\frac{T^{(1)}_{k_1}\times T^{(2)}_{k_2}}{G_{k_1 + k_2}}$.}
\label{fig:slabs3}
\end{figure}

This claim is supported by a simple manipulation of a Neumann boundary condition enriched by the combination of two theories $T^{(1)}_{k_1} \times T^{(2)}_{k_2}$ with chiral Kac-Moody symmetry, which we assumed to produce the $\frac{T^{(1)}_{k_1} \times T^{(2)}_{k_2}}{G_{k_1+k_2}}$ coset. We can bring the $T^{(1)}_{k_1}$ factor away from the boundary to build a slab resolution:
\begin{equation}
    \frac{T^{(1)}_{k_1} \times T^{(2)}_{k_2}}{G_{k_1+k_2}} \simeq  \frac{T^{(1)}_{k_1} \times G_{k_2}}{G_{k_1+k_2}} \times_{G_{k_2}} \frac{T^{(2)}_{k_2}}{G_{k_2}} \, ,
\end{equation}
consistent with the presence of $\frac{T^{(1)}_{k_1} \times G_{k_2}}{G_{k_1+k_2}}$ at the interface. See Figure \ref{fig:slabs3}.

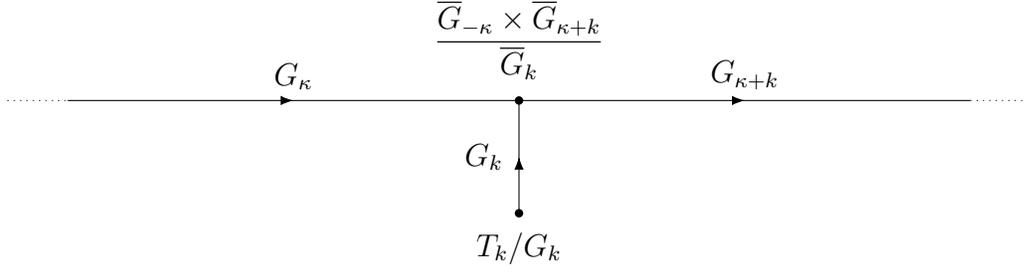
\begin{figure}[ht]
\centering
\begin{tikzpicture}[
  >=Latex,
  decoration={markings, mark=at position 0.5 with {\arrow{Latex}}},
  every node/.style={font=\normalsize}
]
  \def\L{6.0}   
  \def\H{1.5}   

  \draw[dotted] (-\L-0.8,0) -- (-\L,0);
  \draw[postaction={decorate}] (-\L,0) -- (0,0)
      node[midway,above] {$G_{\kappa}$};

  \draw[postaction={decorate}] (0,0) -- (\L,0)
      node[midway,above] {$G_{\kappa + k}$};
  \draw[dotted] (\L,0) -- (\L+0.8,0);

  \draw[postaction={decorate}] (0,-\H) -- (0,0)
      node[midway,left=2pt] {$G_k$};

  \fill (0,0) circle[radius=1.6pt]
        node[above=4pt] {$\displaystyle \frac{\overline G_{-\kappa}\times \overline G_{\kappa+k}}{\overline G_{k}}$};

  \fill (0,-\H) circle[radius=1.6pt]
        node[below=3pt] {$T_k/G_k$};
\end{tikzpicture}
\caption{A conjectural description of the interface between $G_\kappa$ and $G_{\kappa+k}$ Chern-Simons theories defined by coupling the 3d gauge fields to $T_k$ for the case $0>\kappa>-k$ here.}
\label{fig:slabs4}
\end{figure}

A reflection produces another useful claim: for negative $\kappa+k$, we propose a coset 
\begin{equation}
    \frac{G_{-\kappa-k} \times T_k}{G_{-\kappa}} \, ,
\end{equation}
We are left with the ``strongly coupled'' range $0>\kappa>-k$. The only candidate we find somewhat plausible is the ``heterotic'' combination 
\begin{equation}
    \frac{\overline G_{\kappa+k} \times \overline G_{-\kappa}}{\overline G_{k}} \times_{G_k} \frac{T_k}{G_k}
\end{equation}
employing a junction between $G_{\kappa+k} \times G_{-\kappa}$ and $G_k$ CS theories supporting the anti-chiral coset. See Figure \ref{fig:slabs4}.

\begin{figure}[ht]
\centering
\begin{tikzpicture}[
  >=Latex,
  decoration={markings, mark=at position 0.5 with {\arrow{Latex}}},
  every node/.style={font=\normalsize}
]
  \def\L{6.0}   
  \def\H{2.5}   

  \fill (-\H,0) circle[radius=1.6pt]
        node[below=3pt] {$T_{k}/G_{k}$};
  \draw[postaction={decorate}] (-\H,0) -- (0,0)
      node[midway,above] {$G_{k}$};

  \draw[postaction={decorate}] (0,0) -- (\L,0)
      node[midway,above] {$G_{k-\bar k}$};
  \draw[dotted] (\L,0) -- (\L+0.8,0);

  \draw[postaction={decorate}] (0,-\H) -- (0,0)
      node[midway,left=2pt] {$G_{-\bar k}$};
  \fill (0,-\H) circle[radius=1.6pt]
        node[below=3pt] {$\overline T_{\bar k}/\overline G_{\bar k}$};

  \fill (0,0) circle[radius=1.6pt]
        node[above=4pt] {$\displaystyle \frac{\overline G_{k-\bar k}\times \overline G_{\bar k}}{\overline G_{k}}$};
\end{tikzpicture}
\caption{A conjectural description of an enriched Neumann b.c. for $G_{k- \bar k}$ CS theory coupled to $T_{k}\times \overline T_{\bar k}$, for the case $k > \bar k$.}
\label{fig:boutwo}
\end{figure}
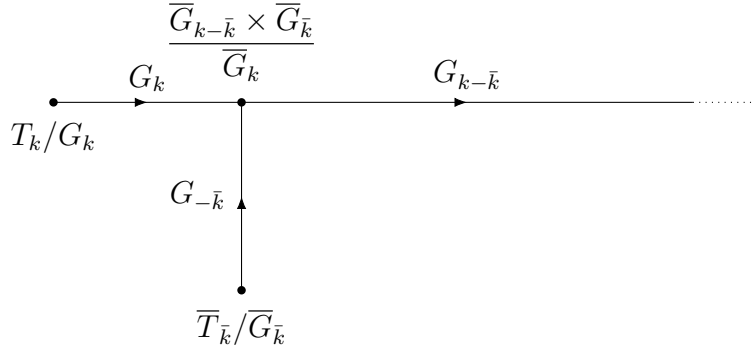

This leads to a proposal for non-chiral boundary conditions. Consider a Neumann b.c. for $G_{k-\bar k}$ CS theory coupled to 2d matter $T_k \times {\overline T}_{\bar k}$ with both chiral and anti-chiral Kac-Moody $G$ symmetry. Although we write this as a product, our analysis would extend easily to a theory $T_{k, \bar k}$ with both symmetries. We get boundary theories 
\begin{align}
    &\frac{\overline G_{k- \bar k} \times {\overline T}_{\bar k}}{\overline G_{k}} \times_{G_k} \frac{T_k}{G_k} \qquad \qquad k>\bar k \cr
     &\frac{{\overline T}_{\bar k}}{\overline G_{\bar k}} \times_{G_{\bar k}} \frac{G_{ \bar k- k} \times  T_k}{G_{\bar k}} \qquad \qquad \bar k > k  \, .
\end{align}
See Figure \ref{fig:boutwo}. 

\subsection{Universal RG flows}
As a motivating example, consider now an interface where we couple CS gauge fields to $T_k \times {\overline T}_{\bar k}$, with a level much greater than $\bar k$ and $k$. We can resolve this configuration by separating the two sets of matter fields. Depending on their relative order, we get two alternatives:
\begin{align}
    &\frac{G_{\kappa} \times T_k}{G_{\kappa + k}} \times_{G_{\kappa + k}} \frac{\overline G_{\kappa+ k - \bar k} \times \overline T_{\bar k}}{\overline G_{\kappa + k}} \cr 
    &\frac{\overline G_{\kappa- \bar k} \times \overline T_{\bar k}}{\overline G_{\kappa}} \times_{G_{\kappa - \bar k}}  \frac{G_{\kappa- \bar k} \times T_k}{G_{\kappa  + k - \bar k}} 
\end{align}
See Figures \ref{fig:uns} and \ref{fig:sta}.

The first option has a weakly relevant coset operator labeled by the adjoint representation of $G_{\kappa + k}$, of dimension
\begin{equation}
    1-\frac{C_2(\mathrm{Adj})}{\kappa + k + h}\,.
\end{equation}
The second has a weakly irrelevant coset operator labeled by the adjoint 
representation of $G_{\kappa -\bar k}$, of dimension
\begin{equation}
    1+\frac{C_2(\mathrm{Adj})}{\kappa -\bar k + h}\,.
\end{equation}
Both are fully renormalized versions of a marginal current bilinear operator $J \cdot \bar J$ in $T_k \times {\overline T}_{\bar k}$. It is natural to assume that coupling to the CS gauge fields will turn on a $\beta$ function for the coupling of $J \cdot \bar J$. We expect the above exactly conformal interfaces to be a stable and unstable fixed point for that RG flow.

\begin{figure}[ht]
\centering
\begin{tikzpicture}[
  >=Latex,
  decoration={markings, mark=at position 0.5 with {\arrow{Latex}}},
  every node/.style={font=\normalsize}
]
  \def\L{6.0}   
  \def\Dx{3.0}  
  \def\H{1.5}   

  \draw[dotted] (-\L-0.8,0) -- (-\L,0);
  \draw[postaction={decorate}] (-\L,0) -- (-\Dx/2,0)
      node[midway,above] {$G_{\kappa}$};

  \draw[postaction={decorate}] (-\Dx/2,0) -- (\Dx/2,0)
      node[midway,above] {$G_{\kappa + k}$};

  \draw[postaction={decorate}] (\Dx/2,0) -- (\L,0)
      node[midway,above] {$G_{\kappa + k - \bar k}$};
  \draw[dotted] (\L,0) -- (\L+0.8,0);

  \draw[postaction={decorate}] (-\Dx/2,-\H) -- (-\Dx/2,0)
      node[midway,left=2pt] {$G_{k}$};
  \fill (-\Dx/2,0) circle[radius=1.6pt]
        node[above=4pt] {$\displaystyle \frac{G_{\kappa}\times G_{k}}{G_{\kappa + k}}$};
  \fill (-\Dx/2,-\H) circle[radius=1.6pt]
        node[below=3pt] {$T_{k}/G_{k}$};

  \draw[postaction={decorate}] (\Dx/2,0) -- (\Dx/2,-\H)
      node[midway,left=2pt] {$G_{\bar k}$};
  \fill (\Dx/2,0) circle[radius=1.6pt]
        node[above=4pt] {$\displaystyle \frac{\overline G_{\kappa + k- \bar k}\times \overline G_{\bar k}}{\overline G_{\kappa + k}}$};
  \fill (\Dx/2,-\H) circle[radius=1.6pt]
        node[below=3pt] {$\overline T_{\bar k}/\overline G_{\bar k}$};

\end{tikzpicture}
\caption{A 3d resolution of the unstable fixed point for an interface defined by coupling 3d CS theory to $T_k \times \overline T_{\overline k}$.}
\label{fig:uns}
\end{figure}
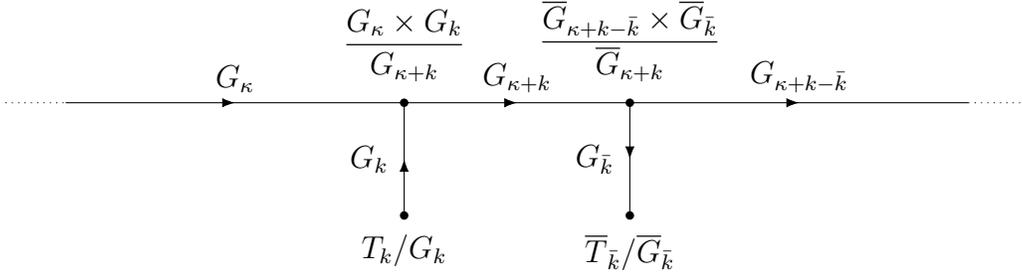
\begin{figure}[ht]
\centering
\begin{tikzpicture}[
  >=Latex,
  decoration={markings, mark=at position 0.5 with {\arrow{Latex}}},
  every node/.style={font=\normalsize}
]
  \def\L{6.0}   
  \def\Dx{3.0}  
  \def\H{1.5}   

  \draw[dotted] (-\L-0.8,0) -- (-\L,0);
  \draw[postaction={decorate}] (-\L,0) -- (-\Dx/2,0)
      node[midway,above] {$G_{\kappa}$};

  \draw[postaction={decorate}] (-\Dx/2,0) -- (\Dx/2,0)
      node[midway,above] {$G_{\kappa - \bar k}$};

  \draw[postaction={decorate}] (\Dx/2,0) -- (\L,0)
      node[midway,above] {$G_{\kappa + k - \bar k}$};
  \draw[dotted] (\L,0) -- (\L+0.8,0);

  \draw[postaction={decorate}] (\Dx/2,-\H) -- (\Dx/2,0)
      node[midway,left=2pt] {$G_{k}$};
  \fill (\Dx/2,0) circle[radius=1.6pt]
        node[above=4pt] {$\displaystyle \frac{G_{\kappa- \bar k}\times G_{k}}{G_{\kappa + k-\bar k}}$};
  \fill (\Dx/2,-\H) circle[radius=1.6pt]
        node[below=3pt] {$T_{k}/G_{k}$};

  \draw[postaction={decorate}] (-\Dx/2,0) -- (-\Dx/2,-\H)
      node[midway,left=2pt] {$G_{\bar k}$};
  \fill (-\Dx/2,0) circle[radius=1.6pt]
        node[above=4pt] {$\displaystyle \frac{\overline G_{\kappa - \bar k}\times \overline G_{\bar k}}{\overline G_{\kappa}}$};
  \fill (-\Dx/2,-\H) circle[radius=1.6pt]
        node[below=3pt] {$\overline T_{\bar k}/\overline G_{\bar k}$};

\end{tikzpicture}
\caption{A 3d resolution of the stable fixed point for an interface defined by coupling 3d CS theory to $T_k \times \overline T_{\overline k}$.}
\label{fig:sta}
\end{figure}
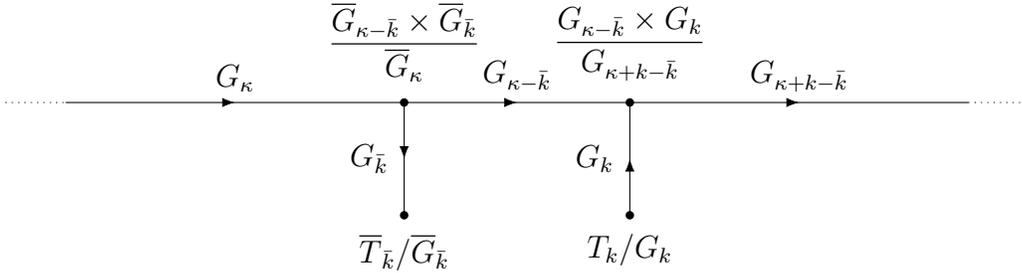

When $k = \bar k$, that RG flow is essentially the well-known RG flow between diagonal coset RCFTs \cite{Zamolodchikov:1991vg,Ravanini:1992fs}
\begin{equation}
    \left|\frac{G_{\kappa} \times G_k}{G_{\kappa + k}}\right|^2 \to \left|\frac{G_{\kappa-k} \times G_k}{G_{\kappa}}\right|^2 
\end{equation}
triggered by the $[1,1;\mathrm{Adj}]$ coset primary, generalizing the RG flow between consecutive minimal models. That RG flow is actually integrable. Here we propose a generalization 
\begin{equation}
    \frac{G_{\kappa} \times G_k}{G_{\kappa + k}} \times_{G_{\kappa + k}} \frac{\overline G_{\kappa+ k - \bar k} \times \overline G_{\bar k}}{\overline G_{\kappa + k}} \to 
    \frac{\overline G_{\kappa- \bar k} \times \overline G_{\bar k}}{\overline G_{\kappa}} \times_{G_{\kappa - \bar k}}  \frac{G_{\kappa- \bar k} \times G_k}{G_{\kappa  + k - \bar k}} \,,
\end{equation}
which applies to interfaces between $G_\kappa \times G_k$ and $G_{\kappa + k - \bar k} \times G_{\bar k}$ 3d Chern-Simons theories.\footnote{It should be possible to demonstrate its integrability by a lift to 4d Chern-Simons theory \cite{Costello:2019tri}.}

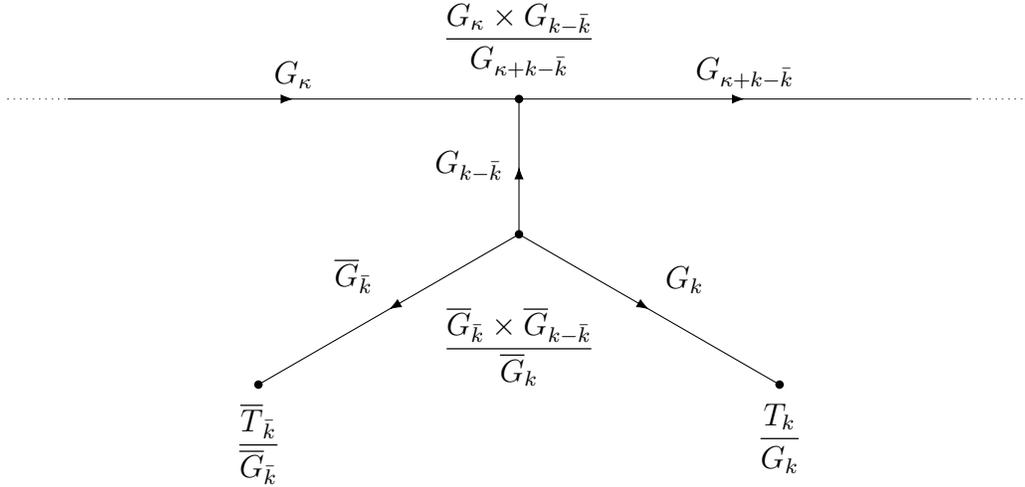
\begin{figure}[ht]
\centering
\begin{tikzpicture}[
  >=Latex,
  decoration={markings, mark=at position 0.5 with {\arrow{Latex}}},
  every node/.style={font=\normalsize}
]
  \def\L{6.0}    
  \def\H{1.8}    
  \def\ang{30}   
  \def\BL{4.0}   
  \def\BR{4.0}   

  \draw[dotted] (-\L-0.8,0) -- (-\L,0);
  \draw[postaction={decorate}] (-\L,0) -- (0,0)
       node[midway,above] {$G_{\kappa}$};
  \draw[postaction={decorate}] (0,0) -- (\L,0)
       node[midway,above] {$G_{\kappa + k - \bar k}$};
  \draw[dotted] (\L,0) -- (\L+0.8,0);

  \fill (0,0) circle[radius=1.6pt]
        node[above=4pt] {$\displaystyle 
           \frac{G_{\kappa}\times G_{k-\bar k}}{G_{\kappa + k - \bar k}}$};

  \draw[postaction={decorate}] (0,-\H) -- (0,0)
       node[midway,left=2pt] {$G_{k-\bar k}$};

  \fill (0,-\H) circle[radius=1.6pt]
        node[below=8pt,yshift=-3ex] {$\displaystyle
           \frac{\overline G_{\bar k}\times \overline G_{k-\bar k}}{\overline G_{k}}$};

  \path (0,-\H) ++(180+\ang:\BL) coordinate (Lend);
  \draw[postaction={decorate}] (0,-\H) -- (Lend)
       node[midway,above left=2pt] {$\overline G_{\bar k}$};
  \fill (Lend) circle[radius=1.6pt]
        node[below=3pt] {$\displaystyle \frac{\overline T_{\bar k}}{\overline G_{\bar k}}$};

  \path (0,-\H) ++(-\ang:\BR) coordinate (Rend);
  \draw[postaction={decorate}] (0,-\H) -- (Rend)
       node[midway,above right=2pt] {$G_k$};
  \fill (Rend) circle[radius=1.6pt]
        node[below=3pt] {$\displaystyle \frac{T_k}{G_k}$};

\end{tikzpicture}
\caption{A 3d resolution of an alternative stable fixed point for an interface defined by coupling 3d CS theory to $T_k \times \overline T_{\overline k}$, with a ``strong'' $J \cdot \bar J$ deformation.}
\label{fig:str}
\end{figure}

Curiously, there is a third fixed point one may consider, associated to a 3d picture with a $G_{k-\bar k}$ intermediate leg. This gives a junction
\begin{equation}
    \frac{G_\kappa \times G_{k-\bar k}}{G_{\kappa + k - \bar k}}\times_{G_{k-\bar k}}\frac{\overline T_{\bar k} \times \overline G_{k-\bar k}}{\overline G_k} \times_{G_k} \frac{T_k}{G_k}\,,
\end{equation}
whose status is less intuitive. This is strongly coupled even at large $\kappa$, with lower central charges than for $T_k \times \overline T_{\bar k}$. We interpret it as a fixed point for a $J \cdot \bar J$ asymptotically free deformation of $T_k \times \overline T_{\bar k}$, weakly deformed by the coupling to the 3d CS gauge fields. See Figure \ref{fig:str}.

We can abstract the RG flow problem a bit by removing the extra degrees of freedom and looking at junctions between $G_{\kappa_1} \times G_{\kappa_2}$
CS theory and $G_{\kappa_3} \times G_{\kappa_1+ \kappa_2 - \kappa_3}$. Assume e.g. that all of the couplings are positive and large and $\kappa_1\ll\kappa_3\ll\kappa_2$. We can split such a junction into elementary trivalent junctions in three ways:
\begin{align}
    s:& \qquad \frac{G_{\kappa_1}\times G_{\kappa_2}}{G_{\kappa_1 + \kappa_2}} \times_{G_{\kappa_1 + \kappa_2}} \frac{\overline G_{\kappa_3}\times \overline G_{\kappa_1+\kappa_2-\kappa_3}}{\overline G_{\kappa_1 + \kappa_2}} \cr
    t:& \qquad\frac{G_{\kappa_1}\times G_{\kappa_3-\kappa_1}}{G_{\kappa_3}} \times_{G_{\kappa_1 - \kappa_3}} \frac{\overline G_{\kappa_3-\kappa_1}\times \overline G_{\kappa_1+\kappa_2- \kappa_3}}{\overline G_{\kappa_2}} \cr
    u:& \qquad\frac{G_{\kappa_1}\times G_{\kappa_2- \kappa_3}}{G_{\kappa_1 + \kappa_2- \kappa_3}} \times_{G_{\kappa_2 - \kappa_3}} \frac{\overline G_{\kappa_3}\times \overline G_{\kappa_2-\kappa_3}}{\overline G_{\kappa_2}} \cr
\end{align}
The $s$ fixed point interface has a weakly relevant operator labeled by the adjoint of $G_{\kappa_1 + \kappa_2}$. The other two fixed points have weakly relevant operators labeled by the adjoint of $G_{\kappa_3 - \kappa_1}$ and labeled by the adjoint of $G_{\kappa_2 - \kappa_3}$ respectively. We expect the $s$ fixed point to flow to the other two depending on the sign of the relevant deformation.\footnote{It would be interesting to assess the integrability of these flows with the help of 4d CS theory \cite{Costello:2019tri}.}

\subsection{Multi-factor interfaces}

These considerations generalize to coupling CS gauge fields to a combination 
\begin{equation}
    \prod_{a=1}^n T^{(a)}_{k_a} \times \prod_{\bar a =1}^{\bar n} {\overline T}^{\bar a}_{\bar k_{\bar a}}\,.
\end{equation}
We now have $n \bar n$ marginal current bilinears $J_a \cdot J_{\bar a}$ leading to a potentially intricate phase diagram. 

We can produce $(n+ \bar n)!$ potentially distinct RG fixed points by separating the interfaces in some order along the topological direction. If $\kappa \gg k,\bar k$ so that the level in all intervals is positive, we can freely permute two consecutive chiral or two consecutive anti-chiral interfaces, as 
\begin{equation}
    \frac{G_{\kappa} \times T_k}{G_{\kappa + k}} \times_{G_{\kappa + k}} \frac{G_{\kappa+k} \times T'_{k'}}{G_{\kappa + k+k'}} =  \frac{G_{\kappa} \times T_k \times T'_{k'}}{G_{\kappa + k+k'}} 
\end{equation}
reducing somewhat the number of possibilities. 

The ``most stable'' option where all current bilinears are irrelevant is 
\begin{equation}
    \frac{\overline G_{\kappa- \sum_{\bar a}\bar k_{\bar a}} \times \prod_{\bar a} \overline T^{(\bar a)}_{\bar k_{\bar a}}}{\overline G_{\kappa}} \times_{G_{\kappa - \sum_{\bar a}\bar k_{\bar a}}}  \frac{G_{\kappa- \sum_{\bar a}\bar k_{\bar a}} \times \prod_a T^{(a)}_{k_a}}{G_{\kappa  + \sum_{a} k_{a} - \sum_{\bar a}\bar k_{\bar a}}} 
\end{equation}
with all anti-chiral matter followed by all chiral matter. The ``most unstable'' option where all current bilinears are relevant is
\begin{equation}
    \frac{G_{\kappa} \times \prod_a T^{(a)}_{k_a}}{G_{\kappa  + \sum_{a} k_{a}} } \times_{G_{\kappa + \sum_{a}k_{a}}} \frac{\overline G_{\kappa+ \sum_{a}k_{a}- \sum_{\bar a}\bar k_{\bar a}} \times \prod_{\bar a} \overline T^{(\bar a)}_{\bar k_{\bar a}}}{\overline G_{\kappa+ \sum_{a}k_{a}}} 
\end{equation}
with all chiral matter followed by all anti-chiral matter. Individual RG flows can be found permuting pairs of chiral and anti-chiral individual interfaces, building a complicated polyhedron. For example, if $n=2$ and $\bar n=1$ we have two marginal operators, a stable fixed point, an unstable fixed point and two fixed points with a relevant and an irrelevant operator, vertices of a square of RG flows.

Besides these ``perturbative'' fixed points, we can produce a larger collection by considering all possible trees with $T/G$ and $\overline T/\overline G$ leaves and two leaves corresponding to $G_{\kappa}$ and $G_{\kappa+ \sum_{a}k_{a}- \sum_{\bar a}\bar k_{\bar a}}$. We then assign chiral or anti-chiral cosets to all vertices depending on the 
levels of the edges around them. Neighboring pairs of chiral or pairs of anti-chiral vertices can be merged together. Neighboring chiral and anti-chiral vertices can be rearranged by RG flows. 

\subsection{Pure 2d conjectures}
As a small aside, it is entertaining to review and generalize the conjectural RG flows for Gross-Neveu-like systems where a factorized CFT
\begin{equation}
    \prod_{a=1}^n T^{(a)}_{k_a} \times \prod_{\bar a =1}^{\bar n} {\overline T}^{\bar a}_{\bar k_{\bar a}}
\end{equation}
is deformed by $J \cdot \bar J$ current deformations. RG fixed points should have a $G$ symmetry with overall level $k_t = \sum_a k_a - \sum_{\bar a} \bar k_{\bar a}$, which we can take to be non-negative without loss of generality and should be promoted to a Kac-Moody symmetry in the far IR if $k_t>0$, disappear otherwise. 

We can build a collection of potential fixed points from graphs, with $G_{k_t}$, $\frac{T_{k_a}}{G_{k_a}}$ and $\frac{{\overline T}_{\bar k_{\bar a}}}{\overline G_{\bar k_a}}$ leaves. For example, with a single chiral and anti-chiral factors and $k_t=0$ one expects the $J \cdot \bar J$ current bilinear coupling with the correct sign to produce an asymptotically free theory which flows at strong coupling to some topological modification \cite{Gaiotto:2020iye} of 
\begin{equation}
    \frac{T_k}{G_k} \times_{G_k} \frac{\overline T_k}{\overline G_k} \, .
\end{equation}
For $k_t>0$ we expect a flow to some topological modification of 
\begin{equation}
    \frac{T_k}{G_k} \times_{G_k} \frac{\overline T_{\bar k}\times \overline G_{k - \bar k} }{\overline G_{k}} \times_{G_{k-\bar k}} G_{k-\bar k} \, .
\end{equation}

\subsection{Fermion interfaces}
In this paper we are interested in interfaces defined by coupling 3d $G=U(N)$ or $G=SU(N)$ gauge fields to $n$ flavors of 2d fundamental chiral free fermions ($\Ff_{\bC}^n$) and $\bar n$ flavors of fundamental anti-chiral fermions ($\overline \Ff_{\bC}^{\bar n}$). We will denote the chiral fermions as $\psi^a$, $\chi_a$ and the anti-chiral fermions as $\overline \psi^{\bar a}$, $\overline \chi_{\bar a}$, leaving gauge indices implicit. 

As detailed in the rest of this Section, we can build a large variety of 2d fixed points which are plausibly related by a network of RG flows. 
In a perturbative language, these fixed points arise as zeroes for the beta functions of marginal $J \cdot \bar J$ (aka four fermion) operators, with fermion masses tuned to zero.
We are particularly interested in the subset of the fixed points which admits a perturbative description when the level $\kappa$ of the 3d gauge fields is large, and in their 't Hooft expansion.

If we impose $U(n) \times U(\bar n)$ global Kac-Moody symmetry, the only option is to 
group the fermions as individual factors $T_n = \Ff^n$ and $\overline T_{\bar n}=\overline \Ff^{\bar n}$ in the above analysis. The two weakly-coupled fixed points are:
\begin{align}
    &\frac{G_{\kappa} \times \Ff_{\bC}^n}{G_{\kappa + n}} \times_{G_{\kappa + n}} \frac{\overline G_{\kappa+ n - \bar n} \times \overline \Ff_{\bC}^{\bar n}}{\overline G_{\kappa + n}} \cr 
    &\frac{\overline G_{\kappa- \bar n} \times \overline \Ff_\bC^{\bar n}}{\overline G_{\kappa}} \times_{G_{\kappa - \bar n}}  \frac{G_{\kappa- \bar n} \times \Ff_\bC^n}{G_{\kappa  + n - \bar n}} \, .
\end{align}
If we allow breaking the flavour symmetries, we can separate the individual fermion flavors in space and consider any ``comb''-like configuration with individual chiral and anti-chiral interfaces across which the Chern-Simons level jumps by $\pm 1$.

The string theory dual to pure Chern-Simons gauge theory in a 't Hooft expansion is well known: the A-model topological string theory. The 't Hooft combinatorics suggests that the  additional 2d matter will be dual to additional probe D-branes in the A-model background, which we will describe in the remainder of the paper: $n$ ``coisotropic'' D-branes and $\bar n$ ``anti-coisotropic'' D-branes.

The fermion masses couple to the non-chiral fermion bilinear operators 
\begin{align}
    \overline M^{a}_{\bar a} &\equiv \overline \chi_{\bar a} \psi^{a} \cr
    M^{\bar a}_a &\equiv \chi_a \overline \psi^{\bar a} 
\end{align}
which are meson operators and thus holographically dual to an open string field stretched between the coisotropic and anti-coisotropic D-branes. Turning on the masses should give a vev to such open string field and possibly merge the corresponding D-branes. 

The $J \cdot \bar J$ terms are meson bilinears and should thus trigger holographic RG flows between different choices of boundary conditions for the open string fields. We thus expect the RG fixed points considered here to be dual to a collection of possible boundary conditions for open string fields in the ``same'' holographic background of $n + \bar n$ 
probe D-branes.

\section{Chiral interfaces and the 't Hooft expansion.} \label{sec:chiral}
In this section we review the relation between the coset chiral algebra 
\begin{equation}
    \cV_{N,\kappa} \equiv \frac{SU(N)_\kappa \times \Ff_\bC^N}{SU(N)_{\kappa+1}} \, ,
\end{equation}
the two-parameter $\cW_\infty$ chiral algebra and the 't Hooft expansion.

Recall that the coset chiral algebra $\cV_{N,\kappa}$ is defined as the sub-algebra of $SU(N)_{\kappa} \times \mathrm{Ff}_\bC^N$ 
which is local with the total $SU(N)_{\kappa+1}$ currents
\begin{equation}
    \tilde J \equiv J + \psi \chi \, ,
\end{equation}
where $\chi$ and $\psi$ are the (anti)fundamental chiral fermions and $J$ the $SU(N)_{\kappa}$ currents. Equivalently, they must be annihilated by the non-negative modes of $\tilde J$. 

As discussed in the previous Sections, $\cV_{N,\kappa}$ occurs as the algebra of local operators at a chiral 2d interface defined by coupling $N$ complex chiral fermions to  $SU(N)$ Chern-Simons theory \cite{Costello:2016nkh,Gaiotto:2017euk}. This construction implies that $\cV_{N,\kappa}$ should admit a 't Hooft expansion, keeping fixed the 't Hooft coupling 
\begin{equation}
    t = \frac{N}{\kappa + N}
\end{equation}
and expanding in powers of $\hbar = (\kappa + N)^{-1}$. Indeed, the structure constants of the chiral algebra can in principle be computed from Feynman diagrams for the Chern-Simons theory coupled to the chiral matter, which admit a standard 't Hooft expansion. As the modification of a large $N$ gauge theory by (anti)fundamental matter (aka ``fundamental modification''), it formally defines a D-brane in a dual string theory. 

A direct proof of this statement from the definition as a coset is a bit more challenging
and is left to an enthusiastic reader.\footnote{A suggestion is to employ the presentation of the coset as the cohomology of a BRST complex 
\begin{equation}
    (SU(N)_\kappa \times \Ff^N \times SU(N)_{-\kappa-1-2N} \times \mathrm{bc}, Q_{\mathrm{BRST}})
\end{equation}
where $\mathrm{bc}$ are Lie algebra-valued ghosts and $Q_{\mathrm{BRST}}$ the BRST charge for a 2d chiral gauge theory.
} As partial evidence, the central charge has the expected large $N$ behavior: 
\begin{equation}
    c_{N,\kappa} = N + \frac{(N^2-1)\kappa}{\kappa+N}-\frac{(N^2-1)\kappa}{\kappa+N+1}\sim N(1-t^2)
\end{equation}
is linear in $N$, as appropriate for a D-brane.

The gauge theory definition also offers natural candidate operators which should admit a good 't Hooft expansion: meson operators defined classically as gauge-invariant bilinears $D_z^n \chi D_z^m \psi$ built from holomorphic covariant derivatives $D_z = \partial_z + A_z$ of the chiral fermions. We will now sharpen that definition and compare it with the coset presentation. 

Quantum-mechanically, it is useful to consider the bilocal generating function 
\begin{equation}
    \chi(z) \left[\mathrm{Pexp} \int_\gamma A \right] \psi(z+s) 
\end{equation}
for some path $\gamma$ connecting $z$ and $z+s$. Such an open Wilson line operator 
still admits a natural 't Hooft expansion and can be neatly regulated by keeping $\gamma$ off the interface away from the endpoints, so that it can be understood as a topological line defect of either $SU(N)_\kappa$ or $SU(N)_{\kappa+1}$ Chern-Simons theory ending at the interface.  

\begin{figure}[t]
  \centering
  \begin{tikzpicture}[x=1cm,y=1cm,baseline]
    \tikzset{
      midline/.style={line width=0.6pt},
      auxline/.style={line width=0.35pt},
      dot/.style={circle,fill=black,inner sep=1.2pt},
      midarrow/.style={
        postaction={
          decorate,
          decoration={
            markings,
            mark=at position 0.5 with {\arrow{>}}
          }
        }
      }
    }

    \draw[midline] (-5,0) -- (5,0);
    \node[left] at (-5,0) {$\mathrm{Ff}_{\bC}^{N}$};

    \node at (0,2.0) {$SU(N)_{\kappa}$};
    \node at (0,-2.0) {$SU(N)_{\kappa+1}$};

    \node[dot] (L1) at (-4.2,0) {};
    \node[dot] (L2) at (-2.6,0) {};
    \node[below] at (L1) {$\chi(z)$};
    \node[below] at (L2) {$\psi(z+s)$};
    \draw[auxline,midarrow]
      (L1) .. controls (-3.8,0.55) and (-3.0,0.55) .. (L2);

    \node[dot] (R1) at (2.6,0) {};
    \node[dot] (R2) at (4.2,0) {};
    \node[above] at (R1) {$\chi(z)$};
    \node[above] at (R2) {$\psi(z+s)$};
    \draw[auxline,midarrow]
      (R1) .. controls (3.0,-0.55) and (3.8,-0.55) .. (R2);

  \end{tikzpicture}
  \caption{The two gauge-invariant, regulated versions of bilocal fermion bilinear operators. Left: $W_+(z;s)$. Right: $W_-(z;s)$. The thick line represents the 2d chiral interface between 3d CS theories. The thin lines represent fundamental Wilson line defects for either Chern-Simons gauge theory. }
  \label{fig:biops}
\end{figure}
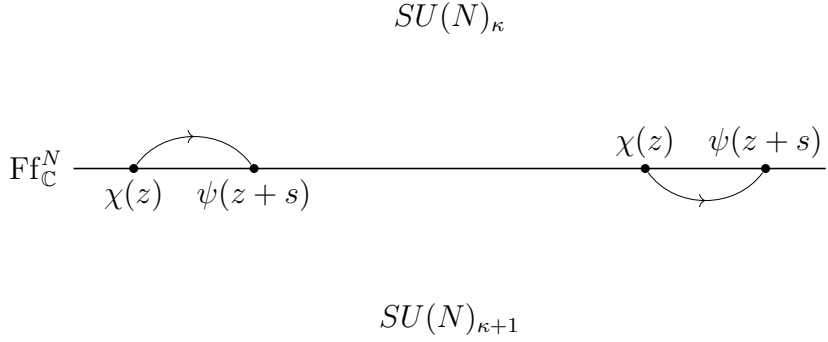

The two possible choices give distinct bilinear coset operators: 
\begin{equation}
	W_+(z;s) = \phi_{\square;1}(z) \phi_{\bar \square;1}(z+s)
\end{equation}
and 
\begin{equation}
	W_-(z;s) = \phi_{1;\bar \square}(z) \phi_{1;\square}(z+s)
\end{equation}
where we recognize the Wilson line endpoints on the interface as coset primaries labeled by the (anti)fundamental representation of either $SU(N)_\kappa$ or $SU(N)_{\kappa+1}$. See Figure \ref{fig:biops}.

The coset primaries $\phi_{\square;1}$ and $\phi_{\bar \square;1}$ are defined by combining an (anti)fundamental primary operator for $SU(N)_\kappa$ (a Wilson line endpoint) with the chiral fermions in an $SU(N)$-invariant way. The resulting scaling dimension 
\begin{equation}
	\Delta_+ \equiv \frac12 + \frac{C_2(\square)}{\kappa + N} \sim \frac{1+t}{2}\, , 
\end{equation}
is finite in the large $N$ limit. The coset primaries $\phi_{1;\square}$ and $\phi_{1;\bar \square}$ are defined by removing from the chiral fermions the contribution of an (anti)fundamental primary operator for $SU(N)_{\kappa+1}$. 
The resulting scaling dimension is
\begin{equation}
	\Delta_- \equiv \frac12 - \frac{C_2(\square)}{\kappa + N+1} \sim \frac{1-t}{2}\, .
\end{equation}
Both are ``dressed'' versions of $\chi$ and $\psi$. 

The $s \to 0$ OPE of the bilocal operators
\begin{equation}
    W_\pm (z;s) \sim \frac{1}{s^{2\Delta_\pm}} + \sum_{n>0} \frac{1}{s^{2\Delta_\pm-n}}W_{\pm;n}(z)
\end{equation}
produces two infinite towers of candidate primary operators which regulate in different ways the classical $\chi D_z^n \psi$ mesons and should still admit a natural 't Hooft expansion. 

The coset  
\begin{equation}
	\frac{SU(N)_{\kappa} \times SU(N)_1}{SU(N)_{\kappa+1}} \, ,
\end{equation}
is known to define a $\cW_N$ chiral algebra minimal model, e.g. a rational chiral algebra generated by $N-1$ currents of dimensions ranging from $2$ to $N$. The extra $U(1)$ current present in $\cV_{N,\kappa}$ combines naturally with the $\cW_N$ currents to give generators of all integer dimension from $1$ to $N$. The $U(1)$ factor in $\cV_{N,\kappa}$ is actually extended to a lattice VOA by $U(1)$ vertex operators coinciding with ``baryons'' built from $N$ factors of $\psi$ or $\chi$. We will mostly ignore this fact in the following.

We will now describe how the two $W_{\pm;n}(z)$ towers relate to each other and to the finite collection of currents which appear in the coset. 

\subsection{Miura operators as mesons}
In order to gain further insight, we invoke the well-known relation between $\cV_{N,\kappa}$ and the two-parameter family of $\cW_\infty$ chiral algebras \cite{Prochazka:2018tlo}.\footnote{We include an $U(1)$ current in the definition of the latter.} The $\cW_\infty$ chiral algebras are parameterized by three complex numbers $\lambda_i$ related as 
\begin{equation}
    \frac{1}{\lambda_1} + \frac{1}{\lambda_2} + \frac{1}{\lambda_3} =0
\end{equation}
and have central charge $c=\lambda_1 + \lambda_2 + \lambda_3+\lambda_1 \lambda_2 \lambda_3$. A triality $S_3$ symmetry permutes the $\lambda_i$ parameters. We will use the notation $\cW_\infty[\lambda_1, \lambda_2]$.

The $\cW_\infty$ chiral algebra admits a truncation whenever the $\lambda_i$ parameters satisfy certain constraints \cite{Prochazka:2015deb,Prochazka:2017qum}. If two constraints are satisfied, it can be further truncated to the chiral algebra of a Minimal Model. For example, if $\lambda_1 =N$ we can truncate $\cW_\infty$ to a $U(1) \times \cW_N$ chiral algebra with generic central charge, and if $\lambda_2 = t = \frac{N}{\kappa + N}$ we can truncate $\cW_N$ further to the Minimal Model chiral algebra. Then $\lambda_3 = -\frac{N}{\kappa+N+1}$. Conversely, $\cW_N$ is defined by replacing the WZW chiral algebra in the definition of the Minimal Model with a Kac-Moody algebra $\fsu(N)_\kappa$ 
\begin{equation}
	\frac{\fsu(N)_\kappa \times SU(N)_1}{\fsu(N)_{\kappa+1}} \, ,
\end{equation}
at generic level $\kappa$. \footnote{It is amusing to use level-rank duality \begin{equation}
    \Ff^{\kappa N} = U(\kappa)_N \times_{SU(N)_\kappa} SU(N)_\kappa \, ,
\end{equation} to derive a second coset description 
\begin{equation}
    \cV_{N,\kappa} = \frac{U(\kappa+1)_N}{U(\kappa)_N}
\end{equation}
which can be extended to non-integer $N$ and integer $\kappa$.}

In the 't Hooft expansion we lose track of integrality of both $\kappa$ and $N$, so there is no difference between $\cV_{N,\kappa}$ and $\cW_\infty[N,t]$. The 't Hooft expansion is an expansion in powers of $\lambda_1^{-1}$ at fixed $\lambda_2$. Triality allows one to 
interpret the 't Hooft expansion as a more conventional semiclassical (e.g. large central charge) expansion of $\cW_\infty[t,N]$, which will have a neat holographic interpretation.

The $\cW_\infty[\lambda_1, \lambda_2]$ algebra is endowed with three ``Miura operators'':
\begin{equation}\label{eq:Miura_diff}
    M^{(i)} \equiv \partial_z^{\lambda_i} + U^{(i)}_1(z) \partial_z^{\lambda_i-1} + U^{(i)}_2(z) \partial_z^{\lambda_i-2} + \cdots
\end{equation}
which allow one to define a certain coproduct on $\cW_\infty$. Either of the three families of currents $U^{(i)}_n(z)$ of scaling dimension $n$ can be taken as generating $\cW_\infty$, with explicit quadratic OPE relations \cite{Prochazka:2014gqa}. The relation between the three presentations is complicated. 

In \cite{Gaiotto:2020dsq}, a precise relation was established between  Miura operators and bilinears of vertex operators $\phi^{(i)}_\square$ which generalize $\phi_{\bar \square;1}(z)$, $\phi_{1;\square}(z)$ and their $\bar \square$ analogues and have scaling dimension \footnote{The physical scaling dimension for the coset fields should be $\frac12 + \frac{N^2-1}{2 N(\kappa+N)}$ and $\frac12 - \frac{N^2-1}{2 N(\kappa+N+1)}$. It is convenient, though, to dress them by appropriate $U(1)$ vertex operators which shift the dimensions to the values in the main text. One can also consider the coset \begin{equation}
    \frac{U(N)_{\kappa,\kappa+N} \times \Ff_\bC^N}{U(N)_{\kappa+1,\kappa+N+1}} \, .
\end{equation}}
\begin{equation}
    \Delta_\square^{(i)} = \frac{\lambda_i+1}{2}\,.
\end{equation}

The relation requires one to identify  
\begin{equation}
	\phi^{(i)}_{\square}(z) \phi^{(i)}_{\bar \square}(w) \sim \frac{1}{(w-z)^{1+\lambda_i}} + \cdots
\end{equation}
as the formal integral kernel for the pseudo-differential operator $M^{(i)}$: 
\begin{equation}\label{eq:intmiura}
	M^{(i)}(z,\partial_z) = \oint ds \phi^{(i)}_{\square}(z) \phi^{(i)}_{\bar \square}(z+s) e^{s \partial_z} \, .
\end{equation}

This identification matches (up to overall normalization) the currents $W_{\pm;n}(z)$ which appear in the expansion of $W_\pm(z;s)$ with the currents $U^{(2)}_n(z)$ and $U^{(3)}_n(z)$ in the quadratic presentation of $\cW_\infty$. 
Our conclusion is that the $U^{(2)}_n(z)$ and $U^{(3)}_n(z)$ currents should behave as mesons in a 't Hooft expansion of interface correlation functions. 
It would be nice to check this directly from the known explicit OPE formulae. 

We will now review some powerful integrability technology which can be used to compute explicitly the $\bC P^1$ correlation functions of any number of mesons.

\subsection{The Global Symmetry Algebra at the planar level}
A notion which proved useful in the study of holography for chiral algebras is that of the ``wedge algebra'' \cite{Bowcock:1991zk,Gaberdiel:2011wb}, aka ``global symmetry algebra'' \cite{Costello:2018zrm}. This is the linearized Lie algebra of modes of mesonic operators which annihilate the vacuum at $0$ and $\infty$ in $\bCP^1$. It is directly identified with the symmetry algebra of the dual holographic background.

In this example, the global symmetry algebra was identified with $\fhs[t]$, the Lie algebra defined by the commutators of elements of an algebra $A_t$, the central quotient of $U(\fsl_2)$ with center $\frac14(t^2-1)$. 
The conventional Minimal Model Holography proposal for an holographic dual description of the 2d CFT essentially employs a three-dimensional Chern-Simons theory with gauge group $\fhs[t]$. 

Here $\fhs[t]$ will appear in a different guise. The expected holographic dual for $SU(N)_\kappa$ Chern-Simons theory on $\bR \times \bCP^1$ is the A-model topological string theory \cite{Gopakumar:1998ki} on a $T^* \bR \times M_t$ geometry \cite{Gaiotto:2025nrd}, where $M_t$ is the deformed $A_1$ singularity with deformation parameter $t$. Recall that $A_t$ is the deformation quantization of the Poisson algebra of holomorphic functions on $M_t$. 
The symmetry algebra of global gauge transformation for a coisotropic D-brane wrapping $\bR \times M_t$ is the Lie algebra of commutators of $A_t$, i.e. $\fhs[t]$. This supports the proposal that the chiral fermion interface is dual to such a coisotropic D-brane in the A-model background.

There are two distinct representations of $A_t$ as the algebra of twisted holomorphic differential operators on $\bC P^1$, i.e. holomorphic differential operators acting on fields of scaling dimension $\frac{1\pm t}{2}$. All such twisted differential operators can be obtained by acting with the $\fsl_2$ generators
\begin{equation} \label{eq:sl2both}
    p^\pm_{2,0} \equiv \partial_z \qquad \qquad p^\pm_{1,1} \equiv z \partial_z +\frac12 (1\pm t)\qquad \qquad p^\pm_{0,2} \equiv z^2 \partial_z + (1\pm t) z
\end{equation}
on $p^\pm_{2n,0} \equiv \partial_z^n$ to define $p^\pm_{a,2n-a}$ with $0\leq a\leq 2n$ for all integer $n$, organized into irreducible $\fsl_2$ representations of  dimension $2n+1$. Geometrically, these quantize two presentations of $M_t$ as a twisted cotangent bundle of $\bC P^1$. They are related by conjugation by $\partial^t$, aka the ``shadow transform''. 

More generally, $\fhs[\lambda]$ admits representations as twisted differential operators on multiple variables $z_r$:
\begin{equation}
    t_{a,b} \to \sum_r p_{a,b}^{\pm;(r)}
\end{equation}
where $p_{a,b}^{\pm;(r)}$ represents $p_{a,b}$ acting on $z_r$. It is no coincidence that these differential operators act naturally on fields with the same scaling dimension as the planar limit of $\phi_{1;\square}(z_r)$ and $\phi_{\square;1}(z_r)$ respectively. We expect them to literally express the action of $\cW_\infty$ modes on a collection of degenerate fields in the planar limit. This is manifest for the $\fsl_2$ modes from the stress tensor, less so for the higher modes. We will come back to this momentarily. 

\subsection{The non-planar wedge algebra}
A remarkable phenomenon which occurs in this specific model is that $\fhs[t]$ admits a deformation $\Lambda[\lambda_1,\lambda_2]$ which effectively functions both as a symmetry of correlation functions at finite $N$ and as a (conjectural) quantum global gauge symmetry algebra for the coisotropic D-brane. This symmetry rigidifies the problem and effectively guarantees a full finite $N$ holographic correspondence for the $\bCP^1$ correlation functions of mesons. 

A concise definition of $\Lambda[\lambda_1,\lambda_2]$ involves the deformation of the above representations of $\fhs[t]$: we keep the same differential operators as before for $t_{a,2-a}$, but deform the $t_{4,0}$ representative to the Calogero Hamiltonian 
\begin{equation}
    t_{4,0} \to \sum_r \partial_{z_r}^2 - \sum_{r,s} \frac{2g(g+1)}{(z_r-z_s)^2}\,.
\end{equation}
We will determine the coupling $g$ in terms of the $\lambda_i$ momentarily. We can anticipate that the $\lambda_i$ coincide with $g^{-1} t$, $t$, and $-(g+1)^{-1} t$, i.e. $g=(\kappa+N)^{-1}$.

The reader can easily verify that repeated commutators of $t_{4,0}$ with $t_{0,2}$ still generate 
an $\fsl_2$ irrep of dimension $5$. It is more challenging to verify that  
\begin{equation}
    t_{2n+2,0} = \frac{1}{3n}[t_{2n,0},t_{3,1}] = \sum_r \partial_{z_r}^{n+1} +\cdots
\end{equation}
produces the infinite tower of higher Calogero Hamiltonians, all commuting with $t_{4,0}$. 
Amazingly, the $[t_{a,b},t_{c,d}]$ commutators still close to polynomials in lower generators and define a two-parameter family of algebras which depends polynomially on $g(g+1)$ and $t^2$. Up to rescaling of the generators, there is a triality symmetry permuting the $\lambda_i$, so we obtain six different representations of the same algebra. We can denote the corresponding generators as $t^{\pm;(i)}_{a,b}$.

These algebraic properties follow from an identification with the symmetry algebra controlling protected correlation functions for the world-volume theory of M2 branes with a transverse $M_t$ geometry: the ADHM quiver gauge theory with two flavours \cite{Gaiotto:2020vqj}. The algebra $\Lambda[\lambda_1,\lambda_2]$ is denoted there as ${\mathcal A}^{(2)}$ and given an alternative presentation as a shifted affine Yangian algebra. The reference also presents two natural embeddings into another algebra $A[\lambda_1,\lambda_2]$ associated to M2 branes with a transverse $\bC^2$ geometry \cite{Costello:2017fbo} and generated from the Calogero Hamiltonian and $\sum_r z_r^n$ generators \cite{Gaiotto:2020dsq}. 

That construction brings $\Lambda[\lambda_1,\lambda_2]$ within the general framework of ``twisted M-theory'' and gives a direct connection to $\cW_\infty$. Namely, the Miura operators and degenerate fields of $\cW_\infty$ satisfy intertwining relations combining the action of $\cW_\infty$ modes and (generalizations of) the Calogero representations of $A[\lambda_1,\lambda_2]$, which control their OPE and free field realizations and give them a five-dimensional geometric interpretation \cite{Gaiotto:2020dsq}. Note that twisted M-theory is described by the same non-commutative five-dimensional Chern-Simons gauge theory which occurs on the worldvolume of the coisotropic D-brane \cite{Costello:2016nkh, Costello:2017fbo}, so these algebraic structures are also directly relevant for the A-model holographic duality we propose.

It would be interesting to extend the analysis of \cite{Gaiotto:2020dsq} and identify the algebraic interplay between $\Lambda[\lambda_1,\lambda_2]$ and $\cW_\infty$ modes relevant for 
correlation functions of degenerate fields in $\bC P^1$. Here we will use a powerful shortcut: the $\bC P^1$ correlation function of a product of any number of Miura operators is known explicitly \cite{Gaiotto:2023ynn}:
\begin{equation} \label{eq:miuco}
    \langle \prod_r M^{(i)}(z_r,\partial_{z_r}) \rangle_{\bC P^1} = \left(\prod_r \partial_{z_r} + \cdots\right)^{\lambda_i}. 
\end{equation}
The expression on the right hand side is the $\lambda_i$-th power of a specific higher Calogero Hamiltonian. In particular, it commutes with $t_{4,0}$!

On the other hand, global conformal transformations act on $\phi^{(i)}_\square$ and $\phi^{(i)}_{\bar \square}$ by $t^{+;(i)}_{a,2-a}$. The integral expression (\ref{eq:intmiura}) for the Miura operators implies that they act on $M^{(i)}(z,\partial_z)$ as the difference of 
$p^+_{a,2-a}$ from the left and $p^-_{a,2-a}$ from the right. In other words, 
the correlation function of Miura operators intertwines the $t^{+;(i)}_{a,b}$ and $t^{-;(i)}_{a,b}$ representations of $\Lambda[\lambda_1,\lambda_2]$! Conversely, this property completely determines the correlation function.

We can give some simple examples. First, 
\begin{equation}
    \langle \phi_\square(z) \phi_{\bar \square}(w) \rangle = \frac{1}{(z-w)^{1+\lambda}}
\end{equation}
which is obviously annihilated by 
\begin{align}
    &\partial_z + \partial_w \cr
    &z \partial_z + w \partial_w + \lambda + 1 \cr
    &z^2 \partial_z + (\lambda + 1) z + w \partial_w + (\lambda + 1) w \cr
    & \partial_z^2 - \partial_w^2
\end{align}
Next, a four point function. Conformal invariance is satisfied by the ansatz
\begin{equation}
    \langle \phi_\square(z_1)\phi_\square(z_2) \phi_{\bar \square}(w_1)\phi_{\bar \square}(w_2) \rangle = \frac{1}{(z_1-z_2)^{\lambda+1}(w_1-w_2)^{\lambda+1}}f\left(\frac{(z_1-w_1)(z_2-w_2)}{(z_1-z_2)(w_1-w_2)}\right)
\end{equation}
and imposing that it is annihilated by 
\begin{equation}
    \partial_{z_1}^2+ \partial_{z_2}^2 - \frac{2 g (g+1)}{(z_1-z_2)^2}  -\partial_{w_1}^2- \partial_{w_2}^2 + \frac{2 g (g+1)}{(w_1-w_2)^2}
\end{equation}
gives the hypergeometric differential equation:
\begin{equation}
    u(1-u) f''(u) + (\lambda+2)(1-2u) f'(u)- (\lambda + g + 2)(\lambda-g+1)=0
\end{equation}
whose solution is ${}_2F_1(a,b;c;u)$ with parameters $a=\lambda + g + 2$, $b=\lambda - g + 1$, $c = \lambda+2$.

The solutions behave as $\frac{1}{(z_1-w_1)^{\lambda+1}}$ or as $1$ in the limit $z_1 \to w_1$. This is as expected for the OPE of a fundamental and an anti-fundamental degenerate fields: the first power is the identity channel and the second is the adjoint channel. The same behavior occurs for $z_1 \to w_2$. For $z_1 \to z_2$ we get $(z_1-z_2)^{g+1}$ or $(z_1-z_2)^{-g}$, corresponding to the symmetric and anti-symmetric channels. 

By expanding out the $z_r \to w_r$ OPE, one can compare the result with the two point functions of Miura operators:
\begin{equation}
    \langle M(z_1,\partial_{z_1})M(z_2,\partial_{z_2}) \rangle = \left( \partial_{z_1} \partial_{z_2} - \frac{g(g+1)}{(z_1-z_2)^2}\right)^\lambda\,.
\end{equation}

We expect analogous statements to hold for correlation functions of any number of degenerate fields of different types, with a generalized Calogero representation $t^+_{a,b}$ acting on the $\square$ fields and another representation $t^-_{a,b}$ acting on $\bar \square$ fields. These generalized Calogero representations are defined in \cite{Gaiotto:2020dsq} with the help of certain meromorphic coproducts. 

It would be interesting to develop these computational tools further. For our purpose, it will be sufficient to identify the same integrable structures in the holographic dual description of the system. We will do so in Section \ref{sec:coiso}.

\subsection{Multiple chiral fermions}
The discussion is readily extended to the case of $K N$ complex fermions. The coset 
\begin{equation}
	\cV^{(K)}_{N,\kappa} \equiv \frac{SU(N)_{\kappa} \times \Ff_\bC^{KN}}{SU(N)_{\kappa+K}} \, .
\end{equation}
is a truncation of a chiral algebra denoted as $\cW^{(K)}_\infty[N,t]$, which is generated by an infinite tower of fields deforming the $K \times K$
\begin{equation}
	\chi^i \partial_z^n \psi_j
\end{equation}
matrices of meson operators. These chiral algebras depend on parameters $\lambda_1 = N$, $\lambda_2 = t$, $\lambda_3 = -\frac{N}{N + K + \kappa}$, but only have a duality symmetry  
$\lambda_2 \leftrightarrow \lambda_3$. 

We can define as before degenerate fields $\phi^i_{\square;1}$ and $\phi_{i;\bar \square;1}$, as well as $\phi_{i;1;\square}$ and $\phi^i_{1;\bar \square}$ with dimensions controlled by $\lambda_2$ and $\lambda_3$ again. We can also define matrix-valued Miura operators $M^{(2)}$ and $M^{(3)}$ from the OPE of degenerate fields. Although we have not proven this, we expect them to coincide with the matrix-valued Miura operators employed in alternative definitions of $\cW^{(K)}_\infty[N,t]$ \cite{Rapcak:2019wzw,Gaiotto:2023ynn}. They contain two alternative sets $U^{(2)}_{n}$ and $U^{(3)}_{n}$ of matrix-valued generators of all integral spins.  

The Global Symmetry Algebra in the planar limit is now the Lie algebra for $\fgl_K[A_t]$, as appropriate for a collection of $K$ coincident coisotropic D-branes. It has a finite $N$ counterpart $\Lambda^{(K)}[\lambda_1,\lambda_2]$ which can be defined e.g. from the theory of M2 branes with a transverse $M_t \times A_{K-1}$ geometry or as a shifted affine Yangian. It admits a representation combining the conformal generators and a $\fgl_K$ spin-Calogero Hamiltonian and the correlation function of Miura operators is a fractional power of a higher Hamiltonian. We refer to \cite{Gaiotto:2023ynn} for details.

\subsection{Symplectic boson interfaces}\label{sec:sym_bos}
We can also discuss a chiral interface where the 3d CS gauge fields are coupled to ``symplectic bosons'', aka bosonic chiral free fields $X(z)$, $Y(z)$ with the same OPE as free fermions. This system is not unitary but it is still interesting. We can denote the corresponding coset as 
\begin{equation}
	\cV^b_{N,\kappa} \equiv \frac{SU(N)_{\kappa+1} \times \Sb_\bC^{N}}{SU(N)_{\kappa}} \, .
\end{equation}
In \cite{Gaiotto:2017euk} it was observed that this coset is still a truncation of the $\cW_\infty$ algebra, with $\lambda_1=-N$. Our whole discussion thus applies almost unchanged and the duality involves a ``parity shifted'' or ``ghost'' coisotropic D-brane. 

We can combine chiral fermions and bosons: 
\begin{equation}
	\cV^{(K|K')}_{N,\kappa} \equiv \frac{SU(N)_{\kappa} \times \Ff_\bC^{KN}\times \Sb_\bC^{K'N}}{SU(N)_{\kappa+K - K'}} \, .
\end{equation}
and work with chiral and Calogero algebras associated to $\fgl_{K|K'}$, see e.g. \cite{Rapcak:2019wzw}. 

\section{Coisotropic D-branes and Holography} \label{sec:coiso}
The  three-dimensional $SU(N)_\kappa$ Chern-Simons theory is expected to be holographic dual to the A-model Topological String Theory \cite{Ooguri:1999bv}.
The CS theory can be defined on any three-manifold, but the details of the 
holographic duality are only understood for a few examples: the three-sphere $S^3$ \cite{Ooguri:1999bv}, discrete quotients thereof \cite{Aganagic:2002wv,Halmagyi:2003mm,Brini:2008ik}, and most recently $\bR \times \bC P^1$ \cite{Gaiotto:2025nrd}. The latter setup is the most convenient for our purposes: it allows for chiral or anti-chiral interfaces wrapping the $\bC P^1$ factor and the dual A-model background has some useful special properties.

Recall that the A-model is naturally defined on real symplectic manifolds. The holographic dual to $SU(N)_\kappa$ Chern-Simons theory on $\bR \times \bC P^1$ conjectured in \cite{Gaiotto:2025nrd} takes a factorized form:
\begin{equation}
    T^*\bR \times M_t
\end{equation}
with the complex symplectic manifold $M_t$ (aka the deformed $A_1$ singularity) seen as a real symplectic manifold by selecting the real part of the complex symplectic form. 

Following the general philosophy of Maldacena duality \cite{Maldacena:1997re, Witten:1998qj}, which presents holography as a limit of open-closed duality, much insight on the holographic dictionary can be gained by engineering the desired QFT on a stack of D-branes and replacing the D-branes by their own back-reaction. In the case at hand, one can start from $N$ D2 branes wrapping $\bR \times \bC P^1$ in $T^* (\bR\times \bC P^1)$ and deform the symplectic form so that it has period $t$ on a two-sphere surrounding the base of the fibration. 

This perspective predicts the dual description of defects in the QFT defined by coupling to extra degrees of freedom which mimic open strings stretched from the original stack to an extra ``probe'' D-brane. The dual description involves a probe D-brane reaching the holographic boundary at the location of the defect. 

Conventional D2 branes in the A-model, which wrap Lagrangian submanifolds, only give rise to topological defects. We will come back to these briefly in Section \ref{sec:Wilson}. Fortunately, the A-model also allows for less conventional ``coisotropic'' D4 branes. The definition of coisotropic D-branes is somewhat intricate, but our setup admits an important collection of ``canonical coisotropic'' D-branes. 

More precisely, the A-model with a four-dimensional target space $X$ equipped with the structure of a complex symplectic manifold admits a canonical coisotropic brane, which we will refer to as $B_\cc[X]$. The space of $B_\cc[X]-B_\cc[X]$ open strings coincides with the deformation quantization of the algebra of holomorphic functions on $X$. We can thus define coisotropic branes in $T^*\bR \times X$ by combining a Lagrangian brane in $T^*\bR$ and $B_\cc[X]$.

A $T^*_x \times B_\cc[T^* \bC P^1]$ D4 brane has a 2d intersection with a D2 brane supported on $\bR \times \bC P^1$ in $T^* (\bR\times \bCP^1)$ and is known to add an open string sector consisting of a complex chiral 2d fermion defined on the $\bC P^1$ intersection (or symplectic bosons, depending on the Grassmann parity of the Chan-Paton bundle). See Appendix \ref{apdx:open_string} for a derivation of this fact. 

After back-reaction we still have a complex symplectic factor in the geometry. It is thus natural to conjecture that our chiral 2d interfaces are dual to a $T^*_x \times B_\cc[M_t]$ D4 brane. We will test this conjecture extensively in this Section. See \cite{Aharony:2019suq} for an alternative holographic application of coisotropic D-branes.


As a first test, observe that the deformation quantization of $M_t$ reproduces the algebra $A_t$, i.e. the central quotient of $U(\fsl_2)$ with Casimir $\frac14(t^2-1)$. Parsing definitions, this identifies $\fhs[t]$ with the classical algebra of global gauge transformations for the $T^*_{x} \times B_\cc[M_t]$ coisotropic D-brane. This is the same symmetry algebra found in the QFT.

\subsection{HT-nc-CS theory on $\mathbb{R}\times M_t$}

In this section, we present more details on the non-commutative Chern-Simons theory on $\mathbb{R}\times M_t$, focusing in particular on its quantization. The work \cite{Costello:2016nkh} discusses in detail how the quantum theory can be defined on $\mathbb{R}\times X_4$, where $X_4$ is a holomorphic-symplectic manifold equipped with a $\mathbb{C}^{\times}$-action. Although the deformed $A_1$ singularity $M_t$ does not admit such a $\mathbb{C}^{\times}$-action, one can still quantize the theory in a suitable sense. 

First, we consider the classical HT-nc-CS theory on $\mathbb{R} \times X_4$, where $X_4$ is a $4d$ holomorphic symplectic manifold. To write down a classical action functional, we need a non-commutative star product $*$ on the sheaf of holomorphic functions $\mathcal{O}_{X_4}$. This star product should make $\mathcal{O}_{X_4}$ into a sheaf of associative algebras. Moreover, we require that each term in the expansion of the $*$-product be a bi-differential operator of finite order. These conditions are satisfied by the quantization of the algebra of holomorphic functions on $X_4 = M_t$. We can then extend the $*$-product to a $*$-product on $\mathcal{A}^{\bullet}_{M_t} = \Omega^{\bullet}(\mathbb{R}) \otimes \Omega^{0,\bullet}(M_t)$.

Thus, we can define a classical field theory on $\mathbb{R}\times M_t$ with action functional 
\begin{equation}
	S = \int_{\mathbb{R}\times M_t} \omega\,\mathrm{Tr}\left( \frac{1}{2}\alpha * d\alpha + \frac{1}{3} \alpha*\alpha *\alpha \right), \quad\quad\alpha\in\mathcal{A}^{\bullet}_{M_t}\otimes\mathfrak{gl}_K[1]
\end{equation}
where $\omega$ is the holomorphic volume form on $M_t$. Here $K$ is the number of coincident D4 branes. Although an individual interface corresponds to $K=1$, it is useful to consider general $K$ at first. This is also relevant for the $\Ff^{K N}_\bC$ 
interface. 

Analyzing the quantization of such a theory requires us to compute the obstruction–deformation complex. See \cite{Costello:2015xsa} for an instructive example. This complex is built from all possible terms one can add to the Lagrangian in different ghost degree, and the differential is given by the classical BRST differential. For the HT-nc-CS theory on $\mathbb{R} \times X_4$, this amounts to computing the Lie algebra (co)homology of $J\mathcal{O}_{X_4} \otimes \mathfrak{gl}_K$, where $J\mathcal{O}_{X^4}$ is the jet bundle of $\mathcal{O}_{X^4}$. Such infinite dimensional Lie algebra cohomology is a difficult task. The solution proposed in \cite{Costello:2015xsa,Costello:2016nkh} is to consider the entire family of theories for all $K$ and imposing a “uniformly in $K$’’ quantization. 

This means that we are only allowed to consider Lagrangians compatible with the inclusions $\mathfrak{gl}_K \to \mathfrak{gl}_{K+k}$. Such Lagrangians consist only of products of single-trace operators. Under this restriction, the problem of computing the obstruction–deformation complex reduces to computing the cyclic cohomology of $J\mathcal{O}_{X_4}$. The “uniformly in $K$’’ obstruction–deformation complex at each loop level has cohomology
\begin{equation}\label{eq:comp_defobs}
	H^{\bullet}(X_4)\otimes (\bC[\eta_3,\dots, \eta_{2k+1},\dots]/\bC)[5]\,,
\end{equation}
where the fermionic generator $\eta_{2k+1}$ lives in degree $2k+1$. The obstruction class, which corresponds to possible anomalies of the quantum theory, lives in total degree $1$ of the above cohomology and is given by
\begin{equation}
	\eta_5 H^1(X_4) \oplus \eta_3 H^3(X_4)\,.
\end{equation}
Fortunately, for $X_4 = M_t$ both $H^1(M_t)$ and $H^3(M_t)$ vanish, so the theory has no anomaly and can be perturbatively quantized. Moreover, we have 
\begin{equation}
	H^{i}(M_t) = \begin{cases}
		\bC, & i = 0,2\\
		0, & \text{others}
	\end{cases}
\end{equation}
Here, $H^2(M_t)$ is generated by the holomorphic symplectic form $\omega$ on $M_t$. The non-vanishing groups $H^0(M_t)$ and $H^2(M_t)$ correspond to the deformation classes $\eta_5 H^0(X_4) \oplus \eta_3 H^2(X_4)$, which lie in the total degree 0 part of the cohomology \eqref{eq:comp_defobs}. This indicates that the HT-nc-CS theory on $\mathbb{R} \times M_t$ may exhibit ambiguities in its quantization. In particular, one may add terms to the action functional at each loop order without generating anomalies. 

The deformation corresponding to $H^0(M_t)$ can be absorbed into a rescaling of the holomorphic volume form, while the deformation associated with a class in $H^2(M_t)$ corresponds to a first-order deformation of the $*$-product. However, the generating class $\omega$ of $H^2(M_t)$ is precisely the symplectic form used to define the original $*$-product on $M_t$. Thus, this quantization ambiguity can be viewed as a scaling of the symplectic form, $\omega \to f(\hbar)\,\omega$, which can be fixed by imposing the normalization condition on the period
\begin{equation}
    \int_{S^2} \omega = t .
\end{equation}
Therefore, perturbative quantization of the HT-nc-CS theory on $\mathbb{R} \times M_t$ is possible, and the only ambiguity is absorbed into the definitions of $t$ and of the overall coupling in the action. 

\subsection{Line defects in 5d nc-HT CS $U(1)$ gauge theory and Calogero representations}
The 5d theory on $T^*_{x} \times M_t$ is topological along the $T^*_{x}$ direction, which we denote as $y$. It is possible to define a variety of topological Wilson line defects, two of which will describe the endpoints of open strings attached to the D4 brane. We will employ such line defects to represent the insertion of $\phi^{(i)}_\square$
and $\phi^{(i)}_{\bar \square}$ degenerate fields in the QFT.\footnote{The 5d theory also includes ``instanton particles'' which are related to the Wilson lines by triality. They should be dual to the insertion of $\phi^{(1)}_\square$ and $\phi^{(1)}_{\bar \square}$ operators, of dimension $\frac{N}{2}$. These likely correspond to ``baryon'' operators defined as the product of all of the $\chi$ fermions or all of the $\psi$ fermions.} 

Although the 5d theory is presented as a gauge theory, it is somewhat gravitational in nature. The gauge field can be also understood as describing deformations of the complex symplectic structure of the holomorphic directions. In a gravitational theory we do not expect to see non-dynamical line defects: line defects should include dynamical degrees of freedom representing their position in spacetime. Something similar happens in the 5d theory: the naive Wilson lines $\mathrm{Pexp} \int A_{y_3} dy_3$ are not gauge invariant 
unless we include extra degrees of freedom representing a position in $M_t$.

Mathematically, the 5d theory is associated to a ``gauge algebra'' $A_{\mathrm{top}}$
and gauge-invariant topological line defects are built from quantum mechanical systems which admit a representation of $A_{\mathrm{top}}$. In flat space $X = \bC^2$ the gauge algebra $A[\lambda_1, \lambda_2]$ was computed in \cite{Costello:2017fbo}, with $\epsilon_i = \lambda_i^{-1}$ (but only depends on the ratios of $\epsilon_i$'s). It admits three distinct algebra maps to the Weyl algebra $\bC[z,\partial_z]$, which define three topological line defects, two of which are ``dynamical'' versions of Wilson lines of charge $\pm 1$. The third represents an instanton particle in the 5d gauge theory. In the presence of multiple defects, one uses the Calogero-like representations mentioned in the previous Section \ref{sec:chiral}.

Based on \cite{Costello:2017fbo,Gaiotto:2019wcc,Gaiotto:2020vqj} and on consistency with $A[\lambda_1, \lambda_2]$ when restricting to $T^*\bC$ patches, we identify $\Lambda[\lambda_1, \lambda_2]$ as the gauge algebra for $X = M_t$. For example, there are  three maps $\Lambda[\lambda_1, \lambda_2]\to A_{\lambda_i}$ to central quotients of $U(\fsl_2)$ with Casimirs $\frac14(\lambda_i^2-1)$, which represent the three types of elementary topological line defects in $T^*_{x^3} \times M_t$. More complicated Calogero representations obtained as coproducts of multiple copies of these maps encode collections of topological line defects. 

In order to relate the topological line defects to $\phi^{(i)}_\square$
and $\phi^{(i)}_{\bar \square}$ boundary insertions, we can use the general philosophy of Maldacena duality: pair up holographically the boundary operators and bulk objects which enter world-volume couplings in the original stack of $N$ D-branes in flat space whose back-reaction creates the holographic geometry. In the case at hand, take the 3d D-branes  to wrap the base of $T^*\bR \times T^* \bC P^1$ and the coisotropic D-brane to sit at $x=0$. 
The 2d intersection supports the $\cW_\infty$ chiral algebra and Wilson lines in the coisotropic worldvolume can end on the intersection from $y>0$ or $y<0$. That creates respectively the $\phi^{(i)}_\square$ or $\phi^{(i)}_{\bar \square}$ degenerate fields.

The back-reaction to $T^*\bR \times M_t$ is described in detail in \cite{Gaiotto:2025nrd} and reviewed in Appendix \ref{app:geometry}. The main feature we use here is that the two presentations of $M_t$ as a twisted cotangent bundle 
of $\bC P^1$ are employed respectively at $y\gg 0$ and $y \ll 0$. Recall that the two presentations are quantized to the two presentations of $A_{\lambda}$ as twisted differential operators on $\bC P^1$. For example, the two quantum presentations are intertwined by $\partial_z^\lambda$, while the two classical presentations are intertwined by $z \to z + \frac{\lambda}{p}$ where $p$ is the fiber coordinate. 

Accordingly, we expect a $\phi^{(i)}_\square(z)$ insertion in the boundary QFT to create a Wilson line of the same type extending along $y\gg 0$, with the gauge algebra represented by the $t^+_{a,b}$ representation. The $\phi^{(i)}_{\bar \square}(z)$ creates a Wilson line extending along $y\ll 0$ and gauge algebra represented by the $t^-_{a,b}$ representation. 

The holographic correlation function is then constrained by the same global symmetry algebra as the QFT correlation function, even at finite $N$. In particular, the correlation function 
\begin{equation}
    \langle \prod_a \phi^{(i)}_{\square}(z_a) \phi^{(i)}_{\bar \square}(w_a) \rangle_{\bC P^1}
\end{equation}
intertwines the $t^{\pm;(i)}_{a,b}$ representations of $\Lambda[\lambda_1,\lambda_2]$ just as we saw in the QFT. An holographic calculation of correlation functions will thus reproduce the QFT answer at all orders in the 't Hooft expansion!

\section{Higher spin gravity and holography}\label{sec:hskk}
In this section, we compare our topological string construction with the minimal model holography proposal in the literature. The minimal model holography predicts a correspondence between $3d$ Vasiliev higher-spin theory, or the Chern–Simons theory based on the higher-spin algebra $\mathfrak{hs}[\lambda]$—and the $2d$ $\mathcal{W}_{\infty}$ minimal model CFT. Our goal here is to explain how the $3d$ Chern–Simons description of higher-spin gravity is related to the 5d HT-nc-CS theory on the deformed geometry obtained from the brane construction.

The first observation is that the holographic dictionary for the 5d HT-nc-CS takes a standard form. 
Recall e.g. the holographic duality for ${\cal N}=4$ SYM: IIB supergravity in AdS${}_5 \times $S${}^5$ is KK-reduced on S${}^5$ to produce a collection of fields in AdS${}_5$, which receive boundary conditions setting to zero the modes which grow towards the boundary. Incidentally, the KK reduction procedure is basically identical in AdS${}_5 \times $S${}^5$ or in flat $\bR^{3,1}\times \bR^6$ and the back-reaction could be studied directly in the KK-reduced theory. 

Analogously, we could consider the KK-reduction of the 5d HT-nc-CS on an $S^2$. Geometrically, we split the ambient A-model $\bR^2 \times \bC^2$ geometry as $(\bR^2 \times \bC) \times \bC$
and KK-reduce the first factor on an $S^3$. At the same time, the $\bR \times \bC$ subspace 
wrapped by the coisotropic brane is reduced on an $S^2$. The sphere KK-reduction of a collection of holomorphic and topological directions always produces a topological theory in the radial direction. See \cite{Zeng:2023qqp} for a purely holomorphic example. The result of the KK reduction is thus an holomorphic-topological 3d theory on $\bR^+ \times \bC$. 

We will compute the KK reduction explicitly and recognize the result as a 3d chiral Poisson sigma model as defined in \cite{khan2025poisson}. A standard boundary condition which sets to zero the ''growing'' modes from the KK reduction will precisely produce a boundary chiral algebra with the correct structure to match $\cW_\infty$. More precisely, the classical action of the 3d model contains the information of a ``Poisson chiral algebra'', which one can compare to the $\lambda_1 \to \infty$ semiclassical limit of $\cW_\infty[\lambda_1=N,\lambda_2=t]$. We will provide a complete proof of this match, including the effect of back-reaction. 

The $\mathfrak{hs}[\lambda]$ Chern-Simons proposal does not quite have the standard form of an holographic correspondence. It is instead a generalization of the well-known realization \cite{Verlinde:1989ua,Gaiotto:2017euk} of $\cW_n$ as an ``oper'' boundary for $SL(n)$ Chern-Simons theory. The basic idea is that $\mathfrak{hs}[t]$ can be seen as an analytic continuation of $\mathfrak{sl}(t)$: when $t$ is an integer, the central quotient of $U(\fsl_2)$ can be truncated to the $n \times n$ matrix algebra. An ``oper'' boundary condition for an $\mathfrak{hs}[t]$ Chern-Simons theory should then reproduce a $\cW_\infty$ algebra with $\lambda_2=t$.

If we stick to a conventional perspective on holography, the $\mathfrak{hs}[t]$ Chern-Simons theory proposal can only be ``right'' if there is some kind of gauge equivalence between that theory and a theory such as the KK-reduced 5d HT-nc-Chern-Simons theory, for which the holographic dictionary takes a standard form.  

This sort of question is interesting even for the simpler context of $\cW_n$ chiral algebras: what is the relation between an $SL(n)$ Chern-Simons theory with oper boundary condition and the 3d chiral Poisson sigma model with action determined by the Poisson limit of the $\cW_n$ chiral algebra? Are the two theories gauge equivalent, or is the relation more subtle? For example, one may envision a scenario where the two theories are {\it not} equivalent, but rather related by a topological interface. 

While this question goes beyond the scope of this paper, we will sketch in Appendix \ref{app:DS} an argument for equivalence, based on a semiclassical limit of the quantum DS reduction relating the $\fsl_n$ Kac-Moody chiral algebra and the $\cW_n$ chiral algebra. 

\subsection{Poisson sigma model of classical $\mathcal{W}_{\infty}$ algebra}
\label{sec:agd-review}
We first briefly review the construction of $3d$ chiral Poisson sigma model defined in \cite{khan2025poisson}, which assigns to any chiral Poisson algebra a three-dimensional gauge theory:
\begin{equation*}
	\{\text{chiral(vertex) Poisson algebras}\} \longrightarrow \{3d \text{ holomorphic topological theories}\}
\end{equation*}
Given a chiral Poisson algebra $\mathbb{C}[\partial^n u_i\mid n \geq 0,i = 1,\dots n]$ equipped with $\lambda$-bracket\footnote{We refer to \cite{kac2017introduction} for more details on the $\lambda$-bracket formulation of chiral Poisson algebra.} $\{u_i~_{\lambda}u_j\} = H_{ji}(u,\partial u,\dots)(\lambda)$, the associated $3d$ theory has scalar fields $\phi_i(x,z,\bar{z})$ and one form fields $\eta^i(x,z,\bar{z}) = \eta^i_xdx + \eta^i_{\bar{z}}d\bar{z}$. The action functional is given by
\begin{equation}
	S = \int_{\mathbb{R}\times\mathbb{C}}dz\left(\eta^i(d_x+\bar{\partial})\phi_i + \frac{1}{2}\eta^iH_{ij}(\phi,\partial\phi,\dots)(\partial)\eta^j \right) \,.
\end{equation}
We have the following gauge transformation on the fields:
\begin{equation}\label{eq:PSM_gauge}
	\begin{aligned}
		&\delta\phi_i = H_{ij}(\partial)\varepsilon^j,\\
		&\delta\eta^i = -(d_x + \bar{\partial})\varepsilon^i-\frac{1}{2}\sum_{n\geq 0}(-1)^n\left(\eta^j\frac{H_{jk}(\partial)}{\partial(\partial^n\phi_i)}\varepsilon^k -\varepsilon^j\frac{H_{jk}(\partial)}{\partial(\partial^n\phi_i)}\eta^k\right) \,.
	\end{aligned}
\end{equation}
It is shown in \cite{khan2025poisson} that the gauge invariance of the action functional is equivalent to the Jacobi identity of the $\lambda$-bracket. 

One important feature of this $3d$ theory is that the boundary algebra associated with the boundary condition $\eta^i|_{\partial} = 0$ naturally provides a quantization of the chiral Poisson algebra to a chiral algebra, provided the three-dimensional theory is anomaly-free.  

We can apply the above construction to the classical $\mathcal{W}_{\infty}$ algebra, which naturally defines a higher spin theory. The boundary chiral algebra of this theory will quantize the classical $\mathcal{W}_{\infty}$ algebra. We will not attempt to show directly that the potential quantum anomalies of the 3d theory can be canceled, but rather focus on classical considerations.

We can discuss the classical $\mathcal{W}_{\infty}$ algebra in the language introduced by Adler and Gelfand–Dickey \cite{Gelfand:1987qu,adler1978trace}, formulated in terms of (pseudo-)differential operators. We follow the presentation in \cite{de2015adler} and focus on the chiral Poisson algebra structure. The standard construction involves $\cW_t$ for integer $t$, but it is readily extended to general $t$. 

The classical $\mathcal{W}_{n}$ algebra is a chiral Poisson algebra structure on the differential polynomials $\mathbb{C}[\partial^k u_i\mid k \geq 0,i = 1,\dots n]$. To describe the corresponding $\lambda$-bracket, it is convenient to introduce the differential operator
\begin{equation}
	L(\partial) = \partial^n + u_1\partial^{n-1} + u_2\partial^{n-2} + \dots + u_{n}\, .
\end{equation}

Associated to $L$ is the Adler map $A^{(L)}(X)=(L\circ X)_+\circ L-L\circ(X\circ L)_+$, where the product $\circ$ of differential operator is defined so that $\partial^n\circ u = \sum_{k\geq0}\binom{n}{k}(\partial^ku)\partial^{n-k}$, and $(-)_+$ extract non-negative powers of $\partial$ in the expression. Equivalently, the Adler map $A^{(L)}$ has the coefficient form $H_{ij}^{(L)}(\partial)$, $1\leq i,j\leq n$. We denote
$H^{(L)}(\partial)(z,w) =\sum_{i,j}H_{ij}^{(L)}(\partial)z^{n-i}w^{n-j}$, which can be expressed as
\begin{equation}\label{eq:Adler_map}
	H^{(L)}(\partial)(z,w) = L(w)i_w(w - z - \partial)^{-1}\circ L(z) - L(z + \partial)i_w(w-z-\partial)^{-1}\circ L^*(-w + \partial)\,.
\end{equation}
The formal adjoint $L^*(\partial)$ is defined by $L^*(\partial) = \sum_{l}(-\partial)^l\circ u_{n-l}$. We also denote $i_w$ the operation of power series expansion around large $w$, so that $ i_w(w- z)^{-1} = \sum_{k\geq 0} z^k/w^{k+1}$. Then the (second type) Adler-Gelfand-Dickey bracket can be written as
\begin{equation}
\begin{aligned}
		&\{L(z)~_{\lambda}L(w)\} = H^{(L)}(\lambda)(w,z)\\
		&= L(z)i_z(z - w -\lambda- \partial)^{-1} L(w) - L(w +\lambda + \partial)i_z(z-w-\lambda-\partial)^{-1} L^*(-z + \lambda)\,.
\end{aligned}
\end{equation}
The above expression can be expanded as follows \cite{de2015adler}:
\begin{equation}\label{eq:GD_bracket}
	\begin{aligned}
			\{u_i~_{\lambda}u_j\} = &\sum_{k,r\geq 0}\binom{k}{r}u_{i-k-1}(\lambda+\partial)^ru_{j+k-r}\\
			&- \sum_{k,r,s\geq 0}(-1)^r\binom{-n-1+j}{r}\binom{-n+i-k-2}{s}u_{j+k-r}(\lambda+\partial)^{r+s}u_{i-k-s-1}\,,
	\end{aligned}
\end{equation}
where we set $u_0 = 1$ and $u_{k} = 0$ for $k<0$.


The $\mathcal{W}_{\infty}$ algebra can be understood as the interpolation of the $\mathcal{W}_{t}$ algebra to complex values of $t$ (or to a formal parameter). This is made possible by the fact that the coefficients of the $\lambda$-brackets are polynomial functions of $t$. Consequently, we can define the chiral Poisson algebra $\mathcal{W}_{\infty}$ as the algebra of differential polynomials $\mathbb{C}[\partial^k u_i \mid k \geq 0, i \geq 1]$, equipped with $\lambda$-brackets as the interpolation of \eqref{eq:GD_bracket}. A more precise definition can be found in \cite{khesin1996universal,riesen2025interpolating}, which we review here. Let $t$ be a complex number or a formal parameter. We denote $\Psi \mathrm{DO}^{t}$ the space of pseudo-differential operators with leading order $t$.	This is the set of formal symbols of the following form
\begin{equation}
	\partial^t + \sum_{i\geq 1}u_i\partial^{t-i}\,.
\end{equation}
Such operators can also be identified with the classical limit of the Miura operators \eqref{eq:Miura_diff}. Given a pseudo-differential operator $L(\partial) = \partial^t + \sum_{i\geq 1}u_i\partial^{t-i}$, the corresponding Adler map $H^{(L)}(\partial)(z,w) = \sum_{i,j}H_{ij}^{(L)}(\partial)z^{t-i}w^{t-j}$ is given by the same expression as \eqref{eq:Adler_map}. The corresponding $\lambda$-bracket is given by $	\{L(z)~_{\lambda}L(w)\} = H^{(L)}(\lambda)(w,z)$, which can be expanded as
\begin{equation}\label{eq:GD_bracket_inf}
	\begin{aligned}
		\{u_i~_{\lambda}u_j\} = &\sum_{k,r\geq 0}\binom{k}{r}u_{i-k-1}(\lambda+\partial)^ru_{j+k-r}\\
		&- \sum_{k,r,s\geq 0}(-1)^s\binom{t-j+r}{r}\binom{t-i+k+s+1}{s}u_{j+k-r}(\lambda+\partial)^{r+s}u_{i-k-s-1}\,.
	\end{aligned}
\end{equation}
The matrix extended classical $\mathcal{W}_\infty$ algebra can be constructed in a similar way. We refer to \cite{de2015adler} for more details.

Now we consider the $3d$ chiral Poisson sigma model associated with the classical $\mathcal{W}_{\infty}$ algebra. This theory has an infinite tower of fields $(\phi_i,\eta^i), i\geq 1$. The kinetic term is the standard one $\int_{\mathbb{R}\times\mathbb{C}}dz \eta^i(d_x+\bar{\partial})\phi_i$. The interaction term is given by
\begin{equation} \label{eq:Winf_matrix}
	\begin{aligned}
		&\sum_{i,j\geq 0}\int_{\mathbb{R}\times\mathbb{C}}dz\sum_{k,r,l\geq 0}\binom{k}{r}\binom{r}{l}\phi_{i-k-1}\partial^{r-l}\phi_{j+k-r}\eta^i\partial^l\eta^j\\
		&- \sum_{k,r,s,l\geq 0}(-1)^s\binom{t-j+r}{r}\binom{t-i+k+s+1}{s}\binom{r+s}{l}\phi_{j+k-r}\partial^{r+s - l}\phi_{i-k-s-1}\eta^i\partial^l\eta^j\,,
	\end{aligned}
\end{equation}
where we set $\phi_0 = 1$ and $\phi_i = 0$ for $i < 0$. As a result, the interaction contains quadratic, cubic, and quartic terms, all of which depend polynomially on $t$. 

It will be helpful to understand the special case of $t = 0$ for the interaction. We can check that the quadratic terms vanish when $t = 0$. The cubic terms simplify to
\begin{equation}\label{eq:cubic_noback}
		\sum_{i,j\geq 0}\int_{\mathbb{R}\times\mathbb{C}}dz\sum_{r,l\geq 0}\binom{i-1}{r}\binom{r}{l}\partial^{r-l}\phi_{j+i-1-r}\eta^i\partial^l\eta^j- \sum_{r\geq 0}(-1)^r\binom{j-1}{r}\phi_{j+i-1-r}\eta^i\partial^r\eta^j\,.
\end{equation}
Later we will compare this action functional with the dimensional reduction of $5d$ HT-nc-CS theory.
	
	\subsection{Dimensional reduction on flat space}
	\label{sec:new_star}
	Before considering the $5d$ HT-nc-CS theory on the deformed geometry, we first analyze the theory on flat space. The relationship between the $5d$ HT-nc-CS theory on $\bR\times\bC^2$ and the $\mathcal{W}_\infty$ algebra has been studied in several works using the method of Koszul duality/universal defects, for instance \cite{Costello:2016nkh,Ishtiaque:2024orn}. The derivation of the $\mathcal{W}_\infty$-algebra coproduct via defect fusion in the $5d$ theory is analyzed in \cite{Oh:2021wes,Ashwinkumar:2024vys}. The Koszul duality setup has also been implemented in the minitwistor correspondence space in \cite{Bittleston:2025gxr}, revealing an interesting connection with the KP hierarchy.

In this work, we take a different approach and relate the $3d$ $\mathcal{W}_\infty$-algebra Poisson sigma model to the dimensional reduction of the $5d$ HT-nc-CS theory to $3d$. We denote the coordinate on $\bR\times\bC^2$ by $(y_3,z_1,z_2)$. The HT-nc-CS theory has the field $\mathcal{A} = \mathcal{A}_{y_3}dy_3 +  \mathcal{A}_{\bar{z}_1}d\bar{z}_1 + \mathcal{A}_{\bar{z}_2}d\bar{z}_2 \in \Omega_{HT}^{1}(\mathbb{R}\times\mathbb{C}^2)\otimes\mathfrak{gl}_K$, where the complex $\Omega_{HT}^{\bullet}(\mathbb{R}\times\mathbb{C}^2)$ can be identified with $C^{\infty}(\mathbb{R}\times \mathbb{C}^2)[dy_3,d\bar{z}_1,d\bar{z}_2]$. 
	The action functional is given by
	\begin{equation}\label{eq:5d_act}
		S(\mathcal{A}) = \int dz_{0}dz_1\Tr\left[(\mathcal{A}\left(\frac{1}{2}d\mathcal{A} + \frac{1}{3}\mathcal{A}*\mathcal{A}\right)\right]\,,
	\end{equation}
	where the star product $*$ is given by the standard Moyal-Weyl formula: 
	\begin{equation}\label{eq:star}
		f*g = fg + \frac{1}{2}\varepsilon_{ij}\frac{\pa f}{\pa z_i} \frac{\pa g}{\pa z_j} + \frac{1}{2!2^2}\varepsilon_{i_1j_1}\varepsilon_{i_2j_2}\frac{\pa^2 f}{\pa z_{i_1}\pa z_{i_2}} \frac{\pa^2 g}{\pa z_{j_1}\pa z_{j_2}} + \dots
	\end{equation}
	To perform the dimensional reduction, we consider the theory on $\mathbb{R}\times\mathbb{C}^2\setminus\{y_3=z_1=0\}\times\mathbb{C}$, where $\{y_3=z_1=0\}\times\mathbb{C}$ is the locus of intersection of the three-dimensional probe D-branes and the coisotropic branes. We apply the isomorphism $\mathbb{R}\times\mathbb{C}^2\setminus\mathbb{C}\cong\mathbb{R}_{>0}\times S^2\times\mathbb{C}$ and compactify the theory on $S^2$. This induces an isomorphism of complexes:
	\begin{equation}
		\Omega_{HT}^{\bullet}(\mathbb{R}\times\mathbb{C}^2) \cong \Omega_{HT}^{\bullet}(\mathbb{R}_{>0}\times \mathbb{C})\otimes \Omega^{\bullet}_{HT}(S^2)\,.
	\end{equation}
We will provide a precise definition of $\Omega^{\bullet}_{HT}(S^2)$ and the above isomorphism in Appendix~\ref{app:KK}. Roughly speaking, $\Omega^{0}_{HT}(S^2)$ consists of all spherical harmonic functions ${Y^{(l)}_m}$ on $S^2$, while $\Omega^{1}_{HT}(S^2)$ can be identified with $\Omega^{0}_{HT}(S^2)\Omega$, where we denote
	\begin{equation}\label{eq:BM_form}
		\Omega = \frac{y_3d\bar{z}_1 - 2\bar{z}_1dy_3}{4\pi(y_3^2 + z_1\bar{z}_1)^{\frac{3}{2}}}\,.
	\end{equation}
$\Omega^{\bullet}_{HT}(S^2)$ is also equipped with a differential that pairs most of these modes, which will be analyzed in detail in Appendix~\ref{app:KK}. Consequently, upon decomposing the five-dimensional field $\mathcal{A}$ into a tower of three-dimensional fields valued in the Kaluza–Klein modes of $\Omega^{\bullet}_{HT}(S^2)$, most modes can be integrated out. Only the following Kaluza–Klein modes survive:
	\begin{equation}\label{eq:KK_modes}
		\begin{aligned}
			&A(y_3,z_2,\bar{z}_2) = \sum_{n \geq 0}A[n](y_3,z_2,\bar{z}_2)z_1^n\,,\\
			&B(y_3,z_2,\bar{z}_2) = \sum_{n \geq 0} B[n](y_3,z_2,\bar{z}_2)\frac{1}{n!}\left(-\frac{\pa}{\pa z_1}\right)^n\Omega \,,
		\end{aligned}
	\end{equation}
	where the fields $A[n](y_3,z_2,\bar{z}_2)$ are $1$-forms in $\Omega_{HT}^{1}(\mathbb{R}_{>0}\times \mathbb{C})$ and $ B[n](y_3,z_2,\bar{z}_2)$ live in $\Omega_{HT}^{0}(\mathbb{R}_{>0}\times \mathbb{C})$. Notice that $\frac{1}{n!}(-\frac{\pa}{\pa z_1})^n\Omega = \frac{(2n+1)!}{(2^nn!)^2} \frac{\bar{z}_1^n}{(y_3^2+z_1\bar{z}_1)^{\frac{n}{2}}}\Omega$. These modes can be identified with the dual of the modes $\{z_1^n\}$ via surface integration on $S^2$:
	\begin{equation}\label{eq:KK_pairing}
		\int_{S^2}z_1^m\frac{(2n+1)!}{(2^nn!)^2} \frac{\bar{z}_1^n}{(y_3^2+z_1\bar{z}_1)^{\frac{n}{2}}}\Omega = \delta_{n,m}\,.
	\end{equation}
	In other words, we can regard the Kaluza–Klein modes $A,B$ as $3d$ fields living in 
	\begin{equation}
			(A,B) \in \Omega_{HT}^{1}(\mathbb{R}_{>0}\times \mathbb{C})\otimes \mathfrak{gl}_K[z_1][1]\oplus \Omega_{HT}^{0}(\mathbb{R}_{>0}\times \mathbb{C})\otimes \mathfrak{gl}_K[z_1]^{\vee}\,.
	\end{equation}
	To write down the action functional for the KK theory, we can simply substitute $\mathcal{A}$ with $A + B$ in the $5d$ action \eqref{eq:5d_act}. This gives us
	\begin{equation}\label{eq:KK_act}
		\int \Tr \left( B(d A + \frac{1}{2}[A,A]_{*}) \right) \,.
	\end{equation}
	However, it turns out that to reproduce the action in \eqref{eq:cubic_noback}, a different $*$-product is required than the one in \eqref{eq:star}. Generally speaking, a $*$-product associated with the antisymmetric tensor $\pi = \frac{1}{2}\varepsilon_{ij} \partial_{z_i} \partial_{z_j}$ can be shifted by any symmetric tensor $\alpha$, yielding an equivalent product. Explicitly, we have the following identity:
	    \begin{equation}\label{eq:star_shift}
        e^{\frac{1}{2}\alpha}(f*_{\pi}g) = (e^{\frac{1}{2}\alpha}f)*_{\pi + \alpha}(e^{\frac{1}{2}\alpha}g)\,.
    \end{equation}
	We consider shifting $\pi = \frac{1}{2}\varepsilon_{ij} \partial_{z_i} \partial_{z_j}$ by $\alpha = \frac{1}{2}(\partial_{z_1} \partial_{z_2} + \partial_{z_2} \partial_{z_1})$, which gives us a new product defined by the tensor $\pi + \alpha = \partial_{z_1}\partial_{z_2}$:
	\begin{equation}
		f(z_1,z_2)*'g(z_1,z_2) = fg + \frac{\pa f}{\pa z_1}\frac{\pa g}{\pa z_2} + \frac{1}{2!}\frac{\pa^2 f}{\pa z_1^2}\frac{\pa^2 g}{\pa z_2^2} + \dots
	\end{equation}
	We can make the following redefinition of the KK fields:
	\begin{equation}\label{eq:redef_field}
			A' = e^{\frac{1}{2} \frac{\partial^2}{\partial z_1 \partial z_2}} A, \quad\quad B' = e^{-\frac{1}{2} \frac{\partial^2}{\partial z_1 \partial z_2}} B\,.
	\end{equation}
	Explicitly, this amounts to redefining
	\begin{equation}
		\begin{aligned}
					A'[n] &= \sum_{k\geq0}\frac{1}{2^k}\binom{n+k}{k}\pa_{z_2}^kA[n+k]\,,\\
					B'[n] & = \sum_{k = 0}^{n}\frac{1}{2^k}\binom{n}{k}\pa_{z_2}^kB[n-k]\,.
		\end{aligned}
	\end{equation}
	Using \eqref{eq:star_shift}, we can check that the action functional retains the same form as \eqref{eq:KK_act},
	\begin{equation}\label{eq:act_new_star}
		\int \operatorname{Tr} \left( B' \left( d A' + \frac{1}{2} [A', A']_{*'} \right) \right),
	\end{equation}
	but with the new product $*'$. Expanding \eqref{eq:act_new_star}, we find that the Kaluza–Klein theory has the kinetic terms $\sum_n\int B'[n]d A[n]$. The interaction terms, up to total $\partial_{z_2}$ derivatives, can be written as follows:
	\begin{equation}
		\sum_{n,m\geq0}\sum_{k = 0}^n \sum_{l = 0}^k (-1)^k\binom{n}{k}\binom{k}{l}\pa^lB'[n+m-k] \pa^{k-l}A'[n] A'[m]   - \sum_{k = 0}^m \binom{m}{k}B'[n+m-k] \pa^kA'[n] A'[m],
	\end{equation}
	 where we denote $\partial = \partial_{z_2}$. In the above expressions, the $\mathfrak{gl}_K$ color indices are suppressed. For $K = 1$, the $k = 0$ terms in the summation vanish, and one can check that it coincides with the cubic terms \eqref{eq:cubic_noback} in the Poisson Sigma model associated with the classical $\mathcal{W}_{\infty}$ algebra, upon identifying $B'[n]$ with $\phi_{n+1}$ and $A'[n]$ with $\eta_{n+1}$.

\subsection{Back-reaction}
In this section, we provide a different perspective on the back-reacted geometry by proposing a field theory description of the closed string sector. Instead of the A-model picture considered in the previous sections, we work with a mixed A/B-model. The duality between these two perspectives is discussed in detail in \cite{Aganagic:2017tvx}. One advantage of the mixed A/B-model description is that it is more compatible with the holomorphic-topological nature of the $5d$ nc-CS theory, and is therefore better suited for a perturbative analysis.

The closed string theory for the mixed A/B-model should be a product of Kodaira-Spencer theory \cite{Bershadsky:1993cx,Costello:2012cy} and K\"{a}hler gravity \cite{Bershadsky:1994sr}. In the BV formalism, the space of field should be the subspace of $\Omega^{\bullet}(\mathbb{R}^2)\otimes\mathrm{PV}^{\bullet,\bullet}(\mathbb{C}^2)[2]$ that live in the kernel of $\partial$, where $\mathrm{PV}^{i,j}(\mathbb{C}^2) = \Omega^{0,j}(\mathbb{C}^2,\wedge^i T^{1,0}_{\mathbb{C}^2})$ is the space of poly-vector fields on $\mathbb{C}^2$. We can identify $\mathrm{PV}^{i,\bullet}(\mathbb{C}^2) \cong \Omega^{2-i,\bullet}(\mathbb{C}^2)$ via the standard volume form $dz_1dz_2$ on $\mathbb{C}^2$. We also have an identification of $(\ker\partial) \cap \Omega^{1,\bullet}$ with $\partial(\Omega^{1,\bullet})$. Therefore, the space of closed string field can be written as
\begin{equation}
	\begin{aligned}
		\alpha \in \Omega^{\bullet}(\mathbb{R}^2)\otimes\Omega^{0,\bullet}(\mathbb{C}^2)[1]\,\\
		\beta dz_1dz_2 \in \Omega^{\bullet}(\mathbb{R}^2)\otimes\Omega^{2,\bullet}(\mathbb{C}^2)[2]\,.
	\end{aligned}
\end{equation}

Before we turn on the non-commutativity, the action functional can be obtained by combining the BCOV action and the K\"{a}hler gravity. We can check that it is the Poisson BF action \cite{Williams:2020fwd}
\begin{equation}
	S = \int_{\mathbb{C}^2\times\mathbb{R}^2}dz_1dz_2\beta(d+\bar{\partial})\alpha + \frac{1}{2}\beta\{\alpha,\alpha\}_{\pa_{z_1}\wedge\pa_{z_2}}\,.
\end{equation}
where $\{-,-\}_{\pa_{z_1}\wedge\pa_{z_2}}$ is the Poisson bracket associated to the Poisson tensor $\pa_{z_1}\wedge\pa_{z_2}$, which is dual to the volume form $dz_1dz_2$. After turning on the non-commutativity, the closed string action is deformed by a term
\begin{equation}
	\int_{\mathbb{C}^2\times\mathbb{R}^2}dz_1dz_2\beta\beta\,.
\end{equation}
The BRST operator then acts on the fields by
\begin{equation}
	\begin{aligned}
		&Q\alpha = (d+\bar{\pa})\alpha  + \frac{1}{2}\{\alpha,\alpha\}_{\pa_{z_1}\wedge\pa_{z_2}}+ \beta\,,\\
		&Q\beta =   (d+\bar{\pa})\beta + \{\alpha,\beta\}_{\pa_{z_1}\wedge\pa_{z_2}}\,.
	\end{aligned}
\end{equation}
It is easy to see that the BRST cohomology vanishes. This is compatible with the pure A-model description of the theory, as the A-model closed string theory on $T^*(\mathbb{R} \times \mathbb{C})$ is perturbatively trivial.

The mixed A/B-model described above provides an alternative perspective on the back-reaction of D3-branes. Specifically, the effect of introducing a stack of $N$ D3-branes supported on $\mathbb{R}_{x_3} \times \mathbb{C}_{z_2}$ is to add a localized source term to the equations of motion. Let us  denote $\eta = \partial\alpha$. We have:
\begin{equation}\label{eq:source_EOM}
	(d+\bar{\pa})\eta  + \frac{1}{2}\{\eta,\eta\} = t \delta_{\mathbb{R}\times\mathbb{C}}\,.
\end{equation}
It has solution \cite{Costello:2016nkh}
\begin{equation}
	\eta_0 = \frac{t(y_3d\bar{z}_1 - 2\bar{z}_1dy_3)}{(y_3^2 + z_1\bar{z}_1)^{\frac{3}{2}}}dz_{1}\,.
\end{equation}

To connect with the back-reacted geometry $M_t$, we consider an alternative, but gauge-equivalent, solution to the sourced EOM \eqref{eq:source_EOM}:
\begin{equation}
	\eta_0' = t\delta_{y_3 \geq 0 }\delta_{z_1,\bar{z}_1}d\bar{z}_1dz_1\,.
\end{equation}
This solution admits a more transparent geometric interpretation, as it gives rise to a Beltrami differential $t\delta_{y_3 \geq 0 }\delta_{z_1,\bar{z}_1}d\bar{z}_1\partial_{z_2}$. It describes a deformation of the complex structure in the region $y_3 \geq 0$, whose holomorphic functions satisfy the following differential equations:
\begin{equation}
	\begin{aligned}
		&\frac{\pa F}{\pa \bar{z}_2} = 0\,,\\
		&\frac{\pa F}{\pa \bar{z}_1} - t\delta_{z_1,\bar{z}_1}\frac{\pa F}{\pa z_2} = 0\,.
	\end{aligned}
\end{equation}
As a result, we have two coordinate patches	$z_{1+}, z_{2+}$ and $z_{1-}, z_{2-}$ in the regions $y_3 \geq 0$ and $y_3 < 0$, respectively. These two patches are related by
\begin{equation}\label{eq:back_gluing}
	z_{1+} = z_{1-},\quad z_{2+} = z_{2-} + \frac{t}{z_{1-}} \, .
\end{equation}
If we instead start with $T^*(\mathbb{R}\times\mathbb{C}\mathbb{P}^1)$ rather than the flat space $\mathbb{R}^2\times\mathbb{C}^2$, the above coordinate transformation will give rise to the deformed $A_1$ singularity $M_t$, which is analyzed carefully in \cite{Gaiotto:2025nrd} and reviewed in Appendix \ref{app:geometry}.

\subsection{Back reaction on KK theory}
\label{sec:backrea}
Given the previous description of the back-reacted geometry as a closed string field, we can describe its effect on the open-string theory on the coisotropic brane. The first-order coupling between the closed string fields and the $5d$ HT-nc-CS field $\mathcal{A}$ takes the following form:
\begin{equation} 
	\int_{\mathbb{C}^2\times\mathbb{R}}\beta\Tr(\mathcal{A}) + \Tr(\mathcal{A}[\alpha,\mathcal{A}]_{*}) \,.
\end{equation}
Therefore, by turning on the background field $\alpha_0$ with $\partial \alpha_0 = \eta_0$, the non-commutative Chern–Simons action is deformed by the term
\begin{equation}
	\int_{\mathbb{C}^2\times\mathbb{R}}\Tr(\mathcal{A}[\alpha_0,\mathcal{A}]_{*})\,.
\end{equation}

As we have discussed, it will be helpful to use the star product $*'$ defined in Section \ref{sec:new_star}. Using \eqref{eq:star_shift}, we can check that by letting $ \mathcal{A}' = e^{\frac{1}{2} \frac{\partial^2}{\partial z_0 \partial z_1}} \mathcal{A}
$, we can equivalently write the deformed action as follows
\begin{equation}
		\int_{\mathbb{C}^2\times\mathbb{R}}\Tr((e^{\frac{\partial^2}{\partial z_1\partial z_2}}\mathcal{A}')[\alpha_0,\mathcal{A}']_{*'})\,.
\end{equation}
Notice that $\eta_0$ already contains the $1$-form $\Omega$, and hence any term involving $B'[n]$ in the expansion of the above action will vanish. Therefore, if we substitute the field $\mathcal{A}'$ by the KK tower $A' = \sum_{n \geq 0} A'[n] z_1^n$, the deformed action becomes
\begin{equation}
	\int\Tr\left( (e^{-\frac{\pa^2}{\pa z_1\pa z_2}}A) [\alpha_0,A']_{\star'} \right)\,.
\end{equation}
We can compute the expansion
\begin{equation}
	[\alpha_0,A']_{\star'} = \sum_{m,k}\frac{(-1)^k}{k+1}t\left( \frac{(-\pa_{z_1})^k}{k!}\Omega\right)  \pa_{z_2}^{k+1}A'[m]z_1^m\,.
\end{equation}
Therefore, we can expand the action as
\begin{equation}
	\sum_{n,m,k,l}(-1)^{l+1}\binom{n}{l}\frac{1}{k+1}\delta_{k,m+n-l}\int t\Tr (\pa_{z_2}^{l+k+1}A'[n]) A'[m]\,.
\end{equation}
We use the identity
\begin{equation}
	\sum_{l=0}^m(-1)^l\binom{n}{l}\frac{1}{m+n+1-l} = \frac{(-1)^mn!m!}{(n+m+1)!}\,.
\end{equation}
Hence we can write the action as
\begin{equation}
	\sum_{n,m\geq 0}\frac{(-1)^{m+1}n!m!}{(n+m+1)!}\int t\Tr (\pa_{z_2}^{n+m+1}A'[n]) A'[m]\,.
\end{equation}

On the other hand, we can check that the terms quadratic in the $\eta$ fields in the action \eqref{eq:Winf_matrix} can be written as\begin{equation}
    \sum_{j,k,s}(-1)^{s+1}\binom{t+k}{j+k}\binom{t}{s}\eta^{k+s+1}\partial^{j+k+s}\eta^j
\end{equation}
We use the expansion $\binom{t+k}{j+k} = t (-1)^{j-1}\frac{k!(j-1)!}{(j+k)!} + \dots$, we find that the term linear in $t$ can be written as 
\begin{equation}
    \sum_{j,k}(-1)^{j}\frac{k!(j-1)!}{(j+k)!}t\eta^{k+1}\partial^{j+k}\eta^{j}
\end{equation}
This matches exactly the expression obtained from KK reduction.

In terms of the boundary chiral algebra, this interaction corresponds to the central extension that is linear in $t$:
\begin{equation}
	B'[n](z)B'[m](0) \sim t\frac{(-1)^nn!m!}{z^{n+m+2}}
\end{equation}

\subsection{Non-linear corrections to the KK theory}
\label{sec:KK_nonlinear}
In performing the KK reduction, we simplified our analysis by omitting a large number of $S^2$ harmonic modes. Roughly speaking, this is because the differential operator of the $5d$ theory acts on these modes as an effective ``mass,'' pairing most of them. We can therefore integrate out these modes, leaving only the modes listed in \eqref{eq:KK_modes} that correspond to the cohomology class $H^{\bullet}_{HT}(S^2)$. However, doing so induces additional corrections to the action functional, even at the semiclassical level. In particular, there are tree-level diagrams mediated by the integrated-out modes. Mathematically, the $5d$ theory can be described by the following dg Lie algebra, viewed as the complex of BV fields:
\begin{equation}\label{eq:KK_BVfull}
    \Omega_{HT}^{\bullet}(\mathbb{R}_{>0}\times \mathbb{C})\otimes \Omega^{\bullet}_{HT}(S^2)\otimes \mathfrak{gl}_K\,.
\end{equation}
Its differential gives rise to the kinetic term, and the Lie bracket $[-,-]_{*'}$ corresponds to the interaction term. Passing from $\Omega^{\bullet}_{HT}(S^2)$ to its cohomology $H^{\bullet}_{HT}(S^2)$ simplifies the field content, but also induces a whole tower of $L_\infty$ operations $\{\ell_n\}_{n\geq 1}$ on
\begin{equation}
    \Omega_{HT}^{\bullet}(\mathbb{R}_{>0}\times \mathbb{C})\otimes H^{\bullet}_{HT}(S^2)\otimes \mathfrak{gl}_K\,.
\end{equation}
Mathematically, this process is called homotopy transfer \cite{berglund2014homological}. With this structure in place, the higher-order corrections to the KK-reduced action functional take the following schematic form:
\begin{equation}\label{eq:KK_higher_sch}
	\sum_{n}\frac{1}{2(n-1)!}\int \Tr A \wedge \ell_{n}(A,B,B,\dots,B)
\end{equation}
As a tautological but useful fact, we can define the twisted differential $\ell^B_1(-) = \sum_{n\geq 1}\ell_n(B^{\otimes n-1},-)$. The entire KK-reduced action functional is then packaged into $\ell^B_1$ as follows:
\begin{equation}\label{eq:KK_higher_sch}
	\frac{1}{2}\int \Tr A \wedge \ell_{1}^B(A)
\end{equation}

Computing the full $L_\infty$ structure $\{\ell_n\}_{n\geq 1}$ directly is quite difficult. In the limit where we ignore backreaction and non-commutativity, we can analyze the homotopy transfer of $\Omega^{\bullet}_{HT}(S^2)$ separately, which induces an $A_\infty$ structure $\{m_{n}\}_{n \geq 2}$ on $H^{\bullet}_{HT}(S^2)$. In this case, we are able to obtain an explicit formula. The details of this analysis are quite technical, so we leave them to Appendix \ref{app:transfer}. We present only the main result in this section.

We choose a basis $\{t_a\}$ of the Lie algebra and let $\{t^a\}$ be the dual basis with respect to the Killing pairing $K$, such that $K(t_a,t^b) = \delta_a^b$. The action functional of the KK theory then has the following higher-order terms:
\begin{equation}\label{eq:KK_higher}
	\begin{aligned}
		&\sum_{n\geq3}\frac{1}{2(n-1)!}\sum_{\substack{l,l'\\j_1,\dots,j_{n-1}}}\sum_{\substack{a,b,\\c_1,\dots,c_{n-1}}}\sum_{\sigma \in S_{n-1}}K(t_a,[\dots[[t_b,t^{c_{\sigma(1)}}],t^{c_{\sigma(2)}}],\dots,t^{c_{\sigma(n-1)}}])\\
		&(m_n)^{l',l}_{j_1,j_2,\dots,j_{n-1}} \int_{\mathbb{C}\times\mathbb{R}_{>0}} A^a[l']A^b[l]B_{c_1}[j_1]B_{c_2}[j_2]\dots B_{c_{n-1}}[j_{n-1}]
	\end{aligned}
\end{equation}
The constants $(m_n)^{l',l}_{j_1,j_2,\dots,j_{n-1}}$ are computed in Appendix~\ref{app:transfer}. They are non-zero only when
\begin{equation}
l + l' = j_1 + j_2 + \dots + j_{n-1} + n - 2.
\end{equation}
The explicit formula is given by
\begin{equation}
	\begin{aligned}
		(m_n)^{l',l}_{j_1,j_2,\dots,j_{n-1}} = &\frac{\prod_{i = 1}^{n-1}(-1)^{j_i}N_{j_i}}{N_lN_{l'}}\sum_{k_{n-2} = |k_{n-3}-j_{n-2}|}^{k_{n-2} + j_{n-1}}\dots\sum_{k_2 = |k_1-j_2|}^{k_1+j_2}\sum_{k_1 = |l-j_1|}^{j_1+l}2^{n-2}\sqrt{\frac{2l+1}{2l'+1}}\\
		&\times\left(\prod_{i = 1}^{n-1}\sqrt{\frac{2j_i+1}{(k_i+l-J_i)(k_i-l+J_i+1)}} C^{k_i,0}_{k_{i-1},0;j_i,0}C^{k_i,l-J_i}_{k_{i-1},l-J_{i-1}-1;j_{i},-j_i}\right) 
	\end{aligned}
\end{equation}
In the above expression, we define
\begin{equation}
	\begin{aligned}
		&k_0 = l\\
		&k_{n-1} = l'\\
		&J_i =  j_1 + j_2+\dots + j_i + i - 1
	\end{aligned}
\end{equation}
By the properties of Clebsch–Gordan coefficients, namely that $C^{k_i,0}_{k_{i-1},0;j_i,0} \neq 0$ only when $k_i-k_{i-1} - j_i \equiv 0 \pmod 2$, we can see that only the operations $m_{2n}$ are non-zero.

For a complete analysis of the non-commutative Chern–Simons theory in the deformed background, the above interactions of the KK theory must be further corrected. This is quite difficult to do while remaining in the basis $A[n], B[n]$. Instead, a different method for a full analysis is provided in the next section.

For $\mathfrak{gl}_K$ with $K \geq 2$, the first non-zero term in the higher-order interaction \eqref{eq:KK_higher} can be written as follows
\begin{equation}
	\sum_{\sigma \in S_3}\frac{1}{3!}\int_{\mathbb{C}\times\mathbb{R}_{>0}} f_{bd}^{c_1}f^{dc_2}_{e}f_{a}^{ec_3}A^a[1]A^b[1]B_{c_{\sigma{(1)}}}[0]B_{c_{\sigma{(2)}}}[0]B_{c_{\sigma{(3)}}}[0]
\end{equation}
From this expression, we can read off the corresponding OPE:
\begin{equation}
	B_a[1](z)B_b[1](0) \sim \frac{1}{z}f_{bd}^{c_1}f^{dc_2}_{e}f_{a}^{ec_3}\sum_{\sigma \in S_3}\frac{1}{3!}B_{c_{\sigma{(1)}}}[0]B_{c_{\sigma{(2)}}}[0]B_{c_{\sigma{(3)}}}[0] 
\end{equation}
which is cubic in the generators. This leads to an apparent discrepancy with the OPE in the (matrix-extended) $\mathcal{W}_{\infty}$ algebra in the Miura basis, where the operator products produce terms that are at most quadratic in the generators.

This discrepancy should not be viewed as a failure of the expected duality. In fact, a similar phenomenon arises in the study of the Yangian in $4d$ Chern–Simons theory \cite{Costello:2017dso}. There, the commutator computed from Feynman diagrams in $4d$ Chern–Simons theory contains cubic terms and is much closer to Drinfeld's $J$-presentation of the Yangian, whereas the commutation relations derived from the RTT presentation of the Yangian are quadratic. This apparent mismatch was resolved in \cite{Costello:2018gyb} by introducing a non-linear transformation between the two sets of generators. From a mathematical perspective, the equivalence between different presentations has been studied and proven in many works, including \cite{Drinfeld:1985rx,Drinfeld:1987sy,guay2019equivalences}. In our case, a similar non-linear transformation is needed to match the results in the basis $B[n]$ and the Miura basis of the $\mathcal{W}_{\infty}$ algebra, which we analyze in detail in the next section.

\subsection{Adler--Gelfand--Dickey structure from raviolo complex}
\label{sec:raviolo-agd}
In this section, we provide a complete proof that the dimensional reduction of $5d$ nc-CS theory is equivalent to the Poisson sigma model of the classical $\mathcal{W}_\infty$ algebra introduced in Section \ref{sec:agd-review}. We establish this using techniques from homological algebra. First, we introduce the raviolo complex $\mathcal{R}av^{\bullet}$, defined as
\begin{equation}
	\mathcal{R}av^0=\left\{a(s)\in\mathbb{C}((\xi))[s] \mid a(0),a(1)\in\mathbb C[\xi]\right\},\quad \mathcal{R}av^1=\mathbb{C}((\xi))[s]ds
\label{eq:scalar-interval-raviolo}
\end{equation}
and the differential is $d\alpha=(\partial_s\alpha)ds$. This complex is introduced in \cite{alfonsi2025raviolo} as an algebraic model for (derived) functions on the raviolo $D\cup_{\mathring{D}}D$. We can construct an algebra quasi-isomorphism $\iota$ from $\Omega_{\mathrm{HT}}^{\bullet}(S^2)$ to $\mathcal{R}av^{\bullet}$. On the generators, the map is defined as
\begin{equation}
	\iota(z_1)=\xi,\quad
	\iota\!\left(\frac{\bar z_1}{y_3^2+z_1\bar z_1}\right)
	=\frac{4s(1-s)}{\xi},\quad
	\iota\!\left(\frac{y_3}{\sqrt{y_3^2+z_1\bar z_1}}\right)
	=1-2s,\quad
	\iota(\Omega)= \frac{ds}{\xi}.
\label{eq:rav-quasi-isom}
\end{equation}
It is easy to see that it can be extended to a morphism from $\Omega_{\mathrm{HT}}^{\bullet}(S^2)$ to $\mathcal{R}av^{\bullet}$ that preserves the dg algebra structure. Moreover, it maps the cohomology $H^n_{HT}(S^2)$ to $H^n(\mathcal{R}av)$ isomorphically:
\begin{equation}
    \iota([z_1^n]) = [\xi^n],\quad \iota\!\left(\left[\frac{1}{n!}(-\partial_{z_1})^n\Omega\right]\right) = [\xi^{-n-1}ds].
\end{equation}

As a result, instead of analyzing the complex $\Omega^{\bullet}_{HT}(\mathbb{R}\times\mathbb{C})\otimes \Omega_{HT}^\bullet(S^2)\otimes\mathfrak{gl}_K$ that arises directly from the KK reduction problem, we consider the following complex:
\begin{equation}
\mathcal{C}^{\bullet} = \mathcal{O}(\mathbb{C})\otimes \mathcal{R}av^{\bullet}\otimes \mathfrak{gl}_K
\end{equation}
It has the differential induced from $\mathcal{R}av^\bullet$ and we equip it with the non-commutative product
\begin{equation}
	F\circ G=\sum_{r\geq0}\frac{1}{r!}
		(\partial_\xi^rF)(\partial_{z}^rG).
\label{eq:ordered-symbol-product}
\end{equation}
For convenience, we define $\Psi_K=\mathcal{O}(\mathbb C)\otimes\mathbb C((\xi))\otimes\mathfrak{gl}_K$ and use the decomposition
\begin{equation}
		(\Psi_K)_+=\mathcal{O}(\mathbb{C})\otimes \mathbb{C}[\xi]\otimes \mathfrak{gl}_K,\quad (\Psi_K)_- =\mathcal{O}(\mathbb{C})\otimes \xi^{-1}\mathbb{C}[[\xi^{-1}]]\otimes \mathfrak{gl}_K\,.
\end{equation} 
Their relation to the space of matrix-valued pseudodifferential operators
$\Psi\mathrm{DO}^{t}_K$ will be explained shortly.

The above complex can be regarded as a replacement for the flat-space KK complex \eqref{eq:KK_BVfull}. To understand the effect of backreaction, note that the sourced background field $\eta_0$ is identified with $t\xi^{-1}ds$ in $\mathcal{R}av^{\bullet}$ under the map $\iota$. It is then natural to formally identify $\alpha_0 = \partial^{-1}\eta_0$ with $t\log\xi\,ds$, leading to the deformed differential
\begin{equation}
	D_{t}F = (\partial_sF-[t\log\xi,F]_{\circ})ds\,.
\end{equation}
This infinitesimal connection can be lift to the finite transformation: $\operatorname{Ad}_{\xi^t}(F) = \xi^t\circ F\circ \xi^{-t}$. In particular, we have
\begin{equation}
	\operatorname{Ad}_{\xi^t}(\xi)=\xi,\qquad
	\operatorname{Ad}_{\xi^t}(z)=z+t\xi^{-1}.
\end{equation}
This is exactly the coordinate gluing equation \eqref{eq:back_gluing} induced by the backreaction.

To obtain the theory after dimensional reduction, we can perform homotopy transfer on the dg Lie algebra $(\mathcal C^\bullet,D_t)$, obtaining an $L_\infty$ structure $\{\ell_n\}_{n\geq1}$ on its cohomology. As explained in Section~\ref{sec:KK_nonlinear}, rather than computing all $\ell_n$ directly, we can compute the twisted differential $(\ell^{\boldsymbol{b}})_1(-)=\sum_{n\geq0}\frac1{n!}\ell_{n+1}(\boldsymbol{b}^{\otimes n},-)$ for $\boldsymbol{b} \in (\Psi_K)_-ds$. Since homotopy transfer commutes with Maurer--Cartan twisting \cite{berglund2014homological}, we can first twist the complex $(\mathcal C^\bullet,D_t)$ by $\boldsymbol{b} = b\,ds\in(\Psi_K)_-ds$, and then perform the homotopy transfer. The twisted differential is given by
\begin{equation}
	D_{t,b}F=\bigl(\partial_sF-[t\log\xi+b,F]_\circ\bigr)ds.
\end{equation}
The advantage of considering $(\mathcal{C}^{\bullet},D_{t,b})$ is that its transferred differential packages all the operations needed for the KK-reduced theory, making it more convenient than dealing with the full $L_\infty$ structure.

Let
\begin{equation}
\label{eq:rav-transport}
	g_{t,b}(s)=\exp_\circ(s(t\log\xi+b)),\qquad
	L(b)=g_{t,b}(1)=\exp_\circ(t\log\xi+b),
\end{equation}
and consider the map
\begin{equation}
	\mathcal U_{t,b}(F)(s) =\operatorname{Ad}_{L(b)\circ g_{t,b}(s)^{-1}}F(s)
\end{equation}
It is easy to check that $\mathcal U_{t,b}$ satisfies $d_s\mathcal U_{t,b}=\mathcal U_{t,b}D_{t,b}$. It therefore gives a chain isomorphism between $(\mathcal C^\bullet,D_{t,b})$ and $(\mathcal C_{L(b)},d_s)$, where
\begin{equation}
	\mathcal C_L^0=\left\{F(s)\in \Psi_K[s] \mid F(0)\in\operatorname{Ad}_L\bigl((\Psi_K)_+\bigr), F(1)\in(\Psi_K)_+\right\},\qquad
	\mathcal C_L^1=\Psi_K[s]ds.
\label{eq:CL-endpoints}
\end{equation}
Thus twisting by $t\log\xi+b$ is locally removed by a finite gauge transformation, while its parallel transport along the interval survives as a nontrivial endpoint conjugation.

For $X\in\xi^{-t}\circ(\Psi_K)_+$, set $r_L(X) =(X\circ L)_+$. The map $r_L:\xi^{-t}\circ(\Psi_K)_+\to(\Psi_K)_+$ is a triangular isomorphism. We give its inverse in Appendix~\ref{app:other}. We also set 
\begin{equation}
	q_L(X) =(L\circ X)_+-\operatorname{Ad}_L(r_L(X))
\end{equation}
We can write $X\circ L=r_L(X)+(X\circ L)_-$. Since conjugation by $L$ preserves $(\Psi_K)_-$, $(L\circ X)_+=(\operatorname{Ad}_L(r_L(X)))_+$ and we can check $q_L(X)\in(\Psi_K)_-$. We also define the Adler map
\begin{equation}
	\mathcal{A}^{(L)}(X) := q_L(X)\circ L =(L\circ X)_+\circ L-L\circ(X\circ L)_+ \in (\Psi_K)_-\circ\xi^t
\end{equation}
We define the two-term complex $\mathcal{K}_L^{\bullet}$ as follows\footnote{Considering the variation of matrix-valued pseudo-differential operators $\Psi\mathrm{DO}^t_K$, we have an identification $T_L\Psi\mathrm{DO}_K^t = (\Psi_K)_-\circ\xi^t = (\Psi_K)_-\circ L$. Then under the trace--residue pairing, $\langle X,Y\rangle := \left(\operatorname{Res}_\xi\operatorname{Tr}(X\circ Y)\right)$, we further have an identification of cotangent space $T_L^*\Psi\mathrm{DO}_K^t \simeq \xi^{-t}\circ(\Psi_K)_+$}
\begin{equation}
	\mathcal{K}_L^{\bullet} =\left[	\xi^{-t}\circ(\Psi_K)_+\xrightarrow{ \mathcal{A}^{(L)} }(\Psi_K)_-\circ\xi^t	\right].
\end{equation}

Now we construct a deformation retract between $(\mathcal{C}_{L}^{\bullet},d_s)$ and $(\mathcal{K}_L^{\bullet},\mathcal{A}^{(L)})$. We first define the map $I_L: \mathcal{K}_L^{\bullet} \to \mathcal{C}_{L}^{\bullet} $. For $X\in\xi^{-t}\circ(\Psi_K)_+$ and $Y\in(\Psi_K)_-\circ\xi^t$, it is given by 
\begin{equation}
	I_L(X)(s)=\operatorname{Ad}_L(r_LX)+s q_L(X), \quad I_L(Y)=(Y\circ L^{-1})ds
\end{equation}
Then we define a map $P_L:\mathcal{C}_{L}^{\bullet}\to \mathcal{K}_L^{\bullet}  $. For $F \in \mathcal{C}_{L}^{0}$ and $G \in \mathcal{C}_{L}^{1}$, we have
\begin{equation}
		P_L(F)=r_L^{-1}\bigl(\operatorname{Ad}_{L^{-1}}F(0)\bigr),\quad  P_L(G) =\left(\int_0^1G(s)\right)_-\circ L
\end{equation}
Finally, the homotopy operator $K_L: \mathcal{C}_{L}^{1} \to \mathcal{C}_{L}^{0}$ is defined as
\begin{equation}
	K_L(G)(s)=\int_0^sG(u)-s\left(\int_0^1G(u)\right)_-\,.
\end{equation}
Direct calculation gives (we also provide detailed computation in Appendix \ref{app:other})
\begin{equation}
\label{eq:rav_retract}
	P_LI_L=1,\qquad d_sK_L+K_Ld_s=1-I_LP_L,
	\qquad P_LK_L=K_LI_L=K_L^2=0.
\end{equation}
As a result, we have a quasi-isomorphism between $(\mathcal{C}_{L}^{\bullet},d_s)$ and $(\mathcal{K}_L^{\bullet},\mathcal{A}^{(L)})$. The physical consequence is that all information we need from the KK-reduced theory is contained in the map $\mathcal{A}^{(L)}$. In fact, writing $X$ in coefficient form $X=\sum_{j\geq1}\xi^{-t+j-1} X_j(z)$, we have 
\begin{equation}
	\mathcal A^{(L)}(X)=\sum_{i,j\geq1} H_{ij}^{(L)}(\partial_{z})(X_j)\,\xi^{t-i},
\end{equation}
We write $\boldsymbol{\eta} = \sum_{i\geq1} \eta^i \xi^{-t + i -1}$, with each $\eta^i \in \Omega^{1}_{HT}(\mathbb{R} \times \mathbb{C})\otimes \mathfrak{gl}_K$ and also write $\boldsymbol{L} = \xi^t + \sum_{j\geq 1} \phi_j \xi^{t-j}$, with each $\phi_j \in \Omega^{0}_{HT}(\mathbb{R} \times \mathbb{C})\otimes \mathfrak{gl}_K$. The complex $(\mathcal{K}_L^{\bullet},\mathcal{A}^{(L)})$ gives the following theory
\begin{equation}
	\frac12\int_{\mathbb R\times\mathbb C_{z}}dz\,\Tr  \operatorname{Res}_{\xi}\left( \boldsymbol{\eta}(d + \bar{\partial})\boldsymbol{L} + \boldsymbol{\eta}\mathcal{A}^{(L)}(\boldsymbol{\eta}) \right)
\end{equation}
This is exactly the chiral Poisson sigma model associated with the AGD $\mathcal{W}_\infty$ algebra that we introduced in Section \ref{sec:agd-review}.

The transformation between the KK basis that we used in the previous sections and the above Miura basis can also be read off from our construction. Under the quasi-isomorphism $\iota$, the cohomology class $[(-\frac{\partial}{\partial z})^n\Omega]$ that corresponds to the $B[n]$ field is mapped to the class $\xi^{-n-1}ds$. Hence we have the identification $b_{n+1} = B'[n]$, where $b=\sum_{j\geq1}b_j(z)\xi^{-j}$. The exponential in \eqref{eq:rav-transport} gives a formal coordinate change from $b$ to the pseudodifferential-operator coefficients:
\begin{equation}
	L(b) =\exp_\circ(t\log\xi+b) =\xi^t+\sum_{j\geq1}u_j\xi^{t-j}
\end{equation}
Expanding the exponential provides a formula 
\begin{equation}
	u_j(b)=b_j+P_j\bigl(t;b_1,\ldots,b_{j-1}\bigr),
\end{equation}
where the explicit form of $P_j$ is discussed in Appendix \ref{app:other}. For example, we have $u_1=b_1$, $u_2=b_2+\frac12b_1^2+\frac t2\partial_{z}b_1$. 

\section{Wilson lines and probe branes} \label{sec:Wilson}
In this Section we discuss briefly how our analysis could be extended to 
correlation functions on a sphere with non-trivial punctures, i.e. insertions of vertex operators transforming in non-trivial modules for $\cW_\infty$. From an holographic perspective, we will limit ourselves to 
insertions which do not change the bulk geometry but may add extra probe D-branes to it. 

In the 3d Chern-Simons theory setup, non-trivial modules are produced by time-like Wilson line defects ending on or passing across the chiral interface. The line defects are associated to irreps of $SU(N)$ with Young Tableaux labels fitting into a $k \times N$ rectangle. If we keep the  
Tableaux $T$ or its conjugate $\bar T$ of fixed size as we send $N \to \infty$, the holographic interpretation involves some collection of strings ending on the Wilson line at the boundary. If we scale the overall size (number of boxes) of the Tableaux linearly in $N$, the holographic interpretation typically involves D2 brane probes. 

The former case can be reduced to a study of finite collections of Wilson lines in the (anti)fundamental representation, which we already analyzed in the rest of the paper. We should thus consider Wilson lines dual to D-brane probes. We will focus on Wilson lines associated to $\Lambda^\bullet \bC^N$, which were the topic of recent work \cite{Gaiotto:2025nrd} adapted to the $T^* \bR \times M_t$ holographic setting. 

The probe D-branes wrap the $\bR$ direction and an holomorphic curve in $M_t$ with asymptotic ends which describe individual Wilson lines. 
We should thus consider how a D-brane wrapping an holomorphic Lagrangian in $M_t$ can cross or end on a canonical coisotropic D-brane. As the two D-branes are transverse in $T^* \bR$, they will cross along the Lagrangian to give a surface defect in the 5d HT-nc-CS world-volume theory on the coisotropic D-brane. 

The literature on the 5d theory and twisted M-theory discusses in detail 
surface defects which wrap a $\bC$ factor in $\bC^2$. The case of a complicated curve in $\bC^2$ or $M_t$ has not been previously studied
and deserves a separate publication. Here we will just sketch some general ideas. 

A simple observation is that $M_t$ is really quantized to $A_t$ in the current context, so the curve should likely be also quantized to 
a module for $A_t$. The reference \cite{Gaiotto:2025nrd} describes the 
relevant probe D-branes as twisted local system $\cL$ on $\bC P^1$ with twist $\exp 2 \pi i t$. A Riemann-Hilbert correspondence can map such a local system to an holomorphic twisted connection $\cD_\cL$ on $\bC P^1$, or better a twisted D-module. This gives immediately a representation of $A_t$ by matrix-valued differential operators analogous to the $p^+_{a,2-a}$, or vice versa. 

The geometric interpretation of this description of these probe D-branes is that they are analogous to the D-branes which support the original 3d Chern-Simons theory, but they have a Chan-Paton bundle of opposite Grassmann parity and have ``spikes'' at the location of the Wilson lines. The number $n$ of such D-branes is related to the total number of boxes in the Wilson lines. See \cite{Gaiotto:2025nrd} for details.

If the probe D-branes cross the coisotropic D-brane, e.g. the Wilson lines cross the interface, the resulting surface defect should support a coset $\cV^b_{n,\kappa+N-n}$ built from symplectic bosons. 

We obtain the following holographic proposal:
\begin{itemize}
    \item Consider $m$ Wilson lines with $\Lambda^\bullet \bC^N$ labels crossing a chiral interface with total number of boxes $n N$. The 3d CS theory in the presence of the lines has an Hilbert space $\cH_{m,n}$ 
    and the large $N$ limit is governed by a phase space $\cP_{m,n}$, see 
    \cite{Gaiotto:2025nrd} for details.
    \item Correlation functions on the interface in the presence of the Wilson lines are organized into conformal blocks, i.e. are valued in $\mathrm{End}(\cH_{m,n})$. At large $N$, we expect that to be described by some Lagrangian correspondence $\cC$ from $\cP_{m,n}$ to itself,
    depending holomorphically on the positions of the punctures. 
    \item The phase space $\cP_{m,n}$ is identified with the moduli space of $\cL$ shapes of the dual probe D-branes. The correspondence $\cC$ maps holographically to a relation between the shapes $\cL_+$ and $\cL_-$ of the probe D-branes on the two sides of the coisotropic D-brane. 
    \item As $n$ is fixed in the large $N$ limit, the coset is weakly coupled/semiclassical. The Kac-Moody currents are identified with the 
    twisted connections $A_+$ and $A_-$. Accordingly, we expect a classical equation 
    \begin{equation}
        A_+ = A_- + \gamma \beta  \, .
    \end{equation}
    This equation actually defines a Lagrangian correspondence to be identified with $\cC$. See \cite{Gaiotto:2016hvd} for an analogous construction for Higgs bundles. 
\end{itemize}
It would be interesting to test this proposal further. 

Beyond the leading order in $N$, the construction predicts the 
presence of an auxiliary $\cW^{\mathrm{aux}}_\infty$ chiral algebra in the holographic setup, with $\lambda_1^{\mathrm{aux}}= -n$, in the presence of $m$ vertex operators which arise from minimal regular singularities in the coset construction. 

We could also study a setup where the numbers $n_+$ and $n_-$ of auxiliary D-branes jump across the interface. This would replace $\cV^b_{n,\kappa+N-n}$ by a $Y^{0,n_+,n_-}$ corner vertex algebra \cite{Gaiotto:2017euk}, defined by a combination of quantum Drinfeld-Sokolov reduction and coset from $SU(n_\pm)$
currents. It is straightforward to use a semiclassical limit to derive the holographic candidate for the correspondence $\cC$ 
between $\cP_{m,n_+}$ and $\cP_{m,n_-}$. For $n_-=0$, we should recover the oper manifold in $\cP_{m,n_+}$.

Extra degenerate insertions in the fundamental representation 
or Miura operators will map to Wilson lines in the 5d theory. 
It should be possible to compose the effect of the back-reaction and the Miura operator for the auxiliary $\cW_\infty$ chiral algebra to give an holographic prediction for the correlation functions, perhaps even at all orders in the 't Hooft expansion. We leave this to future work.

\section*{Acknowledgments}
We would like to thank Kevin Costello for useful conversations. This research was supported in part by a grant from the Krembil Foundation. DG is supported by the NSERC Discovery Grant program and by the Perimeter Institute for Theoretical Physics. Research at Perimeter Institute is supported in part by the Government of Canada through the Department of Innovation, Science and Economic Development and by the Province of Ontario through the Ministry of Colleges and
Universities.

\appendix

\section{The geometry of $M_t$ and its quantization} \label{app:geometry}
The deformed $A_1$ singularity $M_t$ can be defined by the equation 
\begin{equation}
	y \tilde y = w (w-t)
\end{equation}
in $\bC^3$, with complex symplectic form $\omega = dw dy/y$. Turning off $t$ gives the 
$A_1$ singularity, which can be resolved to $T^* \bC P^1$. Away from the singularity, the projection to $\bC P^1$ uses homogeneous coordinates $(w,y) \sim (\tilde y,w)$. 

When $t \neq 0$, the homogeneous coordinates $(w,y)$ and $(\tilde y,w)$ are not equivalent and define two natural projections $\pi_\pm$ to $\bC P^1$. This gives two distinct presentations of $M_t$ as a twisted cotangent bundle:
\begin{align}
	x_+ &= \frac{w}{y} \qquad \qquad \qquad \qquad \qquad y \cr
	x'_+ &= \frac{w-t}{\tilde y} = x_+^{-1} \qquad \qquad - \tilde y = - y x_+^{2} + t x_+
\end{align}
and 
\begin{align}
	x_- &= \frac{w-t}{y} \qquad \qquad \qquad \qquad y \cr
	x'_- &= \frac{w}{\tilde y} = x_-^{-1} \qquad \qquad - \tilde y = - y x_-^{2} - t x_-
\end{align}
``Twisted'' here means that the fiber coordinates in the two patches are related affine-linearly, with an extra shift compared to the standard cotangent bundle definition.

In the holographic setting, the $x_\pm$ coordinates are a back-reacted version of the original complex coordinate $x$ on the holographic boundary,
reducing to it respectively for $y_3\gg 0$ and $y_3\ll 0$. Here $y_3$ is the fiber coordinate on the $T^* \bR$ factor of the full 6d geometry. 

The natural quantization of the algebra of holomorphic functions on $M_t$ is the central quotient $A_t$ of $U(\fsl_2)$. We can realize it in two ways in terms of twisted differential operators on 
$\bCP^1$. The first is 
\begin{equation}
	y = \partial_{x_+} \qquad \qquad w = x_+ \partial_{x_+} + \frac12 \qquad \qquad \tilde y = x_+^2 \partial_{x_+} + (1-t) x_+\,,
\end{equation}
which satisfy
\begin{equation}
	\tilde y y = \left(w - \frac12\right)\left(w - t-\frac12\right) \qquad \qquad y \tilde y = \left(w + \frac12\right)\left(w - t+\frac12\right)\,.
\end{equation}
The second is 
\begin{equation}
	y = \partial_{x_-} \qquad \qquad w = x_- \partial_{x_-} + t + \frac12 \qquad \qquad \tilde y = x_-^2 \partial_{x_-} + (1+t) x_-\,.
\end{equation}
These quantize the two twisted cotangent bundle presentations. 

The relation between the two presentations is easier to understand after a Fourier transform:
\begin{equation}
	y = y \qquad \qquad w = -y \partial_y - \frac12 \qquad \qquad \tilde y = y \partial_y^2  + (1+t) \partial_y
\end{equation}
and 
\begin{equation}
	y = y \qquad \qquad w = -y \partial_y + t - \frac12 \qquad \qquad \tilde y = y \partial_y^2  + (1-t) \partial_y
\end{equation}
are related by conjugation by $y^t$. Before Fourier transform, one can formally use an $(x_+ - x_-)^{-1-t}$ integral kernel (``shadow transform'').

\section{Open string theory on branes}
\label{apdx:open_string}
In this appendix, we review the open-string field theory living on various branes in our topological string setup. The equivalence between the pure $A$-model on $\mathbb{R}^6$ and the mixed $A$/non-commutative $B$-model on $\mathbb{R}^2 \times \mathbb{C}^2$ is carefully discussed in \cite{Aganagic:2017tvx}.

In general, a brane in the mixed $A/B$ model on $X_1 \times X_2$ is specified by a Lagrangian $L$ in $X_1$ and a coherent sheaf $E$ on $X_2$. The open string states stretching between $(L,E)$ and $(L',E')$ in this mixed $A/B$ model are then given by
\begin{equation}
	HF^{\bullet}(L,L')\otimes \mathrm{Ext}^{\bullet}(E,E')\,.
\end{equation}

Given a collection of branes, the corresponding open string states form a cyclic dg algebra $(\mathcal{A}, d, \cdot)$. The open string field theory can then be formulated in the BV formalism using $\mathcal{A}$. The field content is given by $\mathcal{A}[1]$, and the action functional takes the following Chern–Simons type action
\begin{equation}
	S(\alpha) = \frac{1}{2}\Tr(\alpha d\alpha) + \frac{1}{3}\Tr(\alpha^3)\,.
\end{equation}

Now we consider the mixed $A/B$ model with the $A$-model on $\mathbb{R}_x\times\bR_y$ and the non-commutative $B$-model on $\mathbb{C}^2$. The non-commutativity deforms the $B$-model category from coherent $\mathcal{O}_{\mathbb{C}^2}$-modules to coherent $\mathcal{D}_{\mathbb{C}}$-modules. 

First, as an exercise, we review the theory on the $3d$ $D$-branes $(\bR_x\times\{0\},\mathcal{O}_{\bC})^{\oplus N}$. The Floer cohomology part is simply modeled by the algebra $\Omega^{\bullet}(\bR)$. To compute the B-model part $\mathrm{Ext}_{\mathcal{D}_{\bC}}(\mathcal{O}_{\bC},\mathcal{O}_{\bC})$, we use the following resolution of $\mathcal{O}_{\bC}$ as a $\mathcal{D}_{\bC}$-module:\begin{equation}\label{eq:res_O}
	0 \to \mathcal{D}_{\bC} \overset{\partial_z}{\to} \mathcal{D}_{\bC} \to 0\,.
\end{equation}
As a consequence, the $\mathrm{Ext}$ group is computed by $\mathcal{O}_{\bC} \overset{\partial_z}{\to} \mathcal{O}_{\bC}$. To produce an actual field theory, we further take the Dolbeault resolution, so the field content is given by 
\begin{equation}
	\Omega^{0,\bullet}(\bC) \overset{\partial_z}{\to} \Omega^{1,\bullet}(\bC)\,.
\end{equation}
The total complex is simply the de-Rham complex $(\Omega^{\bullet}(\bC),d_{dR})$. Combining with the $A$-model part, and taking a stack of $N$-branes, we get BV theory $A \in \Omega^{\bullet}(\bR\times\bC)\otimes\mathfrak{gl}_N[1]$, with the standard Chern-Simons action. 

Next, we consider the coisotropic branes, given by the Lagrangian $\{0\}\times\mathbb{R}_y$ and $\mathcal{D}_{\mathbb{C}}$ itself, viewed as a $\mathcal{D}_{\mathbb{C}}$-module. The space of open string states from such coisotropic branes to the 3d branes is given by $HF^{\bullet}(\{0\}\times\bR_y,\bR_x\times\{0\})\otimes\mathrm{Ext}_{\mathcal{D}_{\bC}}(\mathcal{D}_{\bC},\mathcal{O}_{\bC})$. The Floer cohomology gives us $\bR$ but in degree $1$. The Ext group is simply $\mathcal{O}_{\bC}$ which has Dolbeault resolution $\Omega^{0,\bullet}(\bC)$. The space of open string states from 3d branes to coisotropic branes is given by $HF^{\bullet}(\bR_x\times\{0\},\{0\}\times\bR_y)\otimes\mathrm{Ext}_{\mathcal{D}_{\bC}}(\mathcal{O}_{\bC},\mathcal{D}_{\bC})$. The Floer cohomology still gives $\bR$ in degree $0$. The Ext group is computed via the resolution \eqref{eq:res_O}, we find $\mathrm{Ext}_{\mathcal{D}_{\bC}}(\mathcal{O}_{\bC},\mathcal{D}_{\bC}) = (\mathcal{D}_{\bC}/\partial\cdot\mathcal{D}_{\bC})[-1] \cong \mathcal{O}_{\bC}[-1]$. After Dolbeault resolution this gives $\Omega^{1,\bullet}(\bC)$. 

To summarize, the open string states stretched between $N$ $3d$ branes and $K$ coisotropic branes gives us the BV field content
\begin{equation}
\begin{aligned}
		\boldsymbol{\psi} &\in \Omega^{0,\bullet}(\bC)\otimes \mathrm{Hom}(\bC^K,\bC^N)\,,\\
	\boldsymbol{\chi} &\in \Omega^{1,\bullet}(\bC)\otimes \mathrm{Hom}(\bC^N,\bC^K)\,.
\end{aligned}
\end{equation}
The action functional is given by $\int\Tr \boldsymbol{\chi}\bar{\partial}_A\boldsymbol{\psi}$.

In \cite{Aganagic:2017tvx}, the authors considered coisotropic branes of the form $(\mathbb{R}_x \times \{0\}, \mathcal{D}_{\mathbb{C}})$, which are parallel to the 3d branes. Such configurations give rise to Chern–Simons theory coupled to 3d holomorphic–topological bifundamental fields.

We also would like to understand anti-coisotropic brane in this setup. We propose the following $\mathcal{D}_{\bC}$ module
\begin{equation}
	M := \bigoplus_{n,m\geq 0}\bC\{z^n\bar{z}^me^{-\frac{|z|^2}{2}}\}\,.
\end{equation}
This is a dense subset of $L^2(\bC)$. The $\mathcal{D}_{\bC}$-module structure is given by 
\begin{equation}
	z\cdot(z^n\bar{z}^me^{-\frac{|z|^2}{2}}) = z^{n+1}\bar{z}^me^{-\frac{|z|^2}{2}},\quad \partial_z\cdot(z^n\bar{z}^me^{-\frac{|z|^2}{2}}) = (nz^{n-1}\bar{z}^m + z^{n}\bar{z}^{m+1})e^{-\frac{|z|^2}{2}}\,.
\end{equation}
We have a $\mathcal{D}_{\bC}$ module map $\varphi: M \to \mathcal{D}_{\bC}$ defined as follows
\begin{equation}
	\varphi(z^n\bar{z}^me^{-\frac{|z|^2}{2}}) = z^n\partial_z^m\,.
\end{equation}
It is easy to check that $\varphi$ is an isomorphism of $\mathcal{D}_{\bC}$-module.

Another important brane is the degree-shifted ghost coisotropic brane. It is specified by the Lagrangian $\{0\}\times\mathbb{R}_y$ together with the $\mathcal{D}_{\mathbb{C}}$-module $\mathcal{D}_{\mathbb{C}}[1]$. The degree shift modifies the grading of the ‘$3$–$5$’ and ‘$5$–$3$’ open strings. The corresponding BV theory is given by
\begin{equation}
	\begin{aligned}
		\boldsymbol{X} &\in \Omega^{0,\bullet}(\bC)\otimes \mathrm{Hom}(\bC^K,\bC^N)\,\\
		\boldsymbol{Y} &\in \Omega^{1,\bullet}(\bC)\otimes \mathrm{Hom}(\bC^N,\bC^K)\,
	\end{aligned}
\end{equation}
with action functional $\int\Tr \boldsymbol{X}\bar{\partial}_A\boldsymbol{Y}$. This $2d$ theory by itself realizes the symplectic boson VOA $\mathrm{Sb}$. As a chiral interface in $3d$ Chern–Simons theory, it produces the coset construction discussed in \ref{sec:sym_bos}.

We can also consider the $5d$ theory on a stack of $K$ coisotropic branes together with $K'$ ‘ghost’ coisotropic branes. As a natural generalization of the result in \cite{Aganagic:2017tvx}, this setup gives rise to the $\mathfrak{gl}_{K|K'}$ 5d non-commutative Chern–Simons theory.

There are close relationships between the free fermion and the symplectic boson interfaces. In fact, by reversing the orientation of $\mathbb{R}^2$ (or equivalently swapping the Lagrangians $\{0\}\times\mathbb{R}_y$ and $\mathbb{R}_x\times\{0\}$), the Floer cohomology contributes a different sign to the ‘$3$–$5$’ and ‘$5$–$3$’ strings. After this change, the coisotropic brane gives rise to the bosonic system. This suggests that there is no essential difference between the free fermion and symplectic boson interfaces, except that they flip the direction of the Chern–Simons level jump. Although free fermions and symplectic bosons define very different VOAs, it was observed in \cite{Gaiotto:2017euk} that the symplectic boson coset also truncates to the $\mathcal{W}_\infty$ algebra.

\section{Homological analysis of the KK theory}
\subsection{More details on the KK reduction}
\label{app:KK}
In this Appendix, we provide more details on the KK reduction performed in the Section \ref{sec:new_star} and \ref{sec:KK_nonlinear}. We consider the isomorphism $\mathbb{R}\times \mathbb{C}^2 \cong \mathbb{R}_{>0}\times S^2\times \mathbb{C}$, and compactify the theory on $S^2$. The first step is the isomorphism of complexes
\begin{equation}
	\Omega_{HT}^{\bullet}(\mathbb{R}\times\mathbb{C}^2) \cong \Omega_{HT}^{\bullet}(\mathbb{R}_{>0}\times \mathbb{C})\otimes \Omega^{\bullet}_{HT}(S^2)\,.
\end{equation}
We denote the coordinates on $\mathbb{R}\times\mathbb{C}^2$ by $(x,z_0,\bar{z}_0,z_1,\bar{z}_1)$ and the coordinates on $\mathbb{R}_{>0}\times \mathbb{C}$ by $r,z_1,\bar{z}_1$. We define $S^2$ by embedding into $\mathbb{R}_t\times\mathbb{C}_{w,\bar{w}}$, cut out by $t^2 +w\bar{w} = 1$.
Then, the above isomorphism can be written as
\begin{equation}
	\begin{aligned}
		&(x,z_0,z_1,\bar{z}_0,\bar{z}_1) \to \\
		&(r = \sqrt{x^2 + z_0\bar{z}_0},z_1 = z_1,\bar{z}_1 = \bar{z}_1, t = \frac{x}{r},w = z_0,\bar{w} = \frac{\bar{z}_0}{r^2})\,,
	\end{aligned}
\end{equation}
and
\begin{equation}
	\begin{aligned}
		&(dx,d\bar{z}_0,d\bar{z}_1) \to \\
		&(dr = \frac{2xdx+z_0d\bar{z}_0}{2r}, d\bar{z}_1 = d\bar{z}_1, dt = \frac{dx}{r} - \frac{xdr}{r^2},d\bar{w}_0 = \frac{d\bar{z}_0}{r^2} - \frac{2\bar{z}_0dr}{r^3})\,.
	\end{aligned}
\end{equation}

As a next step, we analyze the complex $\Omega^{\bullet}_{HT}(S^2)$. It will be convenient to use an algebraic version of $\Omega_{HT}^{\bullet}(S^2)$ given by \cite{Garner:2023zqn}
\begin{equation}
	\begin{aligned}
		&\mathbb{C}[t,w,\bar{w},dw]/(t^2+w\bar{w} = 1,2tdt+wd\bar{w} = 0)\\
		&\cong \mathbb{C}[t,w,\bar{w}]/(t^2+w\bar{w} = 1)\oplus \mathbb{C}[t,w,\bar{w}]/(t^2+w\bar{w} = 1)\Omega\,,
	\end{aligned}
\end{equation}
where $\Omega$ is identified with the $1$-form defined in \ref{eq:BM_form}. This is a dense subspace of $\Omega_{HT}^{\bullet}(S^2)$. The differential of this complex can be written as 
\begin{equation}\label{eq:diff_S2}
	d_{S^2}\bar{w} = t\Omega,\quad d_{S^2}t = -\frac{1}{2}z\Omega\,.
\end{equation}
For convenience we also denote the above complex by $\Omega_{HT}^{\bullet}(S^2)$.

Let us denote $\mathcal{H} = \mathbb{C}[t,w,\bar{w}]/(t^2+w\bar{w} = 1)$. This space has a harmonic decomposition:
\begin{equation}
	\mathcal{H} = \bigoplus_{j\in \mathbb{Z}_{\geq0}}\mathcal{H}_j\,,
\end{equation}
where $\mathcal{H}_j$ is the space of $3d$ harmonic polynomials of degree $j$. This decomposition is compatible with the action of $\mathfrak{sl}_2 = \{\mathbf{e},\mathbf{f},\mathbf{h}\}$. We can identify
\begin{equation*}
	\mathbf{e} = w\pa_t - 2t\pa_{\bar{w}},\quad\mathbf{f} = -\bar{w}\pa_t + 2t\pa_w,\quad\mathbf{h} = 2w\pa_w - 2\bar{w}\pa_{\bar{w}}\,.
\end{equation*}
The space $\mathcal{H}_j$ is isomorphic to the standard $2j+1$ dimensional representation of $\mathfrak{sl}_2$.  By the formula \eqref{eq:diff_S2}, we can identify the differential $d_{S^2} =-\frac{1}{2}\Omega\mathbf{e}$. This immediately implies that the cohomology $H_{HT}^{\bullet}(S^2)$ is given by
\begin{equation*}
	H_{HT}^{n}(S^2) = \begin{cases}
		\mathbb{C}[w], & n = 0;\\
		C[\bar{w}]\Omega, & n = 1.
	\end{cases}
\end{equation*}

This justified the KK tower we picked up in \ref{eq:KK_modes}. For more careful analysis, information about the whole harmonic modes $\mathcal{H}$ is required. We denote the standard $\mathfrak{sl}_2$ orthonormal basis of $\mathcal{H}_j$, normalized using $\int_{S^2}d\sigma_{S^2}$, by 
\begin{equation*}
	\{v^{(j)}_m | -j \leq m \leq j\}\,.
\end{equation*}
The action of the $\mathfrak{sl}_2$ on this basis is given as follows
\begin{equation*}
	\mathbf{e}v^{(j)}_m = \sqrt{(j-m)(j+m+1)}v^{(j)}_{m+1},\quad \mathbf{f}v^{(j)}_m = \sqrt{(j+m)(j-m+1)}v^{(j)}_{m-1},\quad \mathbf{h}v^{(j)}_m = 2m v^{(j)}_{m}\,.
\end{equation*}
The highest and lowest weight vectors are $v^{(j)}_j = \frac{1}{2^jj!}\sqrt{\frac{(2j+1)!}{4\pi}}w^j$, $v^{(j)}_{-j} = \frac{(-1)^j}{2^jj!}\sqrt{\frac{(2j+1)!}{4\pi}}\bar{w}^j$.

We defined $v^{(j)}_m$ as polynomial of $(t,w,\bar{w})$. We can also write them using the spherical coordinate $$(t = \cos\theta,w = \sin\theta e^{i\phi},\bar{w} = \sin\theta e^{-i\phi})$$
Then $\{v^{(j)}_m\}$'s are exactly the $S^2$ spherical harmonics $\{Y^{(j)}_m(\theta,\phi)\}$ (sometimes also written as $Y_j^{m}(\theta,\phi)$).
\subsection{Homotopy Transfer}
\label{app:transfer}
We have seen in Section \ref{sec:hskk} that the cohomology $H_{HT}^{\bullet}(S^2)$ gives rise to the $3d$ KK modes appropriate for comparison with the $\mathcal{W}_\infty$ algebra. As discussed in \cite{Zeng:2023qqp} in a similar setup, non-linear corrections to the $3d$ theory arise from higher operations on the cohomology $H_{HT}^{\bullet}(S^2)$. The goal of this appendix is to compute these higher products using homotopy transfer \cite{Kadeishvili_1980,berglund2014homological}.

The non-commutative multiplication $*$ on the complex $\Omega_{HT}^{\bullet}(\mathbb{R}\times\mathbb{C}^2)$ induce a multiplication on $\Omega_{HT}^{\bullet}(\mathbb{R}_{>0}\times \mathbb{C})\otimes \Omega^{\bullet}_{HT}(S^2)$ via the above isomorphism. However, the explicit form of this multiplication is very complicated as $r$ is also involved in the non-commutative product. To simplify things, we first take the cohomology of $\Omega_{HT}^{\bullet}(\mathbb{R}_{>0}\times \mathbb{C})$. We have $H^{\bullet}(\Omega_{HT}^{\bullet}(\mathbb{R}_{>0}\times \mathbb{C}),d_{HT} )  = \mathcal{O}(\mathbb{C})$, which lies in degree $0$. We have a quasi-isomorphism:
\begin{equation*}
	\Omega_{HT}^{\bullet}(\mathbb{R}\times\mathbb{C}^2) \to \mathcal{O}(\mathbb{C}) \otimes \Omega_{HT}^{\bullet}(S^2)\,.
\end{equation*}
Simple degree argument tells us that no higher product is induced in the above quasi-isomorphism. Therefore, we can focus on the product on $\mathcal{O}(\mathbb{C}) \otimes \Omega^{\bullet}_{HT}(S^2)$, which is easier. 

It is important that we need to perform the homotopy transfer for the whole complex $\mathcal{O}(\mathbb{C}) \otimes \Omega_{HT}^{\bullet}(S^2)$, as the non-commutative product will mix $\mathcal{O}(\mathbb{C})$ with $\Omega_{HT}^{\bullet}(S^2)$. But we first analyze the higher product on $H_{HT}^{\bullet}(S^2)$.

We first construct a special deformation retract between $(\Omega_{HT}^{\bullet}(S^2),d_{S^2})$ and its cohomology $H_{HT}^{\bullet}(S^2)$:
\begin{equation*}
	h\curved (\Omega_{HT}^{\bullet}(S^2),d_{S^2})\overset{p}{\underset{i}\rightleftarrows} (H_{HT}^{\bullet}(S^2) , 0)\,.
\end{equation*}

Here we identify $H_{HT}^{\bullet}(S^2)$ with the subspace of $\Omega_{HT}^{\bullet}(S^2)$, which gives us natural inclusion and projection operators $p:\Omega_{HT}^{\bullet}(S^2) \rightleftarrows H_{HT}^{\bullet}(S^2):i$. They satisfy $p\circ i = 1$.

We define the homotopy operator $h$ to be
\begin{equation}\label{eq:homotopy_h}
	h(v^{(j)}_m\Omega) = \frac{2}{\sqrt{(j+m)(j-m+1)}}v^{(j)}_{m-1}\,.
\end{equation}
It is easy to verify the identity
\begin{equation*}
	i\circ p - 1 = d_{S^2}\circ h + h\circ d_{S^2}\,,
\end{equation*}
and 
\begin{equation*}
	h\circ i = 0,\quad p\circ h = 0,\quad h\circ h = 0\;.
\end{equation*}

Given these data, we can compute the $A_\infty$ structure $\{m_n\}_{n\geq 0}$ on $H_{HT}^{\bullet}(S^2)$. We use the multiplication rule of $\Omega_{HT}^{\bullet}(S^2)$ under the harmonic decomposition, which is the well-known formula for the product of spherical harmonics
\begin{equation*}
	\begin{aligned}
		v^{(j_1)}_{m_1}\cdot v^{(j_2)}_{m_2} & = \sum_{j}\sum_{m}\sqrt{\frac{(2j_1+1)(2j_2+1)}{(2j+1)}}C^{j,m}_{j_1,m_1;j_2,m_2}C^{j,0}_{j_1,0;j_2,0}v^{(j)}_{m}\\
		& = \sum_{j = |j_1-j_2|}^{j_1+j_2}\sqrt{\frac{(2j_1+1)(2j_2+1)}{(2j+1)}}C^{j,m_1+m_2}_{j_1,m_1;j_2,m_2}C^{j,0}_{j_1,0;j_2,0}v^{(j)}_{m_1+m_2}\,.
	\end{aligned}
\end{equation*}

The 2-array product is given by the product on $\mathbb{C}[w]$, which further induce a map $\mathbb{C}[w]\otimes \mathbb{C}[\bar{w}]\Omega \to \mathbb{C}[\bar{w}]\Omega$ via the pairing \eqref{eq:KK_pairing}. We have
\begin{equation*}
	m_2(w^{k},\frac{(2l+1)!}{4\pi(2^ll!)^2}\bar{w}^l\Omega) = \frac{(2l-2k+1)!}{4\pi(2^{l-k}(l-k)!)^2}\bar{w}^{l-k}\Omega\,.
\end{equation*}
By degree reason (the map $m_n$ have degree $2-n$), the $n$-array product $m_n$ is non zero only in the subspace
\begin{equation*}
	\begin{aligned}
		\bigoplus_{\text{perm}}H^{0}_{HT}(S^2)\otimes H^{1}_{HT}(S^2)^{\otimes n-1}\,.\\
		\bigoplus_{\text{perm}}H^{0}_{HT}(S^2)^{\otimes 2}\otimes H^{1}_{HT}(S^2)^{\otimes n-2}\,.
	\end{aligned}
\end{equation*} 
By the $A_\infty$ relation and the pairing, it suffices to compute $m_n$ on $H^{0}_{HT}(S^2)\otimes H^{1}_{HT}(S^2)^{\otimes n-1}$. 

Let us denote
\begin{equation}
	\mu_n(a_0,\bar{a}_1\Omega,\dots,\bar{a}_{n-1}\Omega) = h(\dots h(h(a_0\cdot \bar{a}_1\Omega)\cdot\bar{a}_2\Omega)\dots)\cdot \bar{a}_{n-1}\Omega ,\quad n\geq 2\,,
\end{equation}
where $a_1 \in \mathbb{C}[w]$ and $\bar{a}_i \in \mathbb{C}[\bar{w}]$. The operation $\mu_n$ can also be defined iteratively by
\begin{equation*}
	\begin{aligned}
		&\mu_2(a_0,\bar{a}_1\Omega) = a_0\cdot\bar{a}_1\Omega,\\
		& \mu_n(a_0,\bar{a}_1\Omega,\dots,\bar{a}_{n-1}\Omega) = h(\mu_{n-1}(a_0,\bar{a}_1\Omega,\dots,\bar{a}_{n-2}\Omega))\cdot \bar{a}_{n-1}\Omega\,.
	\end{aligned}
\end{equation*}
The first two terms are
\begin{equation*}
	\begin{aligned}
		\mu_2(v^{l}_{l},v^{j_1}_{-j_1}\Omega) = &\sum_{k = |l-j_1|}^{j_1+l}\sqrt{\frac{(2l+1)(2j_1+1)}{(2k+1)}}C^{k,l-j_1}_{l,l;j_1,-j_1}C^{k,0}_{l,0;j_1,0}v^{(k)}_{l-j_1}\Omega\,,\\
		\mu_3(v^{l}_{l},v^{j_1}_{-j_1}\Omega,v^{j_2}_{-j_2}\Omega) = &\sum_{k_2 = |k_1-j_2|}^{k_1+j_2}\sum_{k_1 = |l-j_1|}^{j_1+l}2\sqrt{\frac{(2l+1)(2j_1+1)(2j_2+1)}{(2k_2+1)(k_1+l-j_1)(k_1-l+j_1+1)}}\\
		&\times C^{k_1,l-j_1}_{l,l;j_1,-j_1}C^{k_1,0}_{l,0;j_1,0}C^{k_2,l-j_1-j_2-1}_{k_1,l-j_1-1;j_2,-j_2}C^{k_2,0}_{k_1,0;j_2,0}v^{(k_2)}_{l-j_1-j_2-1}\Omega\,.
	\end{aligned}
\end{equation*}
More generally we can solve
\begin{equation}\label{eq:higher_mu}
	\begin{aligned}
		&\mu_n(v^{l}_{l},v^{j_1}_{-j_1}\Omega,\dots,v^{j_{n-1}}_{-j_{n-1}}\Omega) \\
		= &\sum_{k_{n-1} = |k_{n-2}-j_{n-1}|}^{k_{n_2} + j_{n-1}}\dots\sum_{k_2 = |k_1-j_2|}^{k_1+j_2}\sum_{k_1 = |l-j_1|}^{j_1+l}2^{n-2}\sqrt{\frac{2l+1}{2k_{n-1}+1}\prod_{i = 1}^{n-1}\frac{2j_i+1}{(k_i+l-J_i)(k_i-l+J_i+1)}}\\
		& \times\left( \prod_{i = 1}^{n-1}C^{k_i,0}_{k_{i-1},0;j_i,0}C^{k_i,l-J_i}_{k_{i-1},l-J_{i-1}-1;j_{i},-j_i}\right) v^{(k_{n-1})}_{l-J_{n-1}}\Omega\,.
	\end{aligned}
\end{equation}
In the above expression we set
\begin{equation}
	\begin{aligned}
		&k_0 = l\,,\\
		&J_i =  j_1 + j_2+\dots + j_i + i - 1\,.
	\end{aligned}
\end{equation}
The relation between $\mu_n$ and $m_n$ is given by
\begin{equation}\label{eq:mu_to_m}
	m_n(a_0,\bar{a}_1\Omega,\dots,\bar{a}_{n-1}\Omega) = p(\mu_n(a_0,\bar{a}_1\Omega,\dots,\bar{a}_{n-1}\Omega))\,.
\end{equation}
Therefore, we simply project the expression \eqref{eq:higher_mu} to the component $k_{n-1} = J_{n-1} - l$. We can express the structure constant $m_n$ in the basis $\{w^l\}$ and $\{ \frac{(2l+1)!}{4\pi(2^{l}l!)^2}\bar{w}^{l}\Omega \}$ of $H_{HT}^{\bullet}(S^2)$. This basis is related to the normalized basis $\{v^{(l)}_l\}$ and $\{v^{(l)}_{-l}\Omega\}$ via the following 
\begin{equation}
	v^{(l)}_l = N_lw^l,\quad v^{(l)}_{-l}\Omega = (-1)^lN_l\bar{w}^l\Omega
\end{equation}
where we define the constant $ N_l = \sqrt{\frac{(2l+1)!}{4\pi(2^{l}l!)^2}}$.

Let us define the constant
\begin{equation}
	\begin{aligned}
		(m_n)^{l',l}_{j_1,j_2,\dots,j_{n-1}} & = (w^{l'},m_n(w^{l},\frac{(2j_1+1)!}{4\pi(2^{j_1}j_1!)^2}\bar{w}^{j_1}\Omega,\dots,\frac{(2j_{n-1}+1)!}{4\pi(2^{j_{n-1}}j_{n-1}!)^2}\bar{w}^{j_{n-1}}\Omega))\\
		& = \frac{\prod_{i = 1}^{n-1}(-1)^{j_i}N_j}{N_lN_{l'}}(v^{(l')}_{l'},m_n(v^{l}_{l},v^{j_1}_{-j_1}\Omega,\dots,v^{j_{n-1}}_{-j_{n-1}}\Omega))
	\end{aligned}
\end{equation}
From the discussion of the previous section, we see that $(m_n)^{l',l}_{j_1,j_2,\dots,j_{n-1}}$ is non-zero only when
\begin{equation}
	l+l' = j_1+j_2+\dots+j_{n-1} + n - 2
\end{equation}
Using \eqref{eq:higher_mu} and \eqref{eq:mu_to_m}, we have
\begin{equation}
	\begin{aligned}
		(m_n)^{l',l}_{j_1,j_2,\dots,j_{n-1}} = &\frac{\prod_{i = 1}^{n-1}(-1)^{j_i}N_{j_i}}{N_lN_{l'}}\sum_{k_{n-2} = |k_{n-3}-j_{n-2}|}^{k_{n-2} + j_{n-1}}\dots\sum_{k_2 = |k_1-j_2|}^{k_1+j_2}\sum_{k_1 = |l-j_1|}^{j_1+l}2^{n-2}\sqrt{\frac{2l+1}{2l'+1}}\\
		&\times\left(\prod_{i = 1}^{n-1}\sqrt{\frac{2j_i+1}{(k_i+l-J_i)(k_i-l+J_i+1)}} C^{k_i,0}_{k_{i-1},0;j_i,0}C^{k_i,l-J_i}_{k_{i-1},l-J_{i-1}-1;j_{i},-j_i}\right) 
	\end{aligned}
\end{equation}
In the above expression we set
\begin{equation}
	\begin{aligned}
		&k_0 = l\\
		&k_{n-1} = l'\\
		&J_i =  j_1 + j_2+\dots + j_i + i - 1
	\end{aligned}
\end{equation}

The Clebsch–Gordan coefficients have the finer property that  $C^{k_i,0}_{k_{i-1},0;j_i,0} \neq 0$ only when $k_i-k_{i-1} - j_i = 0 \mod 2$. Therefore, for $m_n$ to be non-zero we have $k_{n-1} - (j_1+\dots+j_{n-1}) + l = 0 \mod 2$. This implies that only $m_{2n}$'s are non zero.
\subsection{Other supplementary computation}
\label{app:other}
In this appendix, we provide more supplementary computations needed for Section~\ref{sec:raviolo-agd}.

We first analyze the operator $r_L = (-\circ L)_+ : \xi^{-t}\circ (\Psi_K)_+ \to (\Psi_K)_+$. We define the other maps $\bar{r}_L = (\xi^{-t} \circ- \circ L)_+:(\Psi_K)_+ \to (\Psi_K)_+$, and $J = \xi^{-t}\circ- : (\Psi_K)_+ \to \xi^t\circ(\Psi_K)_+$. Then,
\begin{equation}
    \bar{r}_L = r_L\circ J.
\end{equation}
For $X\in(\Psi_K)_+$, we have
\begin{equation}
    \bar{r}_L(X) = (\xi^t\circ X\circ L)_+.
\end{equation}
We can write $\xi^{-t}\circ X \circ L = \xi^{-t}\circ X\circ \xi^t + \xi^t\circ X\circ (\xi^t - L)$. It is easy to check that $\xi^{-t}\circ X\circ \xi^t = X + \text{lower $\xi$-order terms}$ and that $\xi^t\circ X\circ (\xi^t - L)$ also only contains lower $\xi$-order terms. Therefore, we can decompose $\bar{r}_L(X)$ as
\begin{equation}
    \bar{r}_L(X) = X + N_L(X),
\end{equation}
where 
\begin{equation}
    N_L(-) = r_L(\xi^{-t}\circ -) - \mathrm{id}.
\end{equation}
Since $N_L(X)$ has strictly lower $\xi$-order than $X$, it is nilpotent. Therefore, we can write the inverse of $\bar{r}_L$ as 
\begin{equation}
    \bar{r}_L^{-1} = \sum_{m\geq0}(-1)^m N_L^m, 
\end{equation}
which gives the inverse of $r_L$ as
\begin{equation}
    r_L^{-1} = J^{-1} \circ \bar{r}_L^{-1} = \xi^t \circ \sum_{m\geq0}(-1)^m N_L^m. 
\end{equation}

We next verify the deformation retract displayed in \eqref{eq:rav_retract}. The identity $P_LI_L=1_{\mathcal K_L}$ can be checked in each degree.  For $X\in\xi^{-t}\circ(\Psi_K)_+$,
\begin{equation}
	P_LI_L(X) = r_L^{-1}\left(  \operatorname{Ad}_{L^{-1}}I_L(X)(0)\right)=r_L^{-1}(r_L(X)) = X.
\end{equation}
For $Y\in(\Psi_K)_-\circ\xi^t$, the coefficient $Y\circ L^{-1}$ is strictly negative, and hence
\begin{equation}
	P_LI_L(Y) =\left(\int_0^1Y\circ L^{-1}du\right)_-\circ L = (Y\circ L^{-1})\circ L=Y.
\end{equation}

We now verify the retract identity $I_LP_L=1_{\mathcal C_L}-(d_sK_L+K_Ld_s)$. For $F\in\mathcal C_L^0$, we have $I_LP_L(F)(s)=F(0)-sF(0)_-$. On the other hand, using $F(1)_-=0$ and $d_sK_L(F)=0$, we have 
\begin{equation}
\begin{aligned}
	K_Ld_sF + d_sK_LF  =\int_0^s\partial_uF(u)du -s\left(\int_0^1\partial_uF(u)du\right)_- =(1-I_LP_L)F.
\end{aligned}
\end{equation}
For $Gds\in\mathcal C_L^1$, $I_LP_L(Gds)=\left(\int_0^1G(u)du\right)_-ds$, while
\begin{equation}
	d_sK_L(Gds) + K_Ld_s(Gds) =\left(G(s)-\left(\int_0^1G(u)du\right)_-\right)ds = (1-I_LP_L)(Gds).
\end{equation}
Since $d_s$ vanishes on $\mathcal C_L^1$, these two calculations prove
$d_sK_L+K_Ld_s=1-I_LP_L$ in both degrees.

It remains to check the side conditions. $K_L(F) = 0$ for $F \in \mathcal{C}_L^0$ by definition. Since $K_L(G(s)ds)(0)=0$,
\begin{equation}
	P_LK_L(G(s)ds)
	=r_L^{-1}\!\left(
	  \operatorname{Ad}_{L^{-1}}K_L(G(s)ds)(0)\right)=0.
\end{equation}
Moreover, $Y\circ L^{-1}\in(\Psi_K)_-$ is independent of $s$, so
\begin{equation}
\begin{aligned}
	K_LI_L(Y)(s)
	&=s(Y\circ L^{-1})-s(Y\circ L^{-1})_-=0.
\end{aligned}
\end{equation}
Simply by degree reason $K_L^2=0$. Altogether,
\begin{equation}
	P_LI_L=1,\qquad
	d_sK_L+K_Ld_s=1-I_LP_L,\qquad
	P_LK_L=K_LI_L=K_L^2=0.
\end{equation}

Finally, we provide an expression of the Miura basis $\{u_j\}_{j\geq 1}$ in terms of the KK basis $\{b_j\}_{j\geq 1}$. We set
\begin{equation}
	W(s)=g_{t,b}(s)\circ\xi^{-st} =\exp_\circ\bigl(s(t\log\xi+b)\bigr)\circ\xi^{-st}.
\end{equation}
Differentiating $s$ gives
\begin{equation}
\begin{aligned}
	\partial_sW(s)
	&=g_{t,b}(s)\circ(t\log\xi+b)\circ\xi^{-st}
	  -g_{t,b}(s)\circ\xi^{-st}\circ t\log\xi\\
	&=W(s)\circ\operatorname{Ad}_{\xi^{st}}(b).
\end{aligned}
\end{equation}
Since $W(0)=1$, we can integrate
\begin{equation}
	W(s)=1+\int_0^s W(u)\circ\operatorname{Ad}_{\xi^{ut}}(b)\,du.
\end{equation}
Iterated substitution of this equation gives
\begin{equation}
\begin{aligned}
	W(s)=1+\int_0^s\operatorname{Ad}_{\xi^{s_1t}}(b)\,ds_1+\int_{0\leq s_1\leq s_2\leq s}
	\operatorname{Ad}_{\xi^{s_1t}}(b)\circ\operatorname{Ad}_{\xi^{s_2t}}(b)\,ds_1ds_2+\cdots.
\end{aligned}
\end{equation}
This give us the following expression
\begin{equation}
	L(b)\circ\xi^{-t}=1+\sum_{m\geq1}
	\int_{0\leq s_1\leq\cdots\leq s_m\leq1}
	\operatorname{Ad}_{\xi^{s_1t}}(b)\circ\cdots\circ
	\operatorname{Ad}_{\xi^{s_mt}}(b)\,ds_1\cdots ds_m.
\end{equation}
Since $b$ starts in order $\xi^{-1}$, only finitely many values of $m$ contribute to any fixed coefficient of $\xi$. In particular, we can obtain $u_j$ as the $\xi^{-j}$ coefficient of the following 
\begin{equation}
  \sum_{m=1}^{j}\int_{0\leq s_1\leq\cdots\leq s_m\leq1}	\left(	\operatorname{Ad}_{\xi^{s_1t}}(b)\circ\cdots\circ \operatorname{Ad}_{\xi^{s_mt}}(b)\right) ds_1\cdots ds_m.
\end{equation}

\section{The semi-classical limit of a quantum Drinfeld-Sokolov reduction}\label{app:DS}
We refer to \cite{khan2025poisson} for details on the relation between Poisson chiral algebras and 3d HT theories. A Poisson chiral algebra captures the semiclassical limit of a family of chiral algebras which deforms a free chiral algebra with trivial OPE. For example, we can start from a Kac-Moody algebra with standard OPE
\begin{equation}
    J^a(z) J^b(w) \sim \frac{k \delta^{ab}}{(z-w)^2} + \frac{f^{ab}_c}{z-w} J^c(w)
\end{equation}
and rescale the currents by a formal parameter $\hbar \sim k^{-1}$ to $\phi^a(z) = \hbar J^a(z)$, to get 
\begin{equation}
    \phi^a(z) \phi^b(w) \sim \hbar \left[(\hbar k) \frac{\delta^{ab}}{(z-w)^2} + \frac{f^{ab}_c}{z-w} \phi^c(w) \right]
\end{equation}
The right hand side, usually converted to a $\lambda$-bracket, is the data of the Poisson chiral algebra. It may seem identical to the original OPE, but it is extended to composite operators by 
a standard Leibniz rule rather than the intricate associativity relations of a chiral algebra.

The associated 3d HT theory has a BV action of the schematic form
\begin{equation}
    \frac{1}{\hbar} \int dz \left(\eta_a (d_x + \bar \partial) \phi^a + (\hbar k) \delta^{ab} \eta_a \partial \eta_b + f^{ab}_c \eta_a \eta_b \phi^c \right) \, ,
\end{equation}
where $\eta_a$ and $\phi^a$ are forms built from $dx$ and $d \bar z$ only. 
The definition 
\begin{equation}
    A_a = \eta_a + (\hbar k)^{-1} \delta_{ab} \phi^b dz 
\end{equation}
immediately reproduces the standard action of 3d Chern-Simons gauge theory. In particular, this proves that quantum corrections in the 3d theory will reproduce the full Kac-Moody chiral algebra rather than its Poisson limit.

The case of a free chiral algebra, such a $\beta \gamma$ system, is also instructive. 
We can trivially rescale one of the two fields, say $\beta$, to bring the OPE to the form 
\begin{equation}
    \beta \gamma \sim \frac{\hbar}{z-w}
\end{equation}
and build a 3d theory with action:
\begin{equation}
    \frac{1}{\hbar} \int dz \left(\tilde \beta(d_x + \bar \partial) \beta + \tilde \gamma (d_x + \bar \partial) \gamma + \tilde \beta \tilde \gamma \right)\, .
\end{equation}
We can do the same for a $bc$ system, flipping the Grassmann parity of the fields. 

We will need to extend this construction to dg-chiral algebras, i.e. chiral algebras equipped with a current $J_{\mathrm{BRST}}$ with a nilpotent zeromode
\begin{equation}
    Q = \oint J_{\mathrm{BRST}} dz
\end{equation}
used as a differential.

We will assume that the same rescaling which brings the OPE to be proportional to $\hbar$ will also 
rescale 
\begin{equation}
    Q = \frac{1}{\hbar} \oint J_{\mathrm{BRST}} dz
\end{equation}
This equips the resulting Poisson chiral algebra with a differential as well. There are two possible ways to include the effect of $Q$ in the 3d setting:
\begin{itemize}
    \item We can add a $\frac{1}{\hbar} \int_\partial J_{\mathrm{BRST}} dz$ boundary term to the BV action. Essentially, we are first reproducing the underlying chiral algebra and then turning on the differential at the boundary only. This seems a bad choice if our goal is to build a 3d theory which only depends on the Poisson chiral algebra modulo quasi-isomorphism. 
    \item We can add to the bulk action a new term linear in $\eta$, which will modify the BRST transformation of the fields $\phi$ which survive at the boundary. The term mimics the semiclassical BRST action, which will potentially be deformed by the same quantum corrections which build the full chiral algebra from its Poisson limit.
\end{itemize}
The two options are actually related by a field redefinition, but the latter may be more suitable to further redefinitions based on quasi-isomorphisms of Poisson chiral algebras. 

As an example, consider the BRST reduction of a $\fg_{k-h} \oplus \fg_{-k-h}$ Kac-Moody algebra combined with a $\beta\gamma$ system valued in $\fg$. The BRST current has schematic form 
\begin{equation}
    \Tr c(J_1 + J_2) + \frac12 b[c,c]) \, .
\end{equation}
This is known to produce the trivial chiral algebra. The first option above would essentially glue two 3d Chern-Simons theory at a common boundary to produce a trivial interface. The interface does indeed support a trivial chiral algebra, but this is hardly a minimal realization of that. 

The second option combines the two 3d Chern-Simons theories with the $b c$ and $\tilde b \tilde c$ fields and extra interaction of schematic form
\begin{equation}
    \frac{1}{\hbar} \int dz \Tr \tilde b (\phi_1 + \phi_2 + [b,c]) + \tilde c [c,c] + \eta_1 [c,\phi_1]+ \eta_2 [c,\phi_2] + k \hbar (\eta_1 -\eta_2) \partial c
\end{equation}
It would be nice to verify if these interactions trivialize the 3d theory. The first quadratic term acts in the same manner as a quadratic superpotential in a twisted 3d ${\cal N}=2$ theory, and it is likely to eliminate $b$, $\tilde b$, $\phi_1 + \phi_2$ and $\eta_1 + \eta_2$. The rest of the fields, though, still appear to have a non-trivial action.

In general, we expect that two quasi-isomorphic (in an appropriate sense) Poisson chiral algebras give rise to equivalent 3d Poisson sigma model. Mathematically, the 3d theory governs the deformation theory of the Poisson chiral algebra; since quasi-isomorphic Poisson chiral algebras should share the same deformation theory, the resulting 3d theories should be equivalent. In the case of associative algebra, it is known that the Hochschild cohomology of two quasi-isomorphic associative (or $A_\infty$) algebras are quasi-isomorphic \cite{Keller1999IntroductionT,keller2003derived}. This implies that the 2d Poisson sigma models of quasi-isomorphic Poisson algebras are equivalent.

We can apply this strategy to the quantum DS reduction of an $\fsl_n$ Kac-Moody algebra. We can focus on $\fsl_2$ for simplicity. Then we add a single $bc$ system and BRST current
\begin{equation}
    c(J^+ - 1)
\end{equation}
The interactions become $\tilde b (\phi^+-1) + \eta_- (k\hbar \partial c + \phi^0 c) - \eta_0 c \phi^+$. 

The goal of the remainder of this appendix is to demonstrate that there is an isomorphism between this theory and the 3d Poisson sigma model associated with the Virasoro algebra, which is studied in \cite{khan2025poisson}.  We begin by examining the semi-classical DS reduction of the $\mathfrak{sl}_2$ Kac-Moody algebra, which is known to give rise to the Virasoro algebra. In the semi-classical regime, we have a commutative algebra of operators $\mathcal{V}_k(\mathfrak{sl}_2)\otimes \mathrm{bc} = \mathbb{C}[\partial^n\phi^{\pm},\partial^n\phi^0,\partial^nb,\partial^nc]\,.$ The DS differential has the following non-zero components:
\begin{equation}\label{eq:DS_diff}
\begin{aligned}
    Qb& = \phi^+ - 1,\\
    Q\phi^- & = k\hbar \partial c + \phi^0 c,\\
    Q\phi^0 & =  - c \phi^+ \,.
\end{aligned}
\end{equation}
We have a map $i$ from the Virasoro algebra $\mathcal{V}ir = \mathbb{C}[\partial^nT]$ to the above complex given by
\begin{equation}
    i(T) = T_{DS} := \frac{1}{2k\hbar}(\phi^+\phi^- + \frac{1}{2}(\phi^0)^2 + k\hbar \partial \phi^0 + k\hbar \partial b c)
\end{equation}
We can check that $i$ is a chain map, i.e. $Q T_{DS} = 0$. Moreover, $i$ should be a quasi-isomorphism (after a completion). This is a corollary of the quantum DS reduction, and in this $\mathfrak{sl}_2$ case, it is easy to verify directly. The procedure of DS reduction also suggest that we have the following map $p$ from $\mathcal{V}_k(\mathfrak{sl}_2)\otimes \mathrm{bc}$ to $\mathcal{V}ir$ given by
\begin{equation}
    p(\phi^+) = 1,\quad p(\phi^0) = 0,\quad p(\phi^-) = 2 k\hbar T,\quad p(b) = 0,\quad p(c) = 0\,.
\end{equation}
The map $p$ should be a (quasi-)inverse of $i$ in the sense that $p i = 1$, and in principle, there should be a homotopy operator $h$ such that $i p - 1 = Q h + h Q$. Given the above data, we can proceed to construct the quasi-isomorphism of the bulk theory. Before we consider the bulk action responsible for the $\lambda$-bracket, we first focus on the bulk action that is responsible for the DS differential. It induces the differential $Q$ as in \eqref{eq:DS_diff} and on the remaining fields as follows
\begin{equation}
    \begin{aligned}
            Q\eta_+ &= \tilde{b} - \eta_0 c,\\
            Q\eta_0 &= \eta_- c,\\
            Q\tilde{c} & = -k\hbar\partial\eta_- + \eta_-\phi^0 - \eta_0\phi^+\,.
    \end{aligned}
\end{equation}
From the above data, we can construct the following maps between the bulk algebras $\mathbb{C}[\partial^nT,\partial^n\eta]$ and $(\mathbb{C}[\partial^n\phi^{\pm},\partial^n\phi^0,\partial^nb,\partial^nc,\partial^n\tilde{b},\partial^n\tilde{c},\partial^n\eta_{\pm},\partial^n\eta_0],Q)$:
\begin{equation}
    \tilde{i}(T) = T_{DS},\quad \tilde{i}(\eta) = 2k\hbar \mu_-
\end{equation}
It has a (quasi)-inverse
\begin{equation}
    \tilde{p}(\phi^+) = 1,\quad  \tilde{p}(\phi^-) = 2 k\hbar T,\quad \tilde{p}(\eta_-) = \frac{1}{2k\hbar}\eta,\quad \tilde{p}(\eta_0) = -\frac{1}{2}\partial\eta,\quad \tilde{p}(\text{others}) = 0\,.
\end{equation}
At this stage, we have constructed the quasi-isomorphism of chain complexes. In principle, however, this should be extendable to an $L_\infty$ quasi-isomorphism of dg-Lie algebras, equipped with the bracket of the bulk algebra. Specifically, the bracket is given by  $\{T,\eta\} = 1, \{\phi^i,\eta_i\} = 1$, and so forth. By establishing this $L_\infty$ quasi-isomorphism, we can then incorporate the bulk interactions associated with the $\lambda$-bracket. We can already verify that under the map $\tilde{p}$, the Maurer-Cartan elements $\phi^0\eta_+\eta_- + \phi^+\eta_+\eta_0 - \phi^-\eta_-\eta_0 + k\hbar(\eta_0\partial\eta_0 + \eta_{+}\partial\eta_-) + \tilde{b}\tilde{c}$ is mapped to the Maurer-Cartan element $T\eta\partial\eta - \frac{k\hbar}{2}\eta\partial^3\eta$ on $\mathbb{C}[\partial^nT,\partial^n\eta]$. While the map $\tilde{i}$ does not do the inverse, this discrepancy is expected to be resolved by the full data of the $L_\infty$ quasi-isomorphism. From this, we can argue for an equivalence between the 3d chiral Poisson sigma model associated with the Virasoro algebra and the DS reduction.

As discussed above, the relationship between the 3d theory associated with the DS reduction and 3d $\mathfrak{sl}_2$ Chern-Simons theory is more direct. The bulk term $\int \tilde{b}\tilde{c}$ indicates that these fields can be integrated out, effectively reducing the system to the 3d $\mathfrak{sl}_2$ CS theory but pushing the DS reduction to the boundary chiral algebra.

\bibliographystyle{JHEP}

\bibliography{mono}

\end{document}